%% file: main.tex
\documentclass[a4paper,11pt]{article}
\DeclareUnicodeCharacter{2212}{\textminus}
\usepackage{jheppub} 
\usepackage{lineno}

\usepackage{amsmath,amssymb,bm}\allowdisplaybreaks[1]
\usepackage{graphicx}
\usepackage{physics, esint}
\usepackage{xcolor}
\usepackage{hyperref}
\usepackage[normalem]{ulem}
\usepackage{adjustbox}
\usepackage{tikz}
\usepackage{tikz-3dplot}
\usetikzlibrary{arrows.meta}
\usepackage{enumitem}

\usepackage[normalem]{ulem}
\usepackage{soul}
\usepackage{multirow}
\usepackage{orcidlink}

\renewcommand{\figurename}{Fig.}
\renewcommand{\sec}[1]{Sec.~\ref{#1}}
\newcommand{\fig}[1]{Fig.~\ref{#1}}

\newcommand{\eq}[1]{Eq.~\eqref{#1}}
\newcommand{\eqs}[2]{Eqs.~\eqref{#1} and~\eqref{#2}}
\newcommand{\eqss}[3]{Eqs.~\eqref{#1}~\eqref{#2} and~\eqref{#3}}
\newcommand{\refcite}[1]{Ref.~\cite{#1}}
\newcommand{\refs}[1]{Refs.~\cite{#1}}

\newcommand{\pp}[1]{\left(#1\right)}
\newcommand{\bb}[1]{\left[#1\right]}
\newcommand{\cc}[1]{\left\{#1\right\}}
\newcommand{\vv}[1]{\left\langle #1 \right\rangle}
\newcommand{\bigpp}[1]{\big(#1\big)}
\newcommand{\bigbb}[1]{\big[#1\big]}
\newcommand{\bigcc}[1]{\big\{#1\big\}}
\newcommand{\bigvv}[1]{\big\langle #1 \big\rangle}
\newcommand{\Bigpp}[1]{\Big(#1\Big)}

\newcommand{\biggpp}[1]{\bigg(#1\bigg)}
\newcommand{\biggbb}[1]{\bigg[#1\bigg]}

\newcommand{\iv}[1]{\left[#1\right]}
\newcommand{\bigiv}[1]{\big[#1\big]}

\newcommand{\beq}[1][]{\begin{equation}\label{#1}}
\newcommand{\eeq}{\end{equation}}
\newcommand{\bse}[1][]{\begin{subequations}\label{#1}}
\newcommand{\ese}{\end{subequations}}
\newcommand{\nn}{\nonumber}

\newcommand{\qf}{{\Lambda_F}}
\renewcommand{\l}{{\ell}}

\renewcommand{\P}{\mathcal{P}}
\newcommand{\Pb}{\bar{\mathcal{P}}}

\newcommand{\LQCD}{ \Lambda_{\rm QCD} }
\newcommand{\M}{\mathcal{M}}

\newcommand{\C}[3]{%
    \mathcal{C}_{#3} \! \big[#1, #2\big]
}
\newcommand{\Cint}[3]{%
    \mathcal{C}'_{#3} \! \big[#1, #2\big]
}

\def\A{\mathcal{A}}
\def\Q{\mathcal{Q}}

\def\M{\mathcal{M}}

\def\P{\mathcal{P}}

\def\E{\mathcal{E}}

\def\s{\hat{s}}
\def\x{\hat{x}}
\def\z{\hat{z}}

\newcommand{\MS}{{\overline{\rm MS}}}

\newcommand{\fec}{\mathcal{D}}
\newcommand{\unpfec}{\mathcal{D}}
\newcommand{\colfec}{\mathcal{H}}

\title{\boldmath Collins-type fragmentation energy correlator in semi-inclusive deep inelastic lepton-hadron scattering}

\author[a,b,c]{Qing-Hong Cao\,\orcidlink{0000-0003-0033-2665},}
\emailAdd{qinghongcao@pku.edu.cn}

\author[d]{Zhite Yu\,\orcidlink{0000-0003-1503-5364},}
\emailAdd{yuzhite@jlab.org}

\author[e]{C.-P. Yuan\,\orcidlink{0000-0003-3988-5048},}
\emailAdd{yuanch@msu.edu}

\author[a,b]{Shutao Zhang\,\orcidlink{0009-0009-5269-2364},}
\emailAdd{shutaozhang@pku.edu.cn}

\author[a,b]{and Hua Xing Zhu\,\orcidlink{0000-0002-7129-6748}}
\emailAdd{zhuhx@pku.edu.cn}

\affiliation[a]{School of Physics, Peking University, Beijing 100871, China}
\affiliation[b]{Center for High Energy Physics, Peking University, Beijing 100871, China}
\affiliation[c]{School of Physics, Zhengzhou University, Zhengzhou 450001, China}
\affiliation[d]{Theory Center, Jefferson Lab, Newport News, Virginia 23606, USA}
\affiliation[e]{Department of Physics and Astronomy, Michigan State University, East Lansing, Michigan 48824, USA}

\date{\today}
\preprint{\begin{tabular}{r}
    CPTNP-2025-036 \\
    JLAB-THY-25-4544 \\
    MSUHEP-24-023
\end{tabular}}

\abstract{%
We initiate a systematic study of fragmentation energy correlators (FECs), which generalize traditional fragmentation functions and encode non-perturbative information about transverse dynamics in parton fragmentation processes. 
We define boost-invariant, non-perturbative FECs and derive a corresponding collinear factorization formula. A spin decomposition of the FECs is carried out, analogous to that of transverse-momentum-dependent fragmentation functions.
In this work we focus particularly on the Collins-type quark FEC, which is sensitive to chiral symmetry breaking and characterizes the azimuthal asymmetry in the fragmentation of a transversely polarized quark. We perform a next-to-leading-order calculation of the corresponding hard coefficient in semi-inclusive deep-inelastic scattering for the quark non-singlet component, thereby validating the consistency of our theoretical framework.
}

\begin{document}

\maketitle
\flushbottom

\section{Introduction}
\label{sec:intro}

Energy-energy correlators (EEC) have emerged as powerful infrared and collinear (IRC) safe observables for probing microscopic details in particle collisions~\cite{Basham:1979gh,Basham:1978zq,Basham:1978bw,Hofman:2008ar}. In modern language, it measures the correlation $\vv{\E(n_i)\E(n_j)}$ between energy deposits in two detectors along two directions, $n_i = (1, \bm{n}_i)$ and $n_j = (1, \bm{n}_j)$ with angular separation $\cos\theta_{ij} = \bm{n}_i \cdot \bm{n}_j$.
The  asymptotic energy flow operator is 
\begin{equation}\label{eq:eec-def}
\E(n) = \lim_{r\rightarrow\infty} r^2 \int_{-\infty}^\infty dt\; T_{0\vec n}(t,\vec n r),
\end{equation}
where $T_{\mu\nu}$ is the energy-stress tensor~\cite{Tkachov:1995kk, Hofman:2008ar, Kravchuk:2018htv}. 
In the collinear region where $\theta_{ij}\rightarrow 0$, the EEC exhibits a perturbative scaling law $\lim_{n_j \to n_i} \vv{\E(n_i)\E(n_j)} \sim 1/\theta_{ij}^{2-\gamma}$, as derived from the factorization theorem and perturbative collinear resummation~\cite{Dixon:2019uzg} or light-ray operator product expansion (OPE)~\cite{Hofman:2008ar,Kologlu:2019mfz,Chen:2023zzh}, where $\gamma$ is the anomalous dimension of twist-2 spin-3 operator and the constant 2 is from classical scaling~\cite{Dixon:2019uzg,Korchemsky:2019nzm, Chang:2020qpj, Chen:2020adz, Chen:2021gdk}. Non-perturbative effects can then be investigated from analyzing deviations from the perturbative scaling behavior~\cite{Schindler:2023cww,Lee:2024esz,Chen:2024nyc}. 

In recent years, there has been a resurgence of interests in the phenomenology applications of energy correlators. 
These include, for instance, a set of novel event shape observables called transverse energy-energy correlators to investigate deep inelastic scattering~\cite{Li:2020bub,Gao:2023ivm}, a precision determination of the strong coupling constant~\cite{CMS:2024mlf} by extracting the scaling behavior near the collinear limit and calculating the ratio of three-point and two-point projected energy correlators~\cite{Chen:2020vvp,Lee:2022uwt,Chen:2023zlx}, 
a potentially precise extraction of the top quark mass through the measurement of multi-point energy correlators on decaying massive states~\cite{Holguin:2023bjf, Holguin:2022epo, Holguin:2024tkz, Xiao:2024rol}, resolving spin structures of electroweak bosons or fragmenting gluons~\cite{Ricci:2022htc, Song:2025bdj}, 
and revealing the hadronization dynamics in quarkonium production by analyzing neighboring radiation patterns~\cite{Chen:2024nfl}. 
We refer to \refcite{Moult:2025nhu} for a complete review of recent applications of energy correlators.

Combining the concepts of energy correlator and parton distribution function (PDF), 
\refs{Liu:2022wop,Cao:2023oef} introduced the nucleon energy correlator~(NEC), offering a new probe for the internal structure of nucleons.
In this paper, building upon the previous study~\cite{Liu:2024kqt}, 
we extend NECs to fragmentation processes and introduce fragmentation energy correlators (FECs).
Unlike the well-established EEC, which sums over asymptotic energy fluxes from all particles in some given directions, 
the FEC represents an ``exclusive'' extension, with one of the energy fluxes in the original EEC replaced by an observed hadron,
and thus becomes a nonperturbative object.
Compared to the standard PDF or fragmentation function (FF), 
the NEC or FEC tags an additional energy flow around the incoming or outgoing hadron, 
and gives an extra handle to probe the nonperturbative transverse-momentum structure 
intrinsic to the hadron or in the fragmentation process.
As established in \refs{Liu:2024kqt, Chen:2024bpj, Zhu:2025qkx}, 
NECs and FECs can be related to transverse-momentum-dependent (TMD) PDFs and FFs~\cite{Collins:1981uw}, respectively,
in terms of their angular integrals.
They can therefore serve as novel tools to uncover complementary information of QCD tomography.

While TMD PDFs and FFs have played a primary role in the investigation of non-perturbative QCD aspects 
such as confinement and chiral symmetry breaking~\cite{Collins:2011zzd, Aidala:2012mv, Collins:1992kk, Collins:1981uw, Metz:2016swz,Angeles-Martinez:2015sea, Boussarie:2023izj},
their phenomenological extractions from experimental data present significant difficulties.
A major challenge stems from the transverse momentum convolutions between pairs of TMD distributions, %
making a unique one-to-one determination rather difficult.%
\footnote{There are efforts to remove such pairwise convolutions 
by using jets instead of hadrons as final-state observables~\cite{Liu:2018trl, Liu:2020dct},
which has been further improved to be recoiling-free~\cite{Fang:2023thw}.}
Furthermore, the measured transverse momentum distributions are largely dominated 
by the perturbative showering, which greatly dilutes the intrinsic confining details~\cite{Qiu:2020oqr}.
In contrast, NECs and FECs bypass these obstacles by entering the observables through collinear factorization.
The non-perturbative information encoded therein can be directly extracted 
from the correlation of the measured energy flow with the scattered or produced hadron,
without any Sudakov factor contamination.
The physical processes and kinematics for their measurements are 
no different from those for the normal PDFs and FFs,
and thus no new experiments need to be carried out;
one just needs to analyze the same events with an additional energy flow weight.

There have already been intriguing explorations of NECs applied to 
gluon saturation~\cite{Liu:2023aqb} and spin physics~\cite{Li:2023gkh, Guo:2024jch, Guo:2024vpe, Mantysaari:2025mht, Gao:2025cwy}. 
In this paper, we focus on the phenomenological study of one-point quark FECs
in the semi-inclusive deep inelastic lepton-hadron scattering (SIDIS).
The SIDIS has been extensively measured at facilities such as
the HERA at DESY~\cite{HERMES:2012uyd, HERMES:2004mhh, ZEUS:2000esx}, 
Jefferson Lab (with CEBAF)~\cite{HallA:2011Qian, HallC:2012Asaturyan, CLAS:2014Gohn, Accardi:2023chb}, 
and COMPASS at CERN~\cite{COMPASS:2007deuteronCS,COMPASS:2015protonCS,COMPASS:2017multiplicities},
and will be further studied at future high-luminosity facilities
such as the EIC~\cite{Accardi:2012qut, AbdulKhalek:2022hcn, AbdulKhalek:2021gbh, Proceedings:2020eah} 
and EicC~\cite{Anderle:2021wcy}.
At leading twist, the quark FEC has an analogous spin structure to the quark TMD FF,
which, for unpolarized hadrons, yields two fundamental non-perturbative functions: 
an unpolarized FEC, $\unpfec_1$, and a transversely polarized FEC, $\colfec^{\perp}_1$.
It is the latter that is of particular interest to us as it is associated with the quark helicity flipping during the fragmentation process
and thus it serves as a sensitive probe of chiral symmetry breaking effect in QCD.
This feature is shared by the well-known Collins function~\cite{Collins:1992kk} for TMD fragmentation,
so we term it Collins-type FEC.
The FEC observable is analogous, in a sense, to the one-point energy correlator in \refcite{Gao:2025evv}; both yield an azimuthal asymmetry that relies on a transversely polarized nucleon target and enables access to the quark transversity PDF. The difference, however, is that by centering around the observed hadron, 
the FEC observable eliminates the need for jet reconstruction, thereby removing any dependence on jet algorithms.
Formally, FECs can be interpreted as an interpolation in the observable space between EECs and dihadron fragmentation function. Very recently there have been active researches on the connection of EECs and dihadron fragmentation through moment integrals~\cite{Lee:2025okn,Chang:2025kgq,Guo:2025zwb,Herrmann:2025fqy,Kang:2025zto,Zhao:2025ogc}. We expect similar analysis to be performed for FECs, for which the NLO QCD correction to dihadron fragmentation is available~\cite{Ju:2025jxi}.

Since the transverse momentum (\(p_{hT}\)) of the final-state hadron produced in the SIDIS process is typically small, we introduce a novel FEC observable in this paper. This observable is defined by integrating over (\(p_{hT}\)), yielding a more inclusive quantity.
To construct this novel FEC observable, we begin by demonstrating the utility of FEC functions within a rigorous theoretical framework, establishing all-order collinear factorization in energy flow-tagged SIDIS.
Following the Trento convention~\cite{Bacchetta:2004jz}, we work in the Breit frame 
and distinguish a number of kinematic regions according to the magnitude of  $p_{hT}$.
The factorization is mostly easily formulated for a large $p_{hT}$.
To preserve the collinear factorization structure when integrating out the kinematic variable \(p_{hT}\) (in defining the proposed FEC observable), we must modify the original FEC definition— which exhibits explicit \(p_{hT}\) dependence—to derive a new formalism. This new formalism, which uses the particles' plus momenta (\(p_i^+\)) instead of energies as the weight factor in the energy-flow observable, 
is Lorentz boost-invariant along the direction of the moving hadron ($h$). 
Moreover, it allows for precise predictions of various azimuthal distributions that are essential for measuring Collins-type EFC observables associated with energy-tagged SIDIS cross sections, after properly addressing the apparent azimuthal singularity that arises as $p_{hT} \to 0$ during the $p_{hT}$ integration.

The reminder of this paper is structured as follows. 
In \sec{sec:relation-tmds}, we first provide the formal definition of the FEC and elaborate on its relationship to TMD FFs, 
and then introduce the modified boost-invariant definition which is advantageous for phenomenological studies. 
We give a detailed discussion of the collinear factorization in \sec{sec:fsidis} for measuring FECs in SIDIS at a large $p_{hT}$.
We organize the observables according to the azimuthal modulations in the Breit frame, mimicking the well-studied SIDIS formalism,
and present the leading-order (LO) partonic hard coefficients.
\sec{sec:fsidis-int-pt} is devoted to addressing the subtleties in the integration over $p_{hT}$ as $p_{hT} \to 0$. 
We perform a next-to-leading-order (NLO) calculation of the hard coefficients
and demonstrate the cancellation of soft and collinear divergences, 
verifying the consistency of the factorization and our prescription for dealing with the azimuthal singularity. 
Finally, in \sec{sec:summary}, we provide a summary of our findings and a discussion of future prospects.

\section{Fragmentation energy correlators}
\label{sec:relation-tmds}

\subsection{Basic definition}
\label{ssec:def-fec}

As introduced in \refcite{Liu:2024kqt}, the fragmentation energy correlator (FEC)
modifies the normal fragmentation function (FF) by inserting an extra energy flow operator, 
\begin{align}
	\fec^{[\Gamma]}_{h/q, 1}(z, \bm{n}; p_h) 
		&= \frac{z}{2N_c} \sum_X \int \frac{dy^-}{2\pi} e^{i p_h^+ y^- / z} 
			\, \Tr\Big[ \Gamma  
        			\bigvv{ 0 | \, W(\infty, y^-; w) \, \psi(y^-) \, \E(\bm{n}) \, | h, X; {\rm out} } 	\nn\\
        &\hspace{13.5em} \times	
			\bigvv{ h, X; {\rm out} | \, \bar{\psi}(0) \, W^{\dag}(\infty, 0; w) \, | 0 } 	
			\Big],
\label{eq:fec-1}
\end{align}
where the tagged hadron $h$ has its momentum $p_h$ boosted along the $z$-axis direction,
carrying a fraction $z$ of the longitudinal momentum of the fragmenting quark. $\Gamma$ represents a gamma matrix for some spin structure.
We define the lightfront coordinate system using two auxiliary light-like vectors,
\beq
	w = (1, 0, 0, -1) / \sqrt{2}, \quad
	\bar{w} = (1, 0, 0, 1) / \sqrt{2},
\eeq
on which any Lorentz vector $V^{\mu}$ can be decomposed as
\beq[eq:lightfront-coordinates]
	V = (V^+, V^-, \bm{V}_T) \equiv (V \cdot w, V \cdot \bar{w}, V^1, V^2).
\eeq
As for the FF, the FEC quantifies the nonlocal parton correlation on the backward light-cone along $w$.
The two quark fields, $\psi(y^-)$ and $\bar{\psi}(0)$, are connected by the Wilson lines along $w$ for gauge invariance.
In \eq{eq:fec-1}, we defined
\beq
	W(b, x; w) \equiv \pp{ W_{ij}(b, x; w) }
	= \P \exp{- i g \int_0^b d\lambda \, w^{\mu} A_{\mu}^a(x + \lambda w) \, (t^a_{ij})}
\eeq
for the straight Wilson line in the fundamental color representation
going from $x$ to $x + b w$ along the direction $w$. 
We will focus on quark FECs in this paper, with $N_c = 3$ in \eq{eq:fec-1}. 
Similar definitions can be given for antiquark and gluon FECs.

We have marked the hadron states in \eq{eq:fec-1} as ``out'' states as required for FFs~\cite{Collins:2011zzd, Collins:1981uw, Collins:1992kk}.
Any hadrons $X$ alongside $h$ from the fragmentation process are inclusively summed, with their phase space integrated over.
However, we anchor in this sum an energy flow measurement along a fixed direction $\bm{n} = (\sin\theta \cos\phi$, $\sin\theta \sin\phi$, $\cos\theta )$,
\beq[eq:energy-flow]
	\E(\bm{n}) \, | X; {\rm out} \rangle
	= \sum_{i \in X} E_i \, \delta^{(2)}(\bm{n} - \bm{n}_i) | X; {\rm out} \rangle,
\eeq
which counts the energy deposit along the $\bm{n}$ direction, 
and where $\delta^{(2)}(\bm{n} - \bm{n}_i) = \delta(\cos\theta - \cos\theta_i) \, \delta(\phi - \phi_i)$.
The polar and azimuthal angles, $\theta$ and $\phi$, are defined with respect to the explicitly measured hadron $h$;
the two transverse directions $x$ and $y$ will be specified in the discussion of factorization in SIDIS process in \sec{ssec:fsidis-kin}.

The energy flow insertion extracts extra information from the fragmentation process
in addition to the longitudinal momentum fraction $z$ of the hadron.
This is encoded in the $\theta$ and $\phi$ dependence of the FEC in \eq{eq:fec-1},
which obviously maps out the energy deposition structure in the transverse directions,
similarly to the TMD FFs~\cite{Liu:2024kqt}.
We will review this aspect in \sec{ssec:fec-tmd}.
In particular, it is the $\phi$ dependence that is of our main interest in this paper.
By the rotational symmetry around $p_h$, a nontrivial $\phi$ dependence occurs only 
when the fragmenting quark carries a transverse spin.

The spinor projector $\Gamma$ in \eq{eq:fec-1} defines the spin structure of the quark.
At the twist-2 level, we have three possibilities, 
$\Gamma = \gamma^+ / 2 \equiv \gamma \cdot w / 2$ for an unpolarized quark, 
$\gamma^+ \gamma_5 / 2$ for a longitudinally polarized quark, 
or $\gamma^+ \gamma^i \gamma_5 / 2$ for a transversely polarized quark,
with $i = 1$ or $2$ being the transverse Lorentz index.
These define various FECs with different azimuthal dependence and for different polarization states of $h$, 
in a way similar to the TMD FFs~\cite{Boussarie:2023izj}.
Since the mostly measured hadrons in practice are pions, kaons, proton, and antiproton, %
we only consider unpolarized $h$ in this paper, leaving studies beyond to future works.
With this constraint, the number of FECs is much reduced by parity,
giving only two possible FECs.

Using similar notations to the TMD FFs~\cite{Boussarie:2023izj}, we have the unpolarized FEC,
\beq[eq:fec-unp]
	\fec_{h/q, 1}^{[\gamma^+ / 2]}(z, \bm{n}; p_h) 
	= \unpfec_{1, h/q}(z, \theta; p_h),
\eeq
for $\Gamma = \gamma^+ / 2$, and the transversity FEC,
\begin{align}
	\fec_{h/q, 1}^{[\gamma^+ \gamma^i \gamma_5 / 2]}(z, \bm{n}; p_h) 
	= \frac{(\z \times \bm{n}_T)^i}{|\bm{n}_T|} \colfec^{\perp}_{1, h/q}(z, \theta; p_h)
	= (-\sin\phi, \cos\phi)^i \colfec^{\perp}_{1, h/q}(z, \theta; p_h),
\label{eq:fec-sT}
\end{align}
for $\Gamma = \gamma^+ \gamma^i \gamma_5 / 2$.
The longitudinally polarized FEC $\fec_{h/q, 1}^{[\gamma^+ \gamma_5 / 2]}$ vanishes by parity.
In \eq{eq:fec-sT}, we defined $\hat{z} = \bm{P}_h / |\bm{P}_h| = (0, 0, 1)$, 
and $\bm{n}_T = \sin\theta \, (\cos\phi, \sin\phi, 0)$ for the transverse component of $\bm{n}$.
The $\colfec^{\perp}_{1, h/q}(z, \theta; p_h)$ is the FEC analogue of the Collins function.
It induces an azimuthal asymmetry in the energy deposit of the fragmentation of a quark 
that carries a transverse spin $\bm{s}_T = s_T (\cos\phi_S, \sin\phi_S)$,
\begin{align}
	&\fec_{h/q, 1}^{[\gamma^+ / 2]}(z, \bm{n}; p_h) +
	s_T^i \, \fec_{h/q, 1}^{[\gamma^+ \gamma^i \gamma_5 / 2]}(z, \bm{n}; p_h) \nn\\
	&\hspace{4em}
	= \unpfec_{1, h/q}(z, \theta; p_h) - s_T \, \colfec^{\perp}_{1, h/q}(z, \theta; p_h) \, \sin(\phi - \phi_S),
\label{eq:fec-sT-phi}
\end{align}
with the ratio $\colfec^{\perp}_{1, h/q} / \unpfec_{1, h/q}$ playing the role of the spin analyzing power.

While this paper mainly focuses on the one-point FEC in \eq{eq:fec-1}, 
we note that the definition can be easily generalized to $N$-point FECs~\cite{Liu:2024kqt},
\begin{align}
	\fec^{[\Gamma]}_{h/q, N}(z, \{ \bm{n}_i \}; p_h) 
		&= \frac{z}{2N_c} \sum_X \int \frac{dy^-}{2\pi} e^{i p_h^+ y^- / z} \,
			\Tr\Big[ 
				\Gamma \, \bigvv{ 0 | \, W(\infty, y^-; w) \, \psi(y^-)
				\nn\\
				& \hspace{1em} \times 
				\E(\bm{n}_1) \cdots \E(\bm{n}_N) | h, X; {\rm out} }
				\bigvv{ h, X; {\rm out} | \, \bar{\psi}(0) \, W^{\dag}(\infty, 0; w) \, | 0 } 
			\Big],
\label{eq:fec-n}
\end{align}
where the dependence on $\{ \bm{n}_i \}_{i=1}^N$ is symmetrized.
These would certainly allow more complicated spin and azimuthal dependence.
For instance, at $N = 2$, the longitudinally polarized FEC 
$\fec^{[\gamma^+ \gamma_5 / 2]}_{h/q, 2}(z, \{ \bm{n}_1, \bm{n}_2 \}; p_h) 
= \hat{z} \cdot (\bm{n_1} \times \bm{n}_2) \, G_{2, h/q}(z, \theta_1, \theta_2; p_h)$ would not vanish~\cite{Efremov:1992pe}.

\subsection{Relation between TMD FF and FEC}
\label{ssec:fec-tmd}

The relation between TMD FFs and FECs has been constructed in \refcite{Liu:2024kqt}
in such a way that the $N$-th transverse momentum moments of the former are 
directly equal to the angular weighted integral of $N$-point FECs.
The basic observation is that in the so-called {\it hadron frame} 
where the measured hadron $h$ is along the $z$ direction,
the transverse momentum $\bm{k}_T$ of the fragmenting quark in the TMD FF
is identical to the sum of transverse momenta of all the remaining particles (in $X$) in the final state.
This motivates one to rewrite the TMD FF as
\begin{align}
    d^{[\Gamma]}_{h/q}(z, -z \bm{k}_T) 
    &= \frac{1}{2z N_c} \sum_X \int \frac{dy^-}{2\pi} e^{i p_h^+ y^- / z }
		\Tr\bigg[ \Gamma \, \bigvv{ 0 | \, W(\infty, y^-; w) \,\psi(y^-)	\nn\\
    &\hspace{1.12em}	\times
		\delta^{(2)}\bigg( \bm{k}_T - \int d\Omega \, \E(\bm{n}) \bm{n}_{T} \bigg) | h, X; {\rm out} }
		\bigvv{ h, X; {\rm out} | \, \bar{\psi}(0) \, W^{\dag}(\infty, 0; w) \, | 0 } \bigg],
\label{eq:tmd-ff}
\end{align}
where the operator definition on the right-hand side is in the hadron frame with $h$ along the $z$ direction,
but the arguments of the TMD FF $d^{[\Gamma]}_{h/q}(z, -z \bm{k}_T)$ on the left-hand side are defined in the {\it parton frame}
such that the fragmenting quark is moving along the $z$ direction.
We follow the standard normalization that 
$d^{[\Gamma]}_{h/q}(z, \bm{p}_T) \, dz \, d^2\bm{p}_T$ 
stands for the hadron number in the phase space $[z, z+dz] \times [\bm{p}_T, \bm{p}_T + d^2\bm{p}_T]$.
The connection between the two frames is given by $\bm{p}_T = -z \bm{k}_T$.
Clearly, the operator 
$\int d\Omega \, \E(\bm{n}) \bm{n}_{T} \equiv \int d\cos\theta \, d\phi \, \E(\bm{n}) \sin\theta (\cos\phi, \sin\phi)$ 
measures the total transverse momentum of the particles in $X$,
giving meaning to the $\bm{k}_T$ in this new representation of the TMD FF.

Comparing \eqs{eq:tmd-ff}{eq:fec-n}, one can see that 
the $\bm{k}_T$ moments of \eq{eq:tmd-ff} bring out the energy flow factors
and become the angular integrals of \eq{eq:fec-n},
\begin{align}
	z^2 \int d^2 \bm{k}_T \, k_T^{i_1} \cdots k_T^{i_N} \, 
		d_{h/q}^{[\Gamma]}(z, -z \bm{k}_T)	
	= \int d\Omega_1 \cdots d\Omega_N \, n_{1T}^{i_1} \cdots n_{NT}^{i_N} \, 
		\fec^{[\Gamma]}_{h/q, N}(z, \{ \bm{n}_i \}; p_h),
\label{eq:tmd-fec}
\end{align}
where $i_j = 1, 2$ for $j = 1, \cdots, N$ are the transverse index.
We note that this relation is based on the bare definitions of the TMD FFs and FECs
and is susceptible to renormalization.
However, neither side contains rapidity divergences;
the $\bm{k}_T$ integral on the left-hand side cancels the rapidity divergences of the bare TMD FFs,
while the $N$-point FECs on the right-hand side are free of rapidity divergences in the first place.

\eq{eq:tmd-fec} also indicates the difference between TMD FFs and FECs in the transverse momentum structure of the fragmentation.
Taking both in the hadron frame, 
the $\bm{k}_T$ of the TMD FF sums over all the hadrons globally in the phase space,
whereas the energy flow in an FEC operates locally along a certain direction $\bm{n}$.
On the other hand, because of the linear dependence of $\E(\bm{n})$ on the momentum---due to the constraint of IRC safety,
the $k_T$ moments of the same TMD FF at different orders map out 
a whole tower of FECs of different points, $N$, of energy correlators.
For example, taking $N = 2$, we see that the $k_T^2$ moment of a TMD FF
corresponds to a 2-point FEC,
\begin{align}
	z^2 \int d^2 \bm{k}_T \, k_T^2 \, 
		d_{h/q}^{[\Gamma]}(z, -z \bm{k}_T)	
	= \int d\Omega_1 \, d\Omega_2 \, ( \bm{n}_{1T}  \cdot \bm{n}_{2T} ) \, 
		\fec^{[\Gamma]}_{h/q, 2}(z, \{ \bm{n}_1, \bm{n}_2 \}; p_h),
\label{eq:tmd-fec-2}
\end{align}
which cannot be constructed out of a 1-point FEC.

If a TMD FF could be reconstructed from the complete set of its $\bm{k}_T$ moments,
then combining the right-hand side of \eq{eq:tmd-fec} for all values of $N$ would fully recover the TMD FF.
In this sense, the whole set of FECs encapsulates the transverse momentum structure encoded in TMD FFs.
However, in practice, it is impossible to obtain FECs for all $N$.
With only a limited number of finite-point FECs available, the constraints imposed by \eq{eq:tmd-fec} are inherently restricted.
Consequently, FECs and TMD FFs provide complementary information of the fragmentation process.

The $N = 0$ case of \eq{eq:tmd-fec} gives a trivial normalization condition of the TMD FF.
For $N = 1$, there is only one relation,
\begin{align}
	z^2 \int d^2 \bm{k}_T \, k_T^j \, 
		d_{h/q}^{[\Gamma]}(z, -z \bm{k}_T)	
	= \int d\Omega \, n_{T}^j \, 
		\fec^{[\Gamma]}_{h/q, 1}(z, \bm{n}; p_h).
\label{eq:tmd-fec-1}
\end{align}
Both sides are nonvanishing only for $\Gamma = \gamma^+ \gamma^i \gamma_5 / 2$.
Thus it relates the Collins-type FEC to the Collins function, which is defined as
\begin{align}
	d_{h/q}^{[\gamma^+ \gamma^i \gamma_5 / 2]}(z, -z \bm{k}_T)
	= H^{\perp}_{1, h/q}(z, z k_T) \frac{\epsilon^{ij}_{\perp} k_T^j}{m_h},
\label{eq:collins-ff}
\end{align}
with $m_h$ being the mass of $h$.
Combining this and \eq{eq:fec-sT}, we then get the relation implied by \eq{eq:tmd-fec-1},
\beq[eq:relation-collins]
	\int_0^{\pi} d\theta \, \sin^2\theta \, \colfec^{\perp}_{1, h/q}(z, \theta; p_h)
	= - z^2\int \frac{dk_T^2}{2m_h} k_T^2 \, H^{\perp}_{1, h/q}(z, z k_T).
\eeq
In this work, the FECs we define are distinguished from TMD FF by using a script-style symbol. Readers can also differentiate between them based on their respective arguments.
The right-hand side can also be matched onto a twist-3 FF~\cite{Yuan:2009dw}. The definition in the LHS of \eq{eq:relation-collins} is frame dependent. In the next section we shall present a frame-independent definition. 

\subsection{Modified definition with boost invariance}
\label{ssec:mEEC-strategy}

As first pointed out in \refcite{Liu:2022wop}, 
an energy flow can be measured individually in place of a jet
or together with a hadron nearby in the angular region.
In the former case, the operator $\E(\bm{n})$ defined originally in \eq{eq:eec-def} is best suited for $e^+e^-$ collisions in the CM frame,
where energy and angle are suitable kinematic observables~\cite{Bossi:2025xsi, Bossi:2024qeu}. 
When applied to lepton-hadron or hadron-hadron collisions,
it is convenient to replace energy by transverse momentum $p_T$ with respect to the collision axis, and use pseudo-rapidity $\eta$ instead of $\theta$,
as adopted in \refs{CMS:2024mlf, Tamis:2023guc, STAR:2025jut, ALICE:2024dfl}.
In this case, the $p_T$ weight suppresses soft-hadron contributions and justifies the use of perturbative QCD.

In this paper, we consider the FEC as measured in semi-inclusive deep inelastic scattering (SIDIS) of an electron off a proton,
\beq[eq:sidis]
	e(\ell) + p(p) \to e(\ell') + h(p_h) + \E(\bm{n}) + X(p_X).
\eeq
We require the energy flow to be measured near the observed hadron $h$,
so that it can no longer be described by perturbative QCD, but more closely related to nonperturbative fragmentation dynamics.
Then, this process can be factorized~\cite{Liu:2022wop, Liu:2024kqt} into 
an FEC as defined in \eq{eq:fec-1},
which combines the hadron and energy flow in a single nonperturbative object, as will be described in \sec{ssec:fsidis-factorization}.

A disadvantage of the original FEC definition in \eq{eq:fec-1} is its reliance on the energy $E$ and polar angle $\theta$, which introduces an explicit dependence on the magnitude of the hadron momentum $p_h$, as noted in \eq{eq:fec-1}, in addition to its dependence on $z$ and angular variables. This complicates the extraction of the nonperturbative FEC function from a global analysis across experiments conducted at different collider energies. In SIDIS, the reference frame used for analysis typically varies—from the lab frame to the Breit frame—each subject to arbitrary longitudinal boosts. As a result, the FECs differ even when evaluated at the same set of $z$ and $\theta$ values. Furthermore, this definition requires full knowledge of the hadron’s kinematics; partial integration over its transverse momentum $\bm{p}_{hT}$ at the cross-section level would cause mixing among different FECs, thereby obstructing a clean extraction of individual FEC function.

To resolve this issue, we propose a modified definition of the FEC with a nice boost property along the $p_h$ direction,
by replacing the energy flow operator in \eq{eq:energy-flow} with
\begin{align}
	\E_L(\eta, \phi) |X; {\rm out} \rangle &= \sum_{i \in X} p_i^+\, \delta(\eta - \eta_i) \delta(\phi - \phi_i) |X; {\rm out} \rangle,
\label{eq:mEEC}
\end{align}
where $\eta_i = -\ln[ \tan(\theta_i / 2) ]$, $\phi_i$, and $p_i^+ = (E_i + p_i^z) / \sqrt{2}$
are the pseudo-rapidity, azimuthal angle, and plus momentum of the $i$-th particle, respectively, 
with respect to the $h$ direction.
The reason for using $p_i^+$ as the momentum weighting, instead of the transverse momentum $p_{iT}$ as used in \refs{Lee:2006nr, Bauer:2008dt},
is that in the collinear region that we are considering, 
$p_{iT}$ can be of the same order as the hadronic mass scale 
so that nonperturbative corrections might not be negligible.  
In contrast, the $p_i^+$ weight suppresses contributions from soft momenta and hadron masses,
justifying the use of pseudo-rapidity $\eta$ in place of rapidity $y$ in the leading-power accuracy.%
\footnote{Of course, this requires $h$ to be highly boosted, a condition easily satisfied in practice.}
Therefore, by neglecting the masses of hadrons in the energy flow measurement, 
the $\E_L(\eta, \phi)$ transforms under a longitudinal boost $U(\alpha)$ along the $h$ direction as
\beq[eq:EL-boost]
    U(\alpha) \, \E_L(\eta, \phi) \, U^{-1}(\alpha) = e^{-\alpha} \E_L(\eta + \alpha, \phi),
\eeq
where $\alpha$ is defined such that any momentum $p_i$ 
transforms under $U(\alpha)$ by $p_i^+ \to e^{\alpha} p_i^+$.

In this way, we redefine the FEC as
\begin{align}
	\fec^{[\Gamma]}_{h/q, 1}(z, \qf, \phi) 
		&= \frac{z}{2N_c} \sum_X \int \frac{dy^-}{2\pi p_h^+} e^{i p_h^+ y^- / z} 
			\, \Tr\Big[ \Gamma  
        			\big\langle 0 | \, W(\infty, y^-; w) \, \psi(y^-) \nn\\ 
        &\hspace{5em} \times	
        		\E_L(\eta, \phi) \, | h, X; {\rm out} \big\rangle 
			\bigvv{ h, X; {\rm out} | \, \bar{\psi}(0) \, W^{\dag}(\infty, 0; w) \, | 0 } 	
			\Big],
\label{eq:fec-modify}
\end{align}
where an additional factor $1 / p_h^+$ is inserted to compensate for the factor $e^{-\alpha}$ in \eq{eq:EL-boost};
it also makes the FEC dimensionless. 
\eq{eq:fec-modify} is boost invariant along the $h$ direction and therefore depends on $\eta$ only through its difference from the rapidity $y_h$ of $h$.
This dependence is explicitly captured in the newly defined variable,
\beq[eq:qf]
	\qf = m_h \, e^{y_h - \eta} = \sqrt{2} \, p_h^+ e^{-\eta}.
\eeq

Analogous to \eqs{eq:fec-unp}{eq:fec-sT},
we have new definitions of the unpolarized FEC, $\unpfec_{1, h/q}(z, \qf)$, and the Collins-type FEC, $\colfec^{\perp}_{1, h/q}(z, \qf)$,
corresponding to $\Gamma = \gamma^+ / 2$ and $\gamma^+ \gamma^i \gamma_5 / 2$, respectively.
Crucially both functions are now boost invariant and depend only on $z$ and $\qf$.

As noted in \eq{eq:mEEC}, the light-cone momentum component $p_i^+$ is related to energy $E_i$ via
$p_i^+ = E_i (1 + \tanh \eta_i ) / \sqrt{2}$,
and $\E_L(\eta, \phi)$ can be expressed in terms of the original energy flow operator $\E(\bm{n})$ in \eq{eq:eec-def} as
\beq[eq:relate-EL]
    \E_L(\eta, \phi) = \frac{1 + \tanh \eta}{\sqrt{2} \cosh^2\eta} \, \E(\theta, \phi),
\eeq
where the denominator arises from the Jacobian factor associated with the change of variables from 
$\delta(\cos\theta-\cos\theta_i)$ in \eq{eq:energy-flow} to $\delta(\eta-\eta_i)$ in \eq{eq:mEEC}.
Then the $\fec^{[\Gamma]}_{h/q, 1}(z, \qf, \phi)$ in \eq{eq:fec-modify} 
can be related to $\fec^{[\Gamma]}_{h/q, 1}(z, \bm{n}; p_h)$ in \eq{eq:fec-1} \footnote{Although the same symbol is used, the quantities in \eq{eq:fec-modify} and \eq{eq:fec-1} are distinct definitions, related by the non-trivial factor in \eq{eq:relate-two-fec} rather than a simple change of variables. They are distinguished by their arguments.}
by
\beq[eq:relate-two-fec]
	\fec^{[\Gamma]}_{h/q, 1}(z, \qf, \phi )
		= \frac{1 + \tanh \eta}{\sqrt{2} \, p_h^+ \cosh^2\eta} \, \fec^{[\Gamma]}_{h/q, 1}(z, \bm{n}; p_h).
\eeq

It is well known that in the collinear limit, the product of energy and angle, $E_i \theta_i$, is approximately Lorentz invariant.
Thus, we can gain a better insight of the variable $\qf$ by examining the small-$\theta$ limit, which corresponds to large $\eta$.
This gives
\beq
	\theta \simeq \frac{1}{\cosh\eta}, \quad
	\qf \simeq \frac{p_h^+ \theta}{\sqrt{2} }, \quad
    \fec^{[\Gamma]}_{h/q, 1}(z, \qf, \phi )
		\simeq \frac{\theta^3}{\qf} \, \fec^{[\Gamma]}_{h/q, 1}(z, \bm{n}; p_h).
\eeq
Therefore, $\qf$ serves as an effective transverse momentum scale between $h$ and the energy flow,
taking over the role of $\theta$ in probing the transverse structure of parton fragmentation.
The FEC originally defined in \eq{eq:fec-1} transforms as $1/ \theta^3$ under a longitudinal boost, while the new definition in \eq{eq:fec-modify} is boost invariant.

Following the same spirit of \sec{ssec:fec-tmd}, we can connect the boost-invariant FECs to TMD FFs.
This can be simply done by using the relation,
\beq
	\int d\cos\theta \, = \int \frac{d\qf}{\left|\bm{n}_T\right|} \frac{1 + \tanh \eta}{\sqrt{2} \, p_h^+ \cosh^2\eta} ,
\eeq
where $\left|\bm{n}_T\right| = \sin\theta = 1 / \cosh\eta$.
Applying this to \eq{eq:relation-collins} then gives
\beq[eq:relation-collins-2]
	\int d\qf \, \colfec_{1, h/q}^{\perp}(z, \qf) 
		= -z^2\int \frac{dk_T^2}{2m_h} \, k_T^2 \, H^{\perp}_{1, h/q}(z, z k_T),
\eeq
where the left-hand side is the boost-invariant Collins-type FEC. This relation constitutes one of the central results of our work.

\section{FECs from energy-tagged SIDIS: finite-$p_{hT}$ case}
\label{sec:fsidis}

The FECs can be measured in any inclusive hadron production process at lepton-lepton, lepton-hadron, or hadron-hadron collisions,
simply by recording the energy flow surrounding the measured hadron. 
In this section, we consider such new observables in SIDIS at lepton-hadron collision.
We will derive a collinear factorization formalism that differs from that of SIDIS only by replacing the normal FF by the FEC.
The boost-invariant FEC definitions in \sec{ssec:mEEC-strategy} will be employed in our analysis.
Similar studies can be performed for lepton-lepton and hadron-hadron collisions.%

\subsection{Kinematics}
\label{ssec:fsidis-kin}
\begin{figure}[htbp]
	\centering
	\includegraphics[scale=1.1]{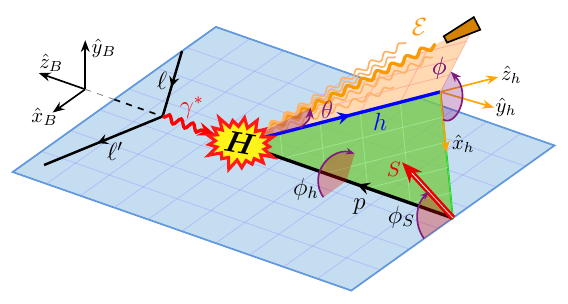}
	\caption{Breit frame for the energy-tagged SIDIS in \eq{eq:sidis}.}
	\label{fig:frame}
\end{figure}

To the leading order in QED, the electron scatters with the hadron in \eq{eq:sidis} by exchanging a virtual photon $\gamma^*(q = \l - \l')$,
\bse[eq:two-stage]\begin{align}
	e(\ell) & \to e(\ell') + \gamma^*(q), \label{eq:two-stage-a} \\
	p(p) + \gamma^*(q) & \to h(p_h) + \E(\bm{n}) + X(p_X), \label{eq:two-stage-b}
\end{align}\ese
where the $\gamma^*$ carries a hard virtuality $Q^2 = -q^2$ much beyond the hadronic mass scale,
and $X$ denotes anything else in the final state.
We consider the most general polarization scenario for both the beam and target, and
describe the kinematics in the Breit frame, as displayed in \fig{fig:frame},
which is defined as the collision frame between 
the proton $p$ and the $\gamma^*$,
with $p$ boosted in the $\hat{z}_B$ direction.
The electron scattering happens in the $\hat{x}_B$-$\hat{z}_B$ plane,
in which we fix the $\hat{x}_B$ axis such that the initial-state electron momentum has a positive $x$ component, $\ell^x > 0$.

Since we work only to the leading power of $1/Q$, we may neglect electron and hadron masses for simplicity
and parametrize the initial-state kinematics as
\begin{align}
	p & = \pp{ p^+,0^-, \bm{0}_T }, \quad
	q = \biggpp{ -x \, p^+, \frac{Q^2}{2 x \, p^+}, \bm{0}_T }, \nn\\
	\ell &= \biggpp{ \frac{x (1 - y) }{y} p^+, \frac{Q^2}{2 x y p^+}, \frac{Q \sqrt{1 - y}}{y}, 0 },
\end{align}
which are written in terms of lightfront coordinates as defined in \eq{eq:lightfront-coordinates} and
where 
\beq[eq:dis-kin-var]
x = \frac{Q^2}{2 p \cdot q}, \quad
y = \frac{p \cdot q}{p \cdot \ell} = \frac{Q^2}{x \, s}.
\eeq
By neglecting the proton mass, the transformation of the azimuthal angle $\phi_S$ of the target transverse spin $\bm{s}_T$ becomes trivial~\cite{Diehl:2005pc};
the $\phi_S$ in the Breit frame varies from event to event and is equivalent to the azimuthal angle of the electron scattering in the lab frame.
The momentum $p_h$ of the final-state hadron $h$ is parametrized using 
its transverse momentum $\bm{p}_{hT} = p_{hT} (\cos\phi_h, \sin\phi_h)$, 
with $p_{hT} = | \bm{p}_{hT} |$ and $\phi_h$ being its azimuthal angle, 
and
\beq[eq:sidis-kin-var]
z = \frac{p \cdot p_h}{p \cdot q} = \frac{p_h^-}{q^-}.
\eeq

These are the same as the SIDIS kinematics.
Our analysis introduces a new observable: the measurement of the energy flow surrounding the final-state hadron $h$. The azimuthal distribution of the energy flow is of primary importance, as it directly probes the polarization of the fragmenting parton.
To quantify this observable, the cross section is weighted by the factor $p_i^+\, \delta(\eta - \eta_i) \delta(\phi - \phi_i)$
in \eq{eq:mEEC}, summed over each final-state particle $i$.
As mentioned there, the $(p_i^+, \eta_i, \phi_i)$ are defined relative to the $h$ direction, 
so we define the hadron coordinate system,
\beq[eq:hadron-frame]
\hat{z}_h = \frac{\bm{p}_h}{|\bm{p}_h|}, \quad
\hat{y}_h = \frac{\hat{z}_B \times \hat{z}_h}{|\hat{z}_B \times \hat{z}_h|}, \quad
\hat{x}_h = \hat{y}_h \times \hat{z}_h,
\eeq
and the energy flow measurements are also performed with respect to this system.
Within this frame, the azimuthal angle $\phi$ is therefore the central variable.
It is crucial to note the distinction in reference frames that this introduces. The light-cone component $p_i^+$ used for the energy flow is defined within this hadron frame. This contrasts with the light-cone coordinates used for the global SIDIS kinematics (e.g., $q^-$ in \eq{eq:sidis-kin-var}), which are defined in the Breit frame.

Without measuring the energy flow, the SIDIS cross section can be expressed in terms of $(x, Q^2, \phi_S, z, p_{hT}^2, \phi_h)$,
\begin{align}
	\frac{d\sigma}{dx \, dQ^2 \, d\phi_S \, dz \, dp_{hT}^2 \, d\phi_h}
	= \frac{y^2}{z Q^2} \frac{\alpha_e^2}{8 (2\pi)^4} \overline{|\M|^2},
	\label{eq:sidis-xsec}
\end{align}
where $\alpha_e = e^2 / (4\pi)$.
Let us denote $\M^X = \M^X_{\lambda_e \lambda_p}$ as the amplitude of 
\beq
e(\ell, \lambda_e) + p(p, \lambda_p) \to e(\ell') + h(p_h) + X(p_X),
\eeq
with the QED coupling $e^2$ factored out
and with $\lambda_e$ and $\lambda_p$ being the helicities of the initial-state electron and proton.
The $\overline{|\M|^2}$ is obtained by integrating out the $X$ phase space in the spin-averaged amplitude squared,
\beq[eq:amp2-sidis]
\overline{|\M|^2}
= \sum_X \sum_{\lambda_e, \lambda_p, \bar{\lambda}_e, \bar{\lambda}_p} 
\int d\Pi_X \, (2\pi)^4 \delta^{(4)}(p + \ell - \ell' - p_h - p_X) 
\bb{ 
	\rho^e_{\lambda_e \bar{\lambda}_e} \rho^p_{\lambda_p \bar{\lambda}_p} 
	\M^X_{\lambda_e \lambda_p} \M_{\bar{\lambda}_e \bar{\lambda}_p}^{X*}
},
\eeq
where the density matrices $\rho^e$ and $\rho^p$ for the beam and target, respectively, 
are defined in the helicity basis and can be written in the Breit frame as
\begin{align}
	\rho^e = \rho^e(P_e)
	= \frac{1}{2} \begin{pmatrix}
		1 + P_e & 0 \\
		0 & 1 - P_e
	\end{pmatrix}, \quad
	\rho^p = \rho^p(P_N, s_T, \phi_S)
	= \frac{1}{2} \begin{pmatrix}
		1 + P_N & s_T \, e^{-i \phi_S} \\
		s_T \, e^{i \phi_S} & 1 - P_N
	\end{pmatrix}.
	\label{eq:den-mtx-ep}
\end{align}
The $P_e$ refers to the longitudinal polarization degree of the beam,
and $P_N$ and $s_T$ the longitudinal and transverse polarization degrees, respectively, of the target;
their numerical values are the same between the lab and the Breit frames in the massless approximation.
We have omitted the helicity indices of final-state particles in \eq{eq:amp2-sidis}, which are implicitly summed over.

Now with the energy flow measured, we replace the $\overline{|\M|^2}$ in \eq{eq:sidis-xsec} by $\overline{|\M_{\E}|^2}$,
which modifies \eq{eq:amp2-sidis} with an additional factor 
\beq[eq:energy-flow-factor]
\sum\nolimits_{i \in X} (p_i^+ / p_h^+) \, \delta(\eta - \eta_i) \, \delta(\phi - \phi_i)
\eeq
inserted in the $\Pi_X$ integral,
where $p_i^+$ and $p_h^+$ are defined as in \eq{eq:mEEC}.
This introduces two additional kinematic observables $(\eta, \phi)$ into \eq{eq:sidis-xsec},
and results in an energy flow-weighted cross section,
\begin{align}
	\frac{d\Sigma}{dx \, dQ^2 \, d\phi_S \, dz \, dp_{hT}^2 \, d\phi_h \, d\eta \, d\phi}
	= \frac{y^2}{z Q^2} \frac{\alpha_e^2}{8 (2\pi)^4} \overline{|\M_{\E}|^2},
	\label{eq:sidis-xsec-e}
\end{align}
where we note that using \eq{eq:qf}, the $d\eta$ can be also written as $d \ln \qf$.

\subsection{Fatorization}
\label{ssec:fsidis-factorization}
We consider the SIDIS in the current fragmentation region, 
with $z$ being of order unity, $z\sim \mathcal{O}(1)$.
When we measure the hadron's transverse momentum $p_{hT}$, as indicated in \eq{eq:sidis-xsec-e},
it needs to be of the same order as $Q$ in order to allow collinear factorization of \eq{eq:sidis} into the FEC.
We focus on this kinematic region in this section and will consider integrating over $\bm{p}_{hT}$ in \sec{sec:fsidis-int-pt}.

Without measuring an energy flow in the final state, 
the standard factorization of SIDIS cross section in \eq{eq:sidis-xsec} has been established
as the convolution of parton distribution functions (PDFs), FFs, and hard coefficients~\cite{Collins:2011zzd},
\begin{align}
	\overline{|\M|^2}
	&\simeq 
	\sum_{a, b} \int \frac{d\xi_1}{\xi_1} \, \int \frac{d\xi_2}{\xi_2^2} \, D_{h/b}(\xi_2, \mu)
	\bb{ f_{a/p}(\xi_1, \mu) \, C_{ab} \pp{ \frac{x}{\xi_1}, \frac{z}{\xi_2}; \frac{\bm{p}_{hT}}{Q}, \frac{Q^2}{\mu^2} }	\right.\nn\\
		&\hspace{14em} \left. {}
		+ P_N \, g_{a/p}(\xi_1, \mu) \, \Delta C_{ab} \pp{ \frac{x}{\xi_1}, \frac{z}{\xi_2}; \frac{\bm{p}_{hT}}{Q}, \frac{Q^2}{\mu^2} }
	} \nn\\
	&\hspace{2em} 
	+ \order{ \LQCD / Q, \LQCD / p_{hT} },
	\label{eq:sidis-factorization}
\end{align}
where the sums over $a$ and $b$ run over parton flavors. 
The unpolarized and polarized quark PDFs, $f_{q/p}$ and $g_{q/p}$, are defined as
\bse[eq:PDFs] \begin{align}
	f_{q/p}(\xi_1) 
	&= \int_{-\infty}^{\infty} \frac{d\lambda}{4\pi} \, e^{-i \lambda \, \xi_1 \, p\cdot u} 
	\bigvv{ p, S | \bar{\psi}(\lambda u) \, \gamma\cdot u \, \psi(0) | p, S } , \\
	P_N \, g_{q/p}(\xi_1)
	&= \int_{-\infty}^{\infty} \frac{d\lambda}{4\pi} \, e^{-i \lambda \, \xi_1 \, p\cdot u} 
	\bigvv{ p, S | \bar{\psi}(\lambda u) \, \gamma\cdot u \gamma_5 \, \psi(0) | p, S },
\end{align} \ese
where $u = (1, - \hat{z}_B) / \sqrt{2}$ is the lightlike vector opposite to the proton momentum direction (cf.~\fig{fig:frame}),
and $S$ is the proton spin vector.
The FF $D_{h/q}$ of a quark is defined similarly to \eq{eq:fec-1}, just with the energy flow operator $\E(\bm{n})$ removed.
Since we consider unpolarized hadron production, only the unpolarized FF is allowed, which sets $\Gamma = \gamma^+ / 2$ in \eq{eq:fec-1}.
Definitions of gluon PDFs and FFs are similarly given in \refcite{Collins:2011zzd}.
The hard coefficients $C_{ab}$ and $\Delta C_{ab}$ correspond to the electron scattering off an unpolarized and polarized parton $a$, respectively,
and inclusively producing an unpolarized parton $b$,
\beq[eq:sidis-hard]
e(\ell) + a(\xi_1 p) \to e(\ell') + b(p_h / \xi_2) + X(p_X).
\eeq

Now we examine the factorization structure when an energy flow is measured around the hadron $h$.
Most of the analysis is identical to the SIDIS~\cite{Collins:2011zzd} so not be repeated.
In the following, we only remark on the differences caused by the energy flow factor in \eq{eq:energy-flow-factor}.

\begin{figure}[htbp]
	\centering
	\begin{tabular}{cc}
		\includegraphics[scale=0.6]{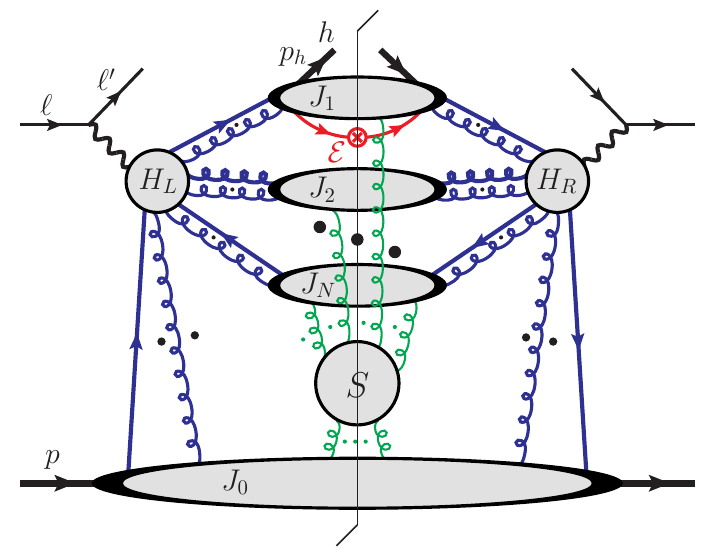} &
		\includegraphics[scale=0.6]{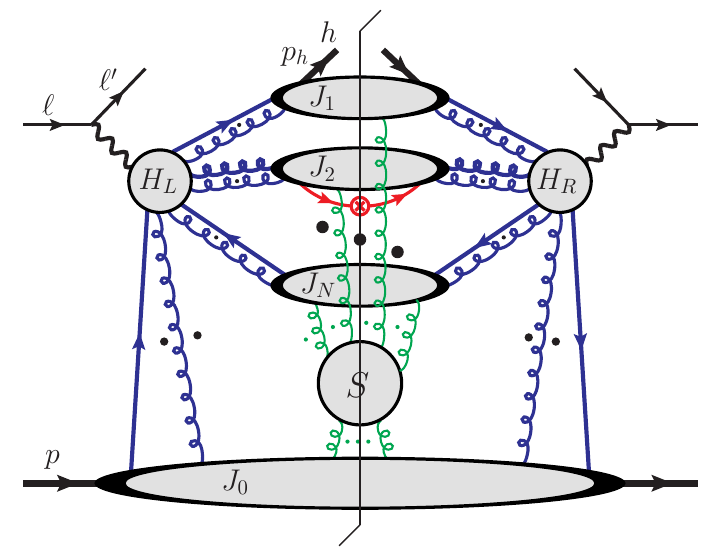} \\
		(a) $R_C$ & (b) $R_H$
	\end{tabular}
	\caption{Leading regions (a) $R_C$ and (b) $R_H$ of the energy-tagged SIDIS in \eq{eq:two-stage}. 
		The cross vertex ($\otimes$) indicates the energy flow,
		which is in the same collinear subgraph $J_1$ as the measured hadron $h$ in $R_C$,
		but is in a separate collinear subgraph $J_2$ in $R_H$.
		We are using cut diagram notation here:
		to the left of the cut is the amplitude $\M^X$, and to the right is its complex conjugate.
		Lines crossed by the cut stand for the final-state particles; 
		they are put on shell with phase space integrated out, except for the observed hadron $h$ 
		and energy flow weighting factor.}
	\label{fig:region}
\end{figure}

Using the terminology of \refcite{Collins:2011zzd}, 
the leading regions contributing to $\overline{|\M_{\E}|^2}$, in the limit of $Q \sim p_{hT} \gg \LQCD$,
are shown in \fig{fig:region} by their corresponding reduced diagrams.
The virtual photon $\gamma^*(q)$ initiates a hard interaction subgraph $H_L$ ($H_R$) to the left (right) of the cut,
which is joined to the collinear subgraph $J_0$ associated with the proton by 
a physically polarized parton line on either side of the cut, along with an arbitrary number of longitudinally polarized gluon lines.
Out of the hard subgraph are produced one or more set of collinear lines, 
collected by the collinear subgraphs $J_1$, $\cdots$, $J_N$ $(N \geq 1)$,
where we choose the observed hadron $h$ to emerge from $J_1$, whose direction is defined by $p_h$.
Depending on whether the energy flow is measured in the same collinear subgraph as $h$, 
we distinguish the leading regions into types $R_C$ and $R_H$.
On top of these, there are arbitrarily many soft gluons, 
collected by the soft subgraph $S$, connecting all these collinear subgraphs.

\subsubsection{Fatorization of $R_C$}
\label{sssec:Rc-factorization}
The factorization of region $R_C$ is exactly the same as of SIDIS
and also shares the similar structure as the dihadron production~\cite{Collins:1993kq}.
For each collinear subgraph, we define two auxiliary lightlike vectors $\bar{w}_i$ and $w_i$,
\beq[eq:ww-J]
\bar{w}_i = (1, \bm{n}_i) / \sqrt{2}, \quad
w_i = (1, -\bm{n}_i) / \sqrt{2}, \quad
\bar{w}_i \cdot w_i = 1,
\eeq
where $\bm{n}_i$ is the unit vector along the direction of $J_i$, for $i = 0, \cdots, N$.
[Compared with \eq{eq:PDFs}, we have defined $w_0 = u$ and $\bar{w}_0 = \bar{u}$.]
Using \eq{eq:ww-J}, we approximate each collinear momentum $k_i$ of $J_i$ by its large component $\hat{k}_i$
when it flows in the hard subgraph,
\beq[eq:p-projection]
k_i^{\mu} \mapsto \hat{k}_i^{\mu} = (k_i \cdot w_i) \, \bar{w}_i^{\mu}.
\eeq
At the leading power, each $J_i$ is connected to the hard subgraph $H_L$ or $H_R$ by only one quark line or physically polarized gluon.
We insert the spinor projector
\beq[eq:spinor-projector]
\P_i = \frac{(\gamma \cdot w_i) (\gamma \cdot \bar{w}_i) }{2} 
\quad 
\biggbb{ \mbox{or} \quad
	\Pb_i = \frac{(\gamma \cdot \bar{w}_i) (\gamma \cdot w_i)}{2} }
\eeq
for each quark line leaving (or entering) the hard subgraph into (or from) the collinear subgraph $J_i$.
This extracts the leading spinor component and projects on shell the quark lines external to the hard part.
Similar projectors are defined for physically polarized gluons~\cite{Collins:2011zzd}.

Along with these can be arbitrarily many collinear gluons, but with scalar polarizations proportional to their momenta.
This fact allows the use of Ward identity to factorize them out of the hard subgraph onto a Wilson line along $w_i$ (on either side of the cut)
for each collinear subgraph $J_i$.
Similarly, the leading polarization components of all soft gluons attached to $J_i$ are proportional to $\bar{w}_i$,
which allows them to be factorized from the collinear subgraphs onto Wilson lines along $\bar{w}_i$.
A simple unitarity argument then shows that the soft Wilson lines are reduced to unity, 
while all the collinear pinch singularities associated with $J_2$, $\cdots$, $J_N$ are cancelled,
such that their integration contours can be deformed to make themselves into the hard region.

This results in a factorization formula that differs from \eq{eq:sidis-factorization} by replacing the FF with the FEC,
\begin{align}
	\overline{|\M_{\E}|^2}
	&\simeq 
	\sum_{a, b} \int \frac{d\xi_1}{\xi_1} \, \int \frac{d\xi_2}{\xi_2^2} \bigg\{ \unpfec_{1, h/b}(\xi_2, \qf, \mu)
	\bb{ f_{a/p}(\xi_1, \mu) \, C_{ab} \pp{ \frac{x}{\xi_1}, \frac{z}{\xi_2}; \frac{\bm{p}_{hT}}{Q}, \frac{Q^2}{\mu^2} }	\right.\nn\\
		&\hspace{16.5em} \left. {}
		+ P_N \, g_{a/p}(\xi_1, \mu) \, \Delta C_{ab} \pp{ \frac{x}{\xi_1}, \frac{z}{\xi_2}; \frac{\bm{p}_{hT}}{Q}, \frac{Q^2}{\mu^2} }
	}	\nn\\
	&\hspace{0.6em}  {}
	+ \colfec^{\perp}_{1, h/b}(\xi_2, \qf, \mu) \, h_{a/p}(\xi_1, \mu)
	\sum_{i, j = 1}^2 \frac{(\hat{p}_h \times \bm{n}_T)^i}{|\bm{n}_T|} 
	\, T_{ab}^{ij}\pp{ \frac{x}{\xi_1}, \frac{z}{\xi_2}; \frac{\bm{p}_{hT}}{Q}, \frac{Q^2}{\mu^2} } s_T^j
	\bigg\} \nn\\
	&\hspace{0.6em}  {}
	+ \order{ \qf / Q, \qf / p_{hT} }.
	\label{eq:Rc-factorize}
\end{align}
Because of the existence of Collins-type FEC, cf.~\eq{eq:fec-sT}, 
there is an extra term in the third line of \eq{eq:Rc-factorize}
which allows the occurrence of the transversity PDF $h_{a/p}$ even for unpolarized hadron production.
It is defined by
\beq[eq:transversity]
s_T^j \, h_{q/p}(\xi_1)
= \int_{-\infty}^{\infty} \frac{d\lambda}{4\pi} \, e^{-i \lambda \, \xi_1 \, p\cdot u} 
\bigvv{ p, S | \bar{\psi}(\lambda u) \, \gamma\cdot u \gamma^j \gamma_5 \, \psi(0) | p, S }.
\eeq
The hard coefficient is now a transverse spin transfer matrix $T_{ab}^{ij}$
that takes the initial-state quark spin along $s_T^j$ to the final-state quark spin along $(\hat{p}_h \times \bm{n}_T)^i$.
We note, however, that because the PDF and FEC are defined using different sets of lightcone vectors%
---$(u, \bar{u})$ and $(w_1, \bar{w}_1)$, respectively---%
the transverse indices $i$ and $j$ are taken in different coordinate systems~\cite{Collins:1993kq}:
$j$ is in the Breit coordinate system $\hat{x}_B$-$\hat{y}_B$-$\hat{z}_B$
while $i$ is in the hadron coordinate system $\hat{x}_h$-$\hat{y}_h$-$\hat{z}_h$,
as shown in \fig{fig:frame}.

The hard coefficients $C_{ab}$, $\Delta C_{ab}$, and $T_{ab}^{ij}$ in \eq{eq:Rc-factorize} are defined with 
amputated partonic cut diagrams traced by certain spinor or Lorentz matrices.
For the quark-to-quark channel, $(a, b) = (q, q)$, they are explicitly given in Appendix~\ref{sec:hard-coef-helicity}.
By a simple Dirac algebra, they can be re-expressed using partonic helicity amplitudes.
The $C_{ab}$ and $\Delta C_{ab}$ in \eq{eq:Rc-factorize} are the same as those in \eq{eq:sidis-factorization} for the SIDIS.
The $T_{ab}^{ij}$ is the same as that for dihadron production in SIDIS~\cite{Collins:1993kq}
and would also be the same as that for transversely polarized hadron production in SIDIS.

\subsubsection{Fatorization of $R_H$}
\label{sssec:Rh-factorization}
Since $R_C$ has the hadron and energy flow in the same collinear subgraph,
\eq{eq:Rc-factorize} holds only when $\qf$ is small, i.e., 
it is accurate up to power corrections of $\qf / p_{hT}$ and $\qf / Q$.
As $\qf$ becomes larger, the region $R_H$ becomes more important,
where the energy flow emerges from a different collinear subgraph $J_2$, 
as shown in \fig{fig:region}(b).

Now that the energy flow and hadron $h$ are independent,
it appears unnatural to define the former with respect to $h$ as in \eq{eq:energy-flow-factor}.
Rather than changing the definition back to \eq{eq:energy-flow},
we argue that as long as we keep $J_2$ away from the back-to-back region with $J_1$,
the energy weight factor in \eq{eq:energy-flow-factor} can be approximated as
\beq
\frac{p_i^+}{p_h^+}
= \frac{p_i \cdot w_1}{p_h \cdot w_1}
\simeq \frac{(p_i \cdot w_2) (\bar{w}_2 \cdot w_1)}{p_h \cdot w_1},
\eeq
up to power suppressed corrections of $\LQCD / Q$.
This requires $\bar{w}_2 \cdot w_1 = \cos^2(\theta / 2)$ to be of order 1, a condition easily satisfied in practice. 

Since $p_i \cdot w_2$ is the leading momentum component in the collinear subgraph $J_2$,
collinear and soft factorization in $R_H$ works in the same way as in $R_C$,
and then unitarity renders the soft factor unity 
and the unobserved collinear subgraphs $J_3$, $\ldots$, $J_N$ into the hard part.
The nontrivial point here is that we are left with two collinear factors for the final state,
one giving the normal FF associated with the hadron $h$,
and the other giving a jet function with the energy flow tagging,
\begin{align}
	J^{\E}(k_2, \eta, \phi)
	&= \frac{1}{N_c} \frac{\bar{w}_2 \cdot w_1}{p_h \cdot w_1} 
	\sum_X \int d^4y \, e^{i k_2 \cdot y} \vv{ 0 | \psi(y) |X} 
	\sum_{i \in X} (p_i \cdot w_2) \delta(\eta - \eta_i)\delta(\phi-\phi_i)\nn\\
	&\hspace{12em} \times 
	\langle X | \bar{\psi}(0)  | 0 \rangle,
	\label{eq:jet-E}
\end{align}
where we suppress the spinor indices and omit the spinor projectors $\P_2$ and $\Pb_2$.
Without the energy flow factor, $J^{\E}$ would reduce to the same jet function as 
$J_3$, $\ldots$, $J_N$, and could be absorbed into the hard part.
In the following, we show that this fact is unaltered by the energy flow factor.

First, on the exact collinear pinch, every line in the collinear subgraph $J_2$ is along the direction specified by $\eta$ and $\phi$.
Then we can write the energy flow factor in \eq{eq:jet-E} as
\beq
\sum_{i \in X} (p_i \cdot w_2) \delta(\eta - \eta_i)\delta(\phi - \phi_i)
\to (k_2 \cdot w_2) \sum_{i \in X} \delta(\eta - \eta_i)\delta(\phi - \phi_i),
\eeq
where $i$ counts every particle in $X$.
This allows us to sum over all possible cuts in \eq{eq:jet-E} to cancel the collinear pinch.

The exact collinear limit, however, exists only in perturbative theory, but neither in the full QCD nor experimental measurement.
For this purpose, it is more appropriate to consider the delta function as being integrated with an angular function peaked around $(\eta, \phi)$.
That is, we replace the energy flow factor in \eq{eq:jet-E} as
\beq[eq:smooth-e-weight-f]
\sum_{i \in X} (p_i \cdot w_2) \delta(\eta - \eta_i)\delta(\phi - \phi_i)
\to \sum_{i \in X} (p_i \cdot w_2) f_R(\eta - \eta_i, \phi - \phi_i),
\eeq
where $f_R$ can be chosen as a Gaussian distribution with a width $R$,
\beq[eq:fR]
f_R(\eta - \eta_i, \phi - \phi_i) = \frac{1}{2\pi R^2} \exp\!\bb{ - \frac{(\eta - \eta_i)^2 + (\phi - \phi_i)^2}{2 R^2} }.
\eeq
Using \eq{eq:smooth-e-weight-f} avoids the difficulty of a sharp delta function 
that involves only particles exactly lying along $(\eta, \phi)$
and makes the sum over cuts more convenient.

Now, as we move the cut in \eq{eq:jet-E}, all the particle lines crossed by the cut line are included in the energy flow weight.
Their angular distances from the center $(\eta, \phi)$ of the collinear subgraph $J_2$ are of the order $\Delta R \sim \l_T / \l^+ \sim \LQCD / Q$,
with $\l_T$ and $\l^+$ being the transverse and plus momenta of any typical line in $J_2$.
As long as $\Delta R / R \ll 1$, we can approximate \eq{eq:fR} by 
\beq[eq:fR-approx]
f_R(\eta - \eta_i, \phi - \phi_i) \simeq \frac{1}{2\pi R^2},
\eeq
with corrections suppressed by $(\Delta R / R)^2 \sim \LQCD^2 / (Q R)^2$.
Then \eq{eq:smooth-e-weight-f} becomes independent of the state $X$,
\beq[eq:energy-flow-approx]
\sum_{i \in X} (p_i \cdot w_2) f_R(\eta - \eta_i, \phi - \phi_i)
\simeq \frac{k_2 \cdot w_2}{2\pi R^2},
\eeq
and the sum-over-cuts argument can be applied to $J_2$ in the same way as $J_3$, $\ldots$, $J_N$.
All the collinear singularities associated with them are cancelled, 
so that they become part of the hard subgraph.

The only left collinear singularities are those associated with the observed hadron $h$,
and are collected by the fragmentation function.
The hard coefficient differs from those of \eq{eq:sidis-factorization} having an energy tagging in the direction around $(\eta, \phi)$.

Taking the limit $R \to 0$ then reduces \eq{eq:smooth-e-weight-f} back to the delta function.
This will not produce any perturbative divergence,
but, as we have mentioned, can induce some nonperturbative effects 
which are cancelled only upon a local average of the angular distribution.

\subsubsection{Subtraction of double counting}
\label{sssec:subtraction}
In Secs.~\ref{sssec:Rc-factorization} and \ref{sssec:Rh-factorization}, 
we have only considered factorizations for the regions $R_C$ and $R_H$ individually.
Each of the two regions contains smaller regions within themselves 
and they also overlap with each other.
To achieve an all-order factorization, one needs to add up contributions from all leading regions of all graphs,
with necessary subtractions made to remove double counting between the different regions.
The subtraction formalism was established systematically in \refs{Collins:1981uk, Collins:2011zzd}.
A detailed discussion of this is beyond the scope of this paper. 
We only make a few remarks that are important to our specific situation.

First, by virtue of the soft cancellation in any of the leading regions, 
we only need to consider subtractions associated with collinear regions.
Subsequently, any leading region will take the form of $R_C$ or $R_H$ in \fig{fig:region},
with the soft subgraph $S$ along with its connections to the collinear subgraphs removed.

For a certain region $L$, a smaller region $L' < L$
is obtained by either
(1) moving lines in the hard subgraph to one of the collinear subgraphs,
or
(2) merging any two or more of the collinear subgraphs into a single one by decreasing their angular separation.
The subtraction term is obtained at the integrand level for a fixed (loop and phase space) momentum configuration 
by successively applying the approximations $\hat{T}_{L'}$ and $\hat{T}_L$, designed for $L'$ and $L$, respectively.
That is, for any graph $G$, the contribution from these two regions, $G_L$ and $G_{L'}$, is given by
\beq
G_L + G_{L'} = \hat{T}_L G + \hat{T}_{L'} G - \hat{T}_L \hat{T}_{L'} G.
\eeq
In more general case, each region would contain chains of smaller regions nested in an ordered way,
and the contribution from this region will be given by a recursive subtraction formula~\cite{Collins:2011zzd}.

The type-(1) subtraction does not change the number or structure of the collinear subgraphs.
It occurs only in gauge theory and involves longitudinally polarized gluons.
Each of these subtraction terms can be factorized in the same way.
The result only affects the collinear factor $J_1$ in both $R_C$ and $R_H$,
modifying the corresponding hard coefficient so as to remove its collinear singularities.

The type-(2) subtraction is more nontrivial and mixes regions of different $N$ as well as the two types $R_C$ and $R_H$.
First, the subtraction associated with the merging of unobserved jets $J_3$, $\ldots$, $J_N$ in either $R_C$ or $R_H$
is simply zero by use of the same unitarity argument as that for the individual collinear cancellation of these jets themselves.
Second, in regions of type $R_H$, merging an unobserved jet with $J_2$ gives a subtraction term that is power suppressed 
by $1 / (QR)^2$ in a way similar to \eq{eq:energy-flow-approx}.
Third, in regions of type $R_C$ or $R_H$, 
merging an unobserved jet with $J_1$ gives a new collinear subgraph associated with $h$.
The subtraction therefore corresponds to removing the collinear singularity from the hard part.

The most nontrivial subtraction term in our case involves merging $J_1$ and $J_2$ in regions of type $R_H$.
This can be perturbatively understood as the following.
While the region $R_H$ is factorized into an FF with a hard coefficient 
in which all collinear singularities associated with $J_2$, $\ldots$, $J_N$ are canceled,
the hard coefficient can contain enhancement associated with small $\qf$ or $\theta$ as $J_2$ becomes close to $J_1$.
However, the latter has already been accounted for by the $J_1$ factor in a corresponding region of type $R_C$.
The subtraction therefore removes this double counting 
and modifies the hard coefficient in the factorization of region $R_H$.

To summarize, the cross section in \eq{eq:sidis-xsec-e} is comprised of two parts.
The first is given by a factorization formula in \eq{eq:Rc-factorize}, which is a good approximation at small $\qf$.
The second part takes a factorization form that is similar to \eq{eq:sidis-factorization}, 
but the hard coefficient contains an energy tag along with a subtraction of the double counted region with \eq{eq:Rc-factorize}.
It is more accurate at moderate $\qf$ values.
The subtraction term serves as a matching between the two parts.
This factorization structure is similar to the transverse momentum dependent factorization in \refs{Collins:1981uk, Collins:1984kg}. 

Since our aim is to extract the FEC from SIDIS,
we will focus on the small $\qf$ region in the rest of this paper,  
and take \eqs{eq:sidis-xsec-e}{eq:Rc-factorize} as our starting point.
A further discussion of the two-part factorization structure will be left for future work.

\subsection{Leading-twist observables for Collins-type quark FEC}
\label{sssec:fsidis-coefs}
The energy-tagged SIDIS cross section in \eq{eq:sidis-xsec-e} contains 
a number of azimuthal observables associated with both $\phi_h$ and $\phi$, cf.\ \fig{fig:frame}.
It remains so even if we consider only up to the leading power in $1/Q$. 
A complete analysis of all the partonic channels of $(a, b)$ in \eq{eq:Rc-factorize} is beyond the scope of this paper.
Rather, we focus on the channels sensitive to the Collins-type quark FEC.
The leading contribution in SIDIS comes from the transition coefficient $T_{qq}$ in \eq{eq:Rc-factorize},
which requires a transversely polarized proton target in the initial state and observation of the $\phi$ distribution in the final state.
In this subsection, we first give a detailed analysis to the $(a, b) = (q, q)$ channel,
for all possible spin configurations of both beam and target,
and then provide a LO result.
Other channels can be analyzed in a similar way.

\subsubsection{Separation of leptonic and partonic subprocesses}
\label{sssec:separation}

In Appendix~\ref{sec:hard-coef-helicity}, we have expressed 
the hard coefficients $C_{ab}$, $\Delta C_{ab}$, and $T_{ab}^{ij}$ 
in terms of quark helicity structures.
To further simplify the results in the Breit frame and especially to make the azimuthal modulations manifest, 
it is convenient to decompose the cross section into leptonic and hadronic parts, 
as indicated by the one-photon exchange approximation in \eq{eq:two-stage}.
This is done in a standard way~\cite{Diehl:2005pc} by separating the lepton-quark scattering amplitude in \eq{eq:partonic-scattering-amp-qq}
at the virtual photon $\gamma^*(q)$ which moves along the $-\hat{z}_B$ direction,
\beq[eq:decompose-A]
\A_{\lambda_e \lambda_e', \lambda_1 \lambda_2}(\ell, k_1, \ell', k_2; X) 
= \frac{1}{Q^2} \sum_{\lambda_{\gamma}} (-1)^{\lambda_{\gamma}} \, 
\mathcal{L}_{\lambda_e \lambda_e'}^{\lambda_{\gamma}}(\ell, q) \,
\Q_{\lambda_1 \lambda_2}^{\lambda_{\gamma}}(k_1, q, k_2; X),
\eeq
where we have restored the helicity indices $\lambda_e$ and $\lambda_e'$ of the initials and final-state electrons, respectively, in $\A$,
as compared to \eq{eq:partonic-scattering-amp-qq}.
Here, $\lambda_{\gamma} = \pm 1$ or $0$ refers to the helicity of $\gamma^*$.
It is defined by rewriting its propagator as a sum over polarization vectors,
\beq
g^{\mu\nu} - \frac{q^{\mu} q^{\nu}}{q^2}
= \sum_{\lambda_{\gamma} = 0, \pm 1} (-1)^{\lambda_{\gamma}} \, \epsilon_{\lambda_{\gamma}}^{\mu}(q) \epsilon_{\lambda_{\gamma}}^{\nu *}(q),
\eeq
where the $(-1)^{\lambda_{\gamma}}$ is due to the spacelike $q$, and the polarization vectors are chosen as
\begin{align}
	\epsilon_{\pm}^{\mu}(q) = \frac{1}{\sqrt{2}} (0, \mp1, i, 0)_{\rm Cartesian}, \quad
	\epsilon_{0}^{\mu}(q) = \frac{1}{Q} (q^{\mu} + 2 x p^{\mu} ) = \frac{1}{Q} \pp{ x p^+, \frac{Q^2}{2x p^+}, \bm{0}_T },
\end{align}	
which satisfy $\epsilon_{\lambda_{\gamma}}(q) \cdot \epsilon^*_{\lambda_{\gamma}'}(q) = (-1)^{\lambda_{\gamma}} \delta_{\lambda_{\gamma} \lambda_{\gamma}'}$
and are specified in Cartesian and lightfront coordinates, respectively.
In \eq{eq:decompose-A}, 
$\mathcal{L}_{\lambda_e \lambda_e'}^{\lambda_{\gamma}}$ is the LO helicity amplitude of \eq{eq:two-stage-a},
\beq
\mathcal{L}_{\lambda_e \lambda_e'}^{\lambda_{\gamma}}(\ell, q)
= \bar{u}_e(\ell + q, \lambda_e') \, \gamma_{\mu} \, u_e(\ell, \lambda_e) \, \epsilon_{\lambda_{\gamma}}^{\mu *}(q),
\eeq
with $u_e$ and $\bar{u}_e$ being the electron spinors.
The $\Q_{\lambda_1 \lambda_2}^{\lambda_{\gamma}}$ is the helicity amplitude of quark-photon scattering,
\beq[eq:hard-qq]
q(k_1, \lambda_1) + \gamma^*(q, \lambda_{\gamma}) \to q(k_2, \lambda_2) + X(k_X),
\eeq
and can be formally defined with the QED current operator $J^{\mu}(x)$ as
\beq[eq:a-q-amplitude]
\Q_{\lambda_1 \lambda_2}^{\lambda_{\gamma}}(k_1, q, k_2; X)
= \bigvv{ q(k_2, \lambda_2), X(k_X); {\rm out} | \, \epsilon_{\lambda_{\gamma}}(q) \cdot J(0) \, | q(k_1, \lambda_1); {\rm in} }.
\eeq
The quark momenta $k_1$ and $k_2$ have been projected on shell and are related to $p$ and $p_h$ by 
\beq[eq:partonic-momenta]
k_1 = \xi_1 \, p, \quad
k_2 = p_h / \xi_2.
\eeq 
Thus the partonic version of the SIDIS variables in \eqs{eq:dis-kin-var}{eq:sidis-kin-var} is 
\beq[eq:partonic-xz]
\x = \frac{Q^2}{2 k_1 \cdot q} = \frac{x}{\xi_1}, \quad
\z = \frac{p \cdot k_2}{p \cdot q} = \frac{z}{\xi_2}.
\eeq
$k_2$ is further described by its transverse component $k_{2T} = p_{hT} / \xi_2$ and azimuthal angle $\phi_h$.

Using \eq{eq:decompose-A}, we can also decompose the hard factor in \eq{eq:hard-H-helicity-space} as
\begin{align}
	H_{\lambda_1 \lambda_2, \lambda_1' \lambda_2'}(\ell, k_1, \ell', k_2)
	= \sum_{\lambda_{\gamma}, \lambda_{\gamma}'} (-1)^{\lambda_{\gamma} + \lambda_{\gamma}'} \,
	L_{\lambda_{\gamma}\lambda_{\gamma}'}(\ell, q) \,
	W_{\lambda_1 \lambda_2, \lambda_1' \lambda_2'}^{\lambda_{\gamma}, \lambda_{\gamma}'}(k_1, q, k_2).
	\label{eq:hard-H-helicity-photon}
\end{align}
This takes the form of an unnormalized photon density matrix $L$
multiplied by the photon quark hard scattering factor $W$.
The former has averaged over the electron beam spin with the density matrix $\rho^e$ in \eq{eq:den-mtx-ep},
\begin{align}
	L_{\lambda_{\gamma}\lambda_{\gamma}'}(\ell, q)
	= \frac{1}{Q^4} \sum_{\lambda_e, \bar{\lambda}_e, \lambda_e'} \rho^e_{\lambda_e \bar{\lambda}_e}(P_e)
	\, \mathcal{L}_{\lambda_e \lambda_e'}^{\lambda_{\gamma}}(\ell, q) 
	\bb{ \mathcal{L}_{\bar{\lambda}_e \lambda_e'}^{\lambda_{\gamma}'}(\ell, q) }^*,
\end{align}
which introduces the electron polarization $P_e$
and can be explicitly calculated to give
\begin{align}
	L_{\pm\pm} 	&= \frac{1}{Q^2(1 - \varepsilon)} \bb{ 1 \pm P_e \sqrt{1 - \varepsilon^2} }, \nn\\
	L_{00} 			&= \frac{1}{Q^2(1 - \varepsilon)} (2\varepsilon), \nn\\
	L_{+-}			&= L_{-+}  = \frac{1}{Q^2(1 - \varepsilon)} (-\varepsilon) = -\frac{1}{2} L_{00}, \nn\\
	L_{\pm 0} 		&= L_{0\pm} = \frac{1}{Q^2(1 - \varepsilon)} \bb{ \pm \sqrt{ \varepsilon (1 + \varepsilon) } + P_e \sqrt{ \varepsilon (1 - \varepsilon) } },
	\label{eq:den-mtx-a}
\end{align}
where we defined $\varepsilon$ as the ratio of longitudinal to transverse photon flux,
\beq
\varepsilon 
= \frac{L_{00}}{L_{++} + L_{--}}
= \frac{1 - y}{1 - y + y^2 / 2},
\eeq
which carries the whole dependence of \eq{eq:hard-H-helicity-photon} on $y$, or the overall c.m.\ energy square $s$ by \eq{eq:dis-kin-var}.
The $W$ in \eq{eq:hard-H-helicity-photon} is defined as
\begin{align}
	W_{\lambda_1 \lambda_2, \lambda_1' \lambda_2'}^{\lambda_{\gamma}, \lambda_{\gamma}'}(k_1, q, k_2)
	&= \sum\nolimits_X \int d\Pi_X \, (2\pi)^4 \delta^{(4)}(q + k_1 - k_2 - k_X) 	\nn\\
	&\hspace{4.5em} \times
	\Q_{\lambda_1 \lambda_2}^{\lambda_{\gamma}}(k_1, q, k_2; X) \,
	\Q_{\lambda_1' \lambda_2'}^{\lambda_{\gamma}' *}(k_1, q, k_2; X),
	\label{eq:W-def}
\end{align}
which includes a sum over $X$, after subtracting the collinear singularities as discussed in \sec{sssec:subtraction}.
Substituting the photon density matrix \eq{eq:den-mtx-a} into \eq{eq:hard-H-helicity-photon}, we have
\begin{align}
	&H_{\lambda_1 \lambda_2, \lambda_1' \lambda_2'}(\ell, k_1, \ell', k_2)
	= \frac{2}{Q^2 (1 - \varepsilon)}
	\bigg[
	\frac{1}{2} \pp{ W^{+, +}_{\lambda_1 \lambda_2, \lambda_1' \lambda_2'} + W^{-, -}_{\lambda_1 \lambda_2, \lambda_1' \lambda_2'} } 
	+ \varepsilon \, W^{0, 0}_{\lambda_1 \lambda_2, \lambda_1' \lambda_2'} 
	\nn\\
	&\hspace{4em}
	- \frac{\varepsilon}{2} \pp{ W^{+, -}_{\lambda_1 \lambda_2, \lambda_1' \lambda_2'} + W^{-, +}_{\lambda_1 \lambda_2, \lambda_1' \lambda_2'} }
	+ \frac{P_e}{2} \sqrt{1 - \varepsilon^2} \pp{ W^{+, +}_{\lambda_1 \lambda_2, \lambda_1' \lambda_2'} - W^{-, -}_{\lambda_1 \lambda_2, \lambda_1' \lambda_2'} }  \nn\\
	& \hspace{4em} 
	- \frac{1}{2} \sqrt{ \varepsilon (1 + \varepsilon) } \pp{ 
		W^{+, 0}_{\lambda_1 \lambda_2, \lambda_1' \lambda_2'} 
		+ W^{0, +}_{\lambda_1 \lambda_2, \lambda_1' \lambda_2'} 
		- W^{-, 0}_{\lambda_1 \lambda_2, \lambda_1' \lambda_2'} 
		- W^{0, -}_{\lambda_1 \lambda_2, \lambda_1' \lambda_2'} 
	}	\nn\\
	& \hspace{4em} 
	- \frac{P_e}{2} \sqrt{ \varepsilon (1 - \varepsilon) } \pp{ 
		W^{+, 0}_{\lambda_1 \lambda_2, \lambda_1' \lambda_2'} 
		+ W^{0, +}_{\lambda_1 \lambda_2, \lambda_1' \lambda_2'} 
		+ W^{-, 0}_{\lambda_1 \lambda_2, \lambda_1' \lambda_2'} 
		+ W^{0, -}_{\lambda_1 \lambda_2, \lambda_1' \lambda_2'} 
	}
	\bigg].
	\label{eq:hard-H-denmtx-W}
\end{align}
This form makes it straightforward to examine the symmetry properties.

First, by the rotational symmetry around the $\hat{z}_B$ axis, 
we can separate the $\phi_h$ dependence from the partonic tensor $W$,
\begin{align}
	W_{\lambda_1 \lambda_2, \lambda_1' \lambda_2'}^{\lambda_{\gamma}, \lambda_{\gamma}'}(k_1, q, k_2)
	&= W_{\lambda_1 \lambda_2, \lambda_1' \lambda_2'}^{\lambda_{\gamma}, \lambda_{\gamma}'}(Q^2, x / \xi_1, z / \xi_2, k_{2T}, \phi_h)	\nn\\
	&= e^{i \bb{ (\lambda_1 - \lambda_1') - (\lambda_{\gamma} - \lambda_{\gamma}') } \phi_h}
	\bar{W}_{\lambda_1 \lambda_2, \lambda_1' \lambda_2'}^{\lambda_{\gamma}, \lambda_{\gamma}'}(Q^2, x / \xi_1, z / \xi_2, k_{2T}).
	\label{eq:W-phi0}
\end{align}
Here, $\bar{W} \equiv W(\phi_h = 0)$ corresponds to the (inclusive) final state 
with $q(k_2)$ moving in the $\hat{x}_B$-$\hat{z}_B$ plane.
Up to the perturbative order we will work in the following, 
the $q \to q$ hard coefficients receive contributions only from the non-singlet (NS) channel,
where quark line does not break in going from $q(k_1)$ to $q(k_2)$ in the cut diagram of \fig{fig:sidis-hard}.
Then the chiral symmetry associated with the massless quark approximation implies that
$\bar{W}_{\lambda_1 \lambda_2, \lambda_1' \lambda_2'}^{\lambda_{\gamma}, \lambda_{\gamma}'}$ is nonzero only when 
$\lambda_1 = \lambda_2$ and $\lambda_1' = \lambda_2'$.
This allows us to simplify the notations by defining
\beq[eq:ns-constraint]
\Q_{\lambda_1 \lambda_2}^{\lambda_{\gamma}}
\overset{\rm NS}{=} \delta_{\lambda_1 \lambda_2} \Q_{\lambda_1}^{\lambda_{\gamma}},
\quad
\bar{W}_{\lambda_1 \lambda_2, \lambda_1' \lambda_2'}^{\lambda_{\gamma}, \lambda_{\gamma}'}
\overset{\rm NS}{=}  \delta_{\lambda_1 \lambda_2} \delta_{\lambda_1' \lambda_2'}
\bar{W}_{\lambda_1\lambda_1'}^{\lambda_{\gamma} \lambda_{\gamma}'},
\quad
H_{\lambda_1 \lambda_2, \lambda_1' \lambda_2'}
\overset{\rm NS}{=}  \delta_{\lambda_1 \lambda_2} \delta_{\lambda_1' \lambda_2'}
H_{\lambda_1 \lambda_1'}.
\eeq
The NS channel also has a simple parity constraint,
\beq[eq:parity-Wbar]
\bar{W}_{\lambda_1 \lambda_2, \lambda_1' \lambda_2'}^{\lambda_{\gamma}, \lambda_{\gamma}'}(Q^2, x / \xi_1, z / \xi_2, k_{2T})
= (-1)^{\lambda_{\gamma} + \lambda_{\gamma}'} 
\bar{W}_{-\lambda_1, -\lambda_2, -\lambda_1', -\lambda_2'}^{-\lambda_{\gamma}, -\lambda_{\gamma}'}(Q^2, x / \xi_1, z / \xi_2, k_{2T}).
\eeq
Together with the Hermiticity property,
$\bar{W}_{\lambda \lambda'}^{\lambda_{\gamma} \lambda_{\gamma}'}
= \bigpp{ \bar{W}_{\lambda' \lambda}^{\lambda_{\gamma}' \lambda_{\gamma}} }^*$,
these reduce $\bar{W}$ down to 18 independent real degrees of freedom,
including 6 real components,
\beq[eq:wbar-real]
\bar{W}^{++}_{++}, \quad
\bar{W}^{++}_{--}, \quad
\bar{W}^{+-}_{+-}, \quad
\bar{W}^{+-}_{-+}, \quad
\bar{W}^{00}_{++}, \quad
\bar{W}^{00}_{+-},
\eeq
and 6 complex ones,
\beq[eq:wbar-complex]
\bar{W}^{++}_{+-}, \quad
\bar{W}^{+-}_{++}, \quad
\bar{W}^{+0}_{++}, \quad
\bar{W}^{+0}_{--}, \quad
\bar{W}^{+0}_{+-}, \quad
\bar{W}^{+0}_{-+}.
\eeq
Using these relations, we can write the hard coefficients 
$C_{qq} $, $\Delta C_{qq} $, and $T^{ij}_{qq} $ in \eqs{eq:C-delC-def-helicity}{eq:T-def-helicity} 
in terms of the $\bar{W}$ components in \eqs{eq:wbar-real}{eq:wbar-complex},
while making the $\phi_h$ dependence explicit.

\subsubsection{Hard coefficients and azimuthal modulations}
\label{sssec:hard-azimuth}

The unpolarized hard coefficient $C_{qq}$ is given by
\begin{align}
	C_{qq}%
	&= \frac{1}{2} \pp{ H_{++} + H_{--} }\nn\\
	&= \frac{2}{Q^2 (1 - \varepsilon)} \bb{ F_{UU, T} + \varepsilon \, F_{UU, L}
		+ \sqrt{2 \varepsilon (1 + \varepsilon)} \, F_{UU}^{(1)} \cos\phi_h
		+ \varepsilon \, F_{UU}^{(2)} \cos(2\phi_h)
		\right.\nn\\
		&\left. \hspace{6.5em} {} 
		+ P_e \sqrt{2 \varepsilon (1 - \varepsilon)} \, F_{LU} \sin\phi_h
	},
	\label{eq:C-in-W}
\end{align}
which is organized by the azimuthal modulations and spin structures. 
We follow the SIDIS convention~\cite{Bacchetta:2006tn} 
to notate the coefficients as $F_{AB} = F_{AB}(Q^2, x/\xi_1, z/\xi_2, p_{hT}^2)$,
using the subscripts $A$ and $B$ to separately indicate the electron and proton polarization states,
with ``$U$'', ``$L$'', and ``$T$'' standing for ``unpolarized'', ``longitudinally polarized'', or ``transversely polarized'', respectively.
When $F_{AB}$ receives contributions from both transverse and longitudinal $\gamma^*$ polarizations, we use a third
subscript $C = T$ or $L$ in $F_{AB, C}$ to further distinguish them.
Clearly, the $F_{AB, L}$ term will have an extra coefficient $\varepsilon$ compared to $F_{AB, T}$.
Furthermore, we use a superscript ``$(n)$'' to indicate the azimuthal modulation $\cos(n \phi_h)$ or $\sin(n \phi_h)$,
which distinguishes $F$'s with the same spin configuration.
The coefficients in \eq{eq:C-in-W} are given by the $\bar{W}$ components as
\begin{align}
	F_{UU, T} &= \frac{1}{2} \pp{ \bar{W}^{++}_{++} + \bar{W}^{++}_{--} }, \quad
	F_{UU, L} = \bar{W}^{00}_{++}, \nn\\
	F_{UU}^{(1)} &= -\frac{1}{\sqrt{2}} \Re\pp{ \bar{W}^{+0}_{++} + \bar{W}^{+0}_{--} }, \quad
	F_{UU}^{(2)} = -\Re\bar{W}^{+-}_{++}, \nn\\
	F_{LU} &= -\frac{1}{\sqrt{2}} \Im\pp{ \bar{W}^{+0}_{++} + \bar{W}^{+0}_{--} }.
	\label{eq:F-coef-U}
\end{align}
The polarized hard coefficient $\Delta C_{qq}$ is given by
\begin{align}
	\Delta C_{qq}%
	&= \frac{1}{2} \pp{ H_{++} - H_{--} }\nn\\
	&= \frac{2}{Q^2 (1 - \varepsilon)} \cc{ 
		\sqrt{2 \varepsilon (1 + \varepsilon)} \, F_{UL}^{(1)} \sin\phi_h 
		+ \varepsilon \, F_{UL}^{(2)} \sin(2\phi_h)
		\right.\nn\\
		&\left.{}  \hspace{6.5em}
		+ P_e \bb{ \sqrt{1 - \varepsilon^2} \, F_{LL} + \sqrt{2 \varepsilon (1 - \varepsilon)} \, F_{LL}^{(1)} \cos\phi_h }
	},
	\label{eq:DeltaC-in-W}
\end{align}
where the coefficients are
\begin{align}
	F_{UL}^{(1)} &= -\frac{1}{\sqrt{2}} \Im\pp{ \bar{W}^{+0}_{++} - \bar{W}^{+0}_{--} }, \quad
	F_{UL}^{(2)} = -\Im\bar{W}^{+-}_{++}, \nn\\
	F_{LL} &= \frac{1}{2} \pp{ \bar{W}^{++}_{++} - \bar{W}^{++}_{--} }, \quad
	F_{LL}^{(1)} = -\frac{1}{\sqrt{2}} \Re\pp{ \bar{W}^{+0}_{++} - \bar{W}^{+0}_{--} }.
	\label{eq:F-coef-L}
\end{align}

As shown in \eq{eq:Rc-factorize}, the hard coefficient $T_{qq}^{ij}$
transfers the initial-state quark's transverse spin $\bm{s}_T$ to that of the final-state quark,
$S_{qT}^i \propto T_{qq}^{ij} s_T^j$.
This transfer is most easily understood by referring to the scattering plane of $q(k_1) + \gamma^*(q) \to q(k_2)$~\cite{Collins:1993kq},
where $\bm{s}_T$ has a component $s_{\parallel} = s_T \cos(\phi_S - \phi_h)$ on the plane 
and a component $s_{\perp} = s_T \sin(\phi_S - \phi_h)$ perpendicular to the plane,
and similarly for $\bm{S}_{qT} = (S_{q\parallel}, S_{q\perp})$ of the final-state quark.
In the simplest case, let us consider the $\gamma^*$ as unpolarized.
Then by performing a parity transformation together with a rotation around this plane by $\pi$, 
one can easily see that the transverse spin components $s_{\perp}$ and $S_{q\perp}$ perpendicular to the plane stay unchanged,
while the components $s_{\parallel}$ and $S_{q\parallel}$ in the plane are flipped.
Hence, if parity is conserved, we should expect these spin components to transfer separately without mixing,
i.e., $(S_{q\parallel}, S_{q\perp}) \propto (s_{\parallel}, s_{\perp})$. 
Now noting that the components $(S_{q\parallel}, S_{q\perp})$ directly align with 
the components $(S_{qT}^x, S_{qT}^y)$ in the hadron frame in \eq{eq:hadron-frame},
we have $\bm{S}_{qT} \propto s_T \pp{ \cos(\phi_S - \phi_h), \sin(\phi_S - \phi_h) }$.
That is, the $T_{qq}^{ij}$ is a rotation matrix of angle $-\phi_h$.

The interference of different $\gamma^*$ helicities can induce additional $(\pm 1) \phi_h$ or $(\pm 2) \phi_h$ phase shift
	relative to that due to the basic rotation of $-\phi_h$.
Therefore, we define the rotation basis for the spin transfer matrix $T_{qq}^{ij}$,
\begin{align}
	t_n(\phi_h) &= R_z\pp{ -(1 + n)\phi_h }
	= \begin{pmatrix}
		\cos(1 + n)\phi_h & \sin(1 + n)\phi_h \\
		-\sin(1 + n)\phi_h & \cos(1 + n)\phi_h
	\end{pmatrix},
	\label{eq:tn-mtx}
\end{align}
which is a rotation around the $z$ axis by $-(1 + n)\phi_h$.
When parity-odd effects are involved, 
which arise due to the electron polarization $P_e$,
the spin components in and perpendicular to the hard scattering plane will be transferred into each other,
inducing a further rotation by $\pi / 2$ on top of $t_n(\phi_h)$.
Thus we also define the parity-odd rotation basis,
\begin{align}
	\tilde{t}_n(\phi) &= R_z(\pi / 2) \, R_z\pp{ -(1 + n)\phi }
	= \begin{pmatrix}
		\sin(1 + n)\phi & -\cos(1 + n)\phi \\
		\cos(1 + n)\phi & \sin(1 + n)\phi
	\end{pmatrix}.
	\label{eq:tn-pv-mtx}
\end{align}
Using these, we can write $T_{qq}^{ij}$ as
\begin{align}
	\pp{ T_{qq}^{ij} }
	&\equiv \begin{pmatrix}
		T_{qq}^{xx} & T_{qq}^{xy} \\
		T_{qq}^{yx} & T_{qq}^{yy}
	\end{pmatrix}
	= \frac{1}{2} \begin{pmatrix}
		H_{+-} + H_{-+} & -i (H_{+-} - H_{-+}) \\
		i (H_{+-} - H_{-+}) & H_{+-} + H_{-+} 
	\end{pmatrix}	\nn\\
	&= \frac{2}{Q^2 (1 - \varepsilon)} 
	\cc{
		\pp{ F_{UT, T} + \varepsilon \, F_{UT, L} } t_0(\phi_h) 
		- \sqrt{2 \varepsilon (1 + \varepsilon)} \, 
		\bb{ F_{UT}^{(1+)} \, t_1(\phi_h) + F_{UT}^{(1-)} \, t_{-1}(\phi_h) } \right.	\nn\\
		& \left.{} \hspace{0.4em}
		- \varepsilon \, 
		\bb{ F_{UT}^{(2+)} \, t_2(\phi_h) + F_{UT}^{(2-)} \, t_{-2}(\phi_h) } \right. \nn\\
		& \left.{} \hspace{0.4em}
		+ P_e \bb{ 
			\sqrt{1 - \varepsilon^2} \, F_{LT} \, \tilde{t}_0(\phi_h)
			+ \sqrt{2 \varepsilon (1 - \varepsilon)} \, 
			\pp{ F_{LT}^{(1+)} \, \tilde{t}_1(\phi_h) 
				+ F_{LT}^{(1-)} \, \tilde{t}_{-1}(\phi_h) 
			}
		}
	},
	\label{eq:T-in-W}
\end{align}
where the coefficients are expressed in $\bar{W}$ as 
\begin{align}
	F_{UT, T} &= \Re \bar{W}^{++}_{+-}, \quad
	F_{UT, L} = \Re \bar{W}^{00}_{+-}, \nn\\
	F_{UT}^{(1+)} &= \frac{1}{\sqrt{2}} \Re \bar{W}^{+0}_{-+}, \quad
	F_{UT}^{(1-)} = \frac{1}{\sqrt{2}} \Re \bar{W}^{+0}_{+-},
	\nn\\
	F_{UT}^{(2+)} &= \frac{1}{2} \Re \bar{W}^{+-}_{-+}, \quad
	F_{UT}^{(2-)} = \frac{1}{2} \Re \bar{W}^{+-}_{+-}, \nn\\
	F_{LT} &= - \Im \bar{W}^{++}_{+-}, \quad
	F_{LT}^{(1+)} = -\frac{1}{\sqrt{2}} \Im \bar{W}^{+0}_{-+}, \quad
	F_{LT}^{(1-)} = \frac{1}{\sqrt{2}} \Im \bar{W}^{+0}_{+-}.
	\label{eq:F-coef-T}
\end{align}
These results show an interesting interplay between the spin transfer and azimuthal $\phi_h$ modulations.
We have labeled coefficients of the rotation matrices $t_{\pm n}(\phi_h)$ or $\tilde{t}_{\pm n}(\phi_h)$
with superscripts ``$(n\pm)$''
to indicate the helicity transfer unit originating from the $\gamma^*$ interference.

Substituting \eqss{eq:C-in-W}{eq:DeltaC-in-W}{eq:T-in-W} back into \eq{eq:Rc-factorize}
exhibits all the azimuthal modulations.
In particular, we note that the $\phi$ distribution of the energy flow around the measured hadron $h$
is given by the transverse spin transfer $T_{qq}^{ij}$,
with the two rotation modes in \eqs{eq:tn-mtx}{eq:tn-pv-mtx} generating, respectively, the modulations,
\begin{align}
	\frac{(\hat{p}_h \times \bm{n}_T)^i}{|\bm{n}_T|} \, t^{ij}_n(\phi_h) \, s_T^j
	& = - s_T \sin\bb{ \phi - \phi_S + (n + 1) \phi_h }, \nn\\
	\frac{(\hat{p}_h \times \bm{n}_T)^i}{|\bm{n}_T|} \, \tilde{t}^{ij}_n(\phi_h) \, s_T^j
	& = s_T \cos\bb{ \phi - \phi_S + (n + 1) \phi_h }.
\end{align}
Combining these with \eq{eq:sidis-xsec-e} then gives the energy flow-weighted cross section,
\begin{align}
	&\bb{ \frac{y^2}{z (1 - \varepsilon)} \frac{\alpha_e^2}{(8\pi^2 Q^2)^2} }^{-1}
	\frac{d\Sigma}{dx \, dQ^2 \, d\phi_S \, dz \, dp_{hT}^2 \, d\phi_h \, d\eta \, d\phi} \nn\\
	&\hspace{1em} 
	= \C{f}{\unpfec}{F_{UU, T}} + \varepsilon \, \C{f}{\unpfec}{F_{UU, L}} \nn\\
	&\hspace{2em}  
	+ \sqrt{2 \varepsilon (1 + \varepsilon)} \, \C{f}{\unpfec}{F_{UU}^{(1)}} \cos\phi_h 
	+ \varepsilon \, \C{f}{\unpfec}{F_{UU}^{(2)}} \cos(2\phi_h)
	\nn\\
	&\hspace{2em}  
	+ P_e \, \sqrt{2 \varepsilon (1 - \varepsilon)} \, \C{f}{\unpfec}{F_{LU}} \sin\phi_h	 \nn\\
	&\hspace{2em} 
	+ P_N \cc{
		\sqrt{2 \varepsilon (1 + \varepsilon)} \, \C{g}{ \unpfec}{F_{UL}^{(1)}} \sin\phi_h 
		+ \varepsilon \, \C{g}{\unpfec}{F_{UL}^{(2)}} \sin(2\phi_h)
	}
	\nn\\
	&\hspace{2em}  
	+ P_e P_N \cc{
		\sqrt{1 - \varepsilon^2} \, \C{g}{\unpfec}{F_{LL}} 
		+ \sqrt{2 \varepsilon (1 - \varepsilon)} \, \C{g}{\unpfec}{F_{LL}^{(1)}} \cos\phi_h 
	}
	\nn\\
	&\hspace{2em}  
	+ s_T \bigg\{
	\bb{
		- \C{h}{\colfec^{\perp}}{F_{UT, T}} 
		- \varepsilon \, \C{h}{\colfec^{\perp}}{F_{UT, L}}
	} \sin(\phi - \phi_S + \phi_h) 
	\nn\\
	& \hspace{5em}
	+ \sqrt{2 \varepsilon (1 + \varepsilon)} \, 
	\Big[ \C{h}{\colfec^{\perp}}{F_{UT}^{(1+)} } \sin(\phi - \phi_S + 2\phi_h)
	+ {\C{h}{\colfec^{\perp}}{F_{UT}^{(1-)} } \sin(\phi - \phi_S) }
	\Big]
	\nn\\
	& \hspace{5em}
	+ \varepsilon \, 
	\Big[ \C{h}{\colfec^{\perp}}{F_{UT}^{(2+)} } \sin(\phi - \phi_S + 3\phi_h)
	+ {\C{h}{\colfec^{\perp}}{F_{UT}^{(2-)} } \sin(\phi - \phi_S - \phi_h) }
	\Big]
	\bigg\}
	\nn\\
	&\hspace{2em}  
	+ P_e s_T \bigg\{ 
	\sqrt{1 - \varepsilon^2} \, \C{h}{\colfec^{\perp}}{F_{LT} } \cos(\phi - \phi_S + \phi_h) 
	\nn\\
	& \hspace{5em}
	+ \sqrt{2 \varepsilon (1 - \varepsilon)} \, 
	\Big[ \C{h}{\colfec^{\perp}}{F_{LT}^{(1+)} } \cos(\phi - \phi_S + 2\phi_h)
	+ \C{h}{\colfec^{\perp}}{F_{LT}^{(1-)} } \cos(\phi - \phi_S) 
	\Big]
	\bigg\} \nn\\
	&\hspace{2em}  
	+ (\mbox{other partonic channels}) + \order{\qf / Q, \qf / p_{hT}},
	\label{eq:fsidis-TRc-result}
\end{align}
where we introduced the notation for the convolution,
\begin{align}
	\C{f}{\fec}{F} 
	&= \C{f}{\fec}{F} (Q^2, x, z, p_{hT}^2; \qf) \nn\\
	&\equiv \int \frac{d\xi_1}{\xi_1} \int \frac{d\xi_2}{\xi_2^2} \, 
	f_{q/p}(\xi_1) \, \fec_{1, h/q}(\xi_2, \qf) \, F(Q^2, x/\xi_1, z/\xi_2, p_{hT}^2 / \xi_2^2),
	\label{eq:conv-fin-pt}
\end{align}
with $f$ being a PDF, $\fec$ being an FEC, and $F$ being a hard coefficient in Eq.~\eqref{eq:F-coef-U}~\eqref{eq:F-coef-L} or \eqref{eq:F-coef-T}.

It is the various $\phi$ modulations in \eq{eq:fsidis-TRc-result} that help to probe the Collins-type FEC.
We re-emphasize the associated physics as follows.
The matrix $T_{qq}$ arising from the hard scattering in \eq{eq:hard-qq}  
transfers the transverse spin from the initial-state quark to the final-state quark,
thus defining the transverse spin magnitude and direction of the latter.
The fragmentation process of the final-state quark then conveys this spin information 
into an observational signal via the azimuthal energy deposition asymmetry by \eq{eq:fec-sT-phi}.
These modulations have been explicitly spelled out in \eq{eq:fsidis-TRc-result}, 
appearing in simple trigonometric forms 
which allow their coefficients to be extracted by standard harmonic analysis.
Clearly, due to chiral symmetry, 
nontrivial $\phi$ modulations and target transverse spin asymmetries (i.e., terms proportional to $s_T$) 
are correlated with each other, such that the $\phi$ dependence
always appears in the form of $\phi - \phi_S$.
Without measuring the $\phi$ distribution, 
all the terms involving $s_T$, Collins-type FEC, and transversity PDF will vanish.

While we have included all the polarization configurations in \eq{eq:fsidis-TRc-result},
in practice, one may have only a subset of those terms. 
For example, we need only $s_T \neq 0$ to probe the Collins-type FEC,
in which case all the terms involving $P_e$ or $P_N$ can be set to zero.

At LO, other partonic channels contributing in \eq{eq:fsidis-TRc-result}
include $\bar{q} \to \bar{q}$, $q(\bar{q}) \to g$, $g \to q(\bar{q})$.
As we will note in \sec{sssec:sidis-lo-qg}, 
the $\bar{q} \to \bar{q}$ channel can be simply obtained from the $q\to q$ channel by a charge conjugation.
The $q(\bar{q}) \to g$ channel will involve the gluon FEC, 
which has a linear polarization component that induces a $\cos(2\phi)$ or $\sin(2\phi)$ modulation.
While this is an interesting signal by itself, 
it can be generated purely from perturbative origin~\cite{Chen:2020adz, Chen:2021gdk, Yu:2022kcj, Li:2023gkh, Guo:2024jch},
not intrinsically sensitive to chiral symmetry breaking;
the modulations can also be readily distinguished from the quark case.
Similarly, for the $g \to q(\bar{q})$ channel, only the unpolarized quark FEC can arise.
Neither of these two channels allows the transversity PDF at the leading twist. 
Including them in a full analysis is left as a future development;
we focus our attention on the $q \to q$ channel in this paper.

\subsubsection{LO results for the $q \to q$ channel}
\label{sssec:sidis-lo-qg}
\begin{figure}[htbp]
	\centering
	\begin{tabular}{cc}
		\includegraphics[scale=0.6]{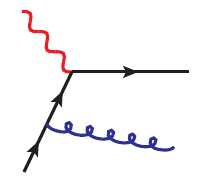} &
		\includegraphics[scale=0.6]{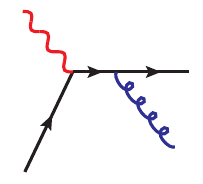}  \\
		(a) & (b)
	\end{tabular}
	\caption{LO diagrams of the $q + \gamma^* \to q + g$ scattering amplitude.}
	\label{fig:q2qg}
\end{figure}

Since we require a finite (and large) $p_{hT}$ for the final-state hadron, 
the LO hard scattering has $X(k_X) = g(k_g)$ in \eq{eq:hard-qq}, given by
\beq[eq:lo-q2q]
q(k_1, \lambda) + \gamma^*(q, \lambda_{\gamma}) \to q(k_2, \lambda) + g(k_g, \lambda_g).
\eeq
The two diagrams are shown in \fig{fig:q2qg}.
We denote its helicity amplitude as $\Q_{\lambda \lambda_g}^{\lambda_{\gamma} }$,
making use of the chiral symmetry given in \eq{eq:ns-constraint}.
The partonic tensor defined in \eq{eq:W-def} is given at this order by
\begin{align}
	W_{\lambda\lambda'}^{\lambda_\gamma\lambda_\gamma'}(k_1, q, k_2)
	&= \sum_{\lambda_g = \pm 1} \int \frac{d^3 \bm{k}_g}{(2\pi)^3 2|\bm{k}_g|} \, (2\pi)^4 \delta^{(4)}(k_1 + q - k_2 - k_g) \, 
	\Q_{\lambda \lambda_g}^{\lambda_{\gamma}} \,
	\Q_{\lambda' \lambda_g}^{\lambda_{\gamma}' *},
	\label{eq:qq-hadronic-tensor-g}
\end{align}
with the gluon spin summed over and phase space integrated out.

The partonic kinematics is described by the variables $\hat{x}$ and $\hat{z}$ defined in \eq{eq:partonic-xz},
$Q^2$, 
$k_{2T} = p_{hT} / \xi_2$,
and $\phi_h$.
We also define the partonic c.m.\ energy squared, 
\beq
\s = (k_1 + q)^2 = \frac{Q^2 (1 - \x)}{\x}.
\eeq
After integrating over $\bm{k}_g$ with the delta function in \eq{eq:qq-hadronic-tensor-g} 
and factoring out the $\phi_h$ phase as in \eq{eq:W-phi0},
we normalize $W_{\lambda\lambda'}^{\lambda_\gamma\lambda_\gamma'}$ by
\begin{align}
	W_{\lambda \lambda'}^{\lambda_\gamma \lambda_\gamma'}(\x, \z, Q^2, k_{2T}^2, \phi_h)
	&= \pp{ 8\pi^2 e_q^2 \alpha_s C_F } \z \, \delta\!\pp{ k_{2T}^2 - \z (1 - \z) \s } \nn\\
	&\hspace{1em}
	\times e^{i [(\lambda - \lambda') - (\lambda_\gamma - \lambda_\gamma')] \phi_h}
	\cdot 
	\bar{w}_{\lambda \lambda'}^{\lambda_\gamma \lambda_\gamma'}(\x, \z),
	\label{eq:qq-hadronic-tensor-phi}
\end{align}
where $e_q$ is the electric charge of $q$, being $2/3$ and $-1/3$ for the up and down quarks, respectively,
and $C_F = 4/3$ is a color factor in QCD.
We see from the remaining delta function that $k_{hT}$ is not independent of $\z$ and $\x$; it is given by
\beq[eq:k2-hT-z]
k^2_{2T} = \z (1 - \z) \s = Q^2 \frac{\z}{\x} \, (1 - \z) (1 - \x).
\eeq
This will not be the case beyond LO.

We present the explicit expressions of $\bar{w}_{\lambda \lambda'}^{\lambda_\gamma \lambda_\gamma'}$ 
in terms of the hard coefficients in \eqss{eq:F-coef-U}{eq:F-coef-L}{eq:F-coef-T}.
Let us use a similar normalization to \eq{eq:qq-hadronic-tensor-phi},
\beq[eq:FX-fX]
F_{X}(Q^2, \x, \z, k_{2T}^2)
= \pp{ 8\pi^2 e_q^2 \alpha_s C_F } \z \, \delta\!\pp{ k_{2T}^2 - \z (1 - \z) \s } 
\cdot f_X(\x, \z),
\eeq
where $X$ refers to the super- and/or sub-scripts of $F$.
Then the hard coefficients for unpolarized target are 
\begin{align}
	f_{UU, T} &= \frac{1}{2} \pp{ \bar{w}^{++}_{++} + \bar{w}^{++}_{--} } 
	&&= 2 \bb{ \frac{1 + \x^2 \z^2}{(1 - \x) (1 - \z)} + (1 - \x) (1 - \z) }, \nn\\
	f_{UU, L} &= \bar{w}^{00}_{++} 
	&&= 8 \x \z, \nn\\
	f_{UU}^{(1)} &= -\frac{1}{\sqrt{2}} \Re\pp{ \bar{w}^{+0}_{++} + \bar{w}^{+0}_{--} } 
	&&= -4 \sqrt{\frac{\x \z}{(1 - \x) (1 - \z)}} \bb{(1 - \x) (1 - \z) + \x \z}, \nn\\
	f_{UU}^{(2)} &= -\Re\bar{w}^{+-}_{++} 
	&&= 4 \x \z, \nn\\
	f_{LU} &= -\frac{1}{\sqrt{2}} \Im\pp{ \bar{w}^{+0}_{++} + \bar{w}^{+0}_{--} } 
	&&= 0.
	\label{eq:fU-coefs-pt}
\end{align}
For polarized target, we have the hard coefficients, 
\begin{align}
	f_{UL}^{(1)} &= -\frac{1}{\sqrt{2}} \Im\pp{ \bar{w}^{+0}_{++} - \bar{w}^{+0}_{--} }
	&&= 0, \nn\\
	f_{UL}^{(2)} &= -\Im\bar{w}^{+-}_{++} 
	&&= 0, \nn\\
	f_{LL} &= \frac{1}{2} \pp{ \bar{w}^{++}_{++} - \bar{w}^{++}_{--} }
	&&= 2 \bb{ \frac{1 + \x^2 \z^2}{(1 - \x) (1 - \z)} - (1 - \x) (1 - \z) }, \nn\\
	f_{LL}^{(1)} &=-\frac{1}{\sqrt{2}} \Re\pp{ \bar{w}^{+0}_{++} - \bar{w}^{+0}_{--} }
	&&= 4 \sqrt{\frac{\x \z}{(1 - \x) (1 - \z)}} \bigbb{(1 - \x) (1 - \z) - \x \z}.
	\label{eq:fL-coefs-pt}
\end{align}
For transversely polarized target, the nonzero hard coefficients are
\begin{align}
	f_{UT, T} &= \bar{w}^{++}_{+-}
	&& = 4, \nn\\
	f_{UT}^{(1-)} &= \frac{1}{\sqrt{2}} \bar{w}^{+0}_{+-}
	&& = 4 \sqrt{\frac{\x \z}{(1 - \x) (1 - \z)}}, \nn\\
	f_{UT}^{(2-)} &= \frac{1}{2} \bar{w}^{+-}_{+-}
	&&= - \frac{4 \x \z}{(1 - \x) (1 - \z)},
	\label{eq:fT-coefs-pt}
\end{align}
while all the other coefficients are 0, 
\beq[eq:fT-coefs-pt-0]
f_{UT, L} = f_{UT}^{(1+)} = f_{UT}^{(2+)} = f_{LT} = f_{LT}^{(1+)} = f_{LT}^{(1-)} = 0.
\eeq

Although we have not explicitly calculated the antiquark-to-antiquark ($\bar{q} \to \bar{q}$) channel, its properties can be inferred from our results for the quark channel. We summarize the key points as follows:
(1) The hard coefficients are invariant under charge conjugation ($C$) and are therefore identical for the $\bar{q} \to \bar{q}$ and $q \to q$ channels.
(2) The FECs are also related by $C$ symmetry. For instance, the FEC for a $\bar{d}$ antiquark fragmenting into a $\pi^0$ is identical to that for a $d$ quark fragmenting into a $\pi^0$.
(3) The most significant difference in the overall contribution stems from the PDFs. The cross section for the antiquark channel is weighted by the antiquark PDF, $f_{\bar{q}/p}(x), g_{\bar{q}/p}(x), h_{\bar{q}/p}(x)$, which is known to be different from the quark PDF, $f_{q/p}(x), g_{q/p}(x), h_{q/p}(x)$, especially at large $x$.

\eq{eq:FX-fX} along with Eqs.~\eqref{eq:fU-coefs-pt}--\eqref{eq:fT-coefs-pt-0}
can be directly inserted into \eq{eq:conv-fin-pt} to give
the convolution with PDF and FEC,
\begin{align}
	\label{eq:conv-fin-pt-1}
	\C{f}{\fec}{F_X} =& \frac{ 8\pi^2 e_q^2 \alpha_s C_F}{z}
	\int_x^1 \frac{d\xi_1}{\xi_1} \int_z^1 \frac{d\xi_2}{\xi_2} \, f_{q/p}(\xi_1) \, \fec_{1, h/q}(\xi_2, \qf) \, 
	f_X\!\pp{ \frac{x}{\xi_1}, \frac{z}{\xi_2} } \nn\\
	&\hspace{12em}
	\times \delta\!\pp{ \frac{p_{hT}^2}{z^2} - Q^2 \frac{(\xi_1 - x)(\xi_2 - z)}{x z} }.
\end{align}
These can be readily integrated numerically, given an input of the PDF and FEC,
to make predictions for the cross section in \eq{eq:fsidis-TRc-result}.

\section{FECs from energy-tagged SIDIS: inclusive with $p_{hT}$}
\label{sec:fsidis-int-pt}
\begin{figure}
	\centering
	\includegraphics[scale=1.1]{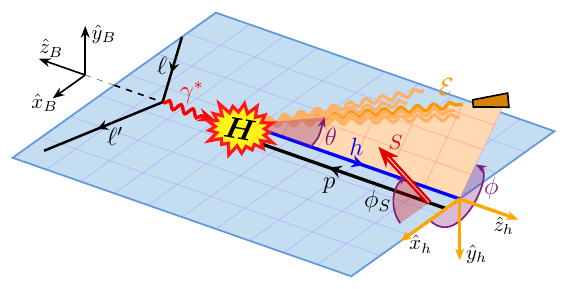}
	\caption{An illustration for the energy-tagged SIDIS with $\bm{p}_{hT}$ integrated. 
		The kinematics shown here corresponds to the LO QCD process, 
		where the recoiled quark moves along $-\z_B$ in the Breit frame.}
	\label{fig:recoil-frame}
\end{figure}

It is advantageous in many cases to only record the longitudinal momentum fraction $z$ in \eq{eq:sidis-kin-var} of the final-state hadron $h$,
while integrating over its transverse momentum $\bm{p}_{hT}$.
First, most of the events lie in the small $p_{hT}$ region, where the factorization in \eq{eq:Rc-factorize} does not apply. 
It needs to be replaced by a TMD factorization involving a TMD PDF of the target and TMD FEC of $h$~\cite{Ji:2004wu, Bacchetta:2006tn}.
As discussed in \sec{ssec:fec-tmd}, we regard FEC as an alternative way to probe transverse fragmentation dynamics.
Following that, it is the {\it collinear} factorization in \eq{eq:sidis-kin-var} that grants the FEC a unique advantage 
as compared to TMD distributions.
Hence, rather than seeking a TMD factorization at small $p_{hT}$, we inclusively integrate over it.
This still leads to a collinear factorization, with the hard scale associated with $h$ being its minus momentum in the proton-$\gamma^*$ c.m.\ frame,
\beq
p_{h, {\rm c.m.}}^- = z \, Q \, \sqrt{\frac{1-x}{2x}} \gg m_h,
\eeq
which requires $z$ not to be too small and $x$ away from $1$.
Second, integrating over $\bm{p}_{hT}$ greatly reduces the differential cross section formula in \eq{eq:fsidis-TRc-result},
making experimental extraction of the azimuthal modulation coefficients much simplified.
The azimuthal angle $\phi$ still serves as a useful observable. 
It is defined on an event-by-event basis for each finite value of $p_{hT}$ 
with respect to the hadron coordinate system in \eq{eq:hadron-frame}.
For $p_{hT} = 0$, we choose the latter as given by the 
backward-scattering limit of \eq{eq:hadron-frame} in the plane $\phi_h = \pi$,
related to the Breit frame system by
\beq[eq:hadron-system-z]
\hat{z}_h = - \hat{z}_B, \quad
\hat{x}_h = \hat{x}_B, \quad
\hat{y}_h = - \hat{y}_B,
\eeq
which is illustrated in \fig{fig:recoil-frame}.
Its justification will be presented below in \eq{eq:hard-qq-X0}.

In this section, we establish the framework for such an ``inclusive'' description. 
Roughly speaking, the cross section is obtained by integrating over $p_{hT}^2$ and $\phi_h$ in \eq{eq:fsidis-TRc-result},
but there are some subtleties to be addressed.
First, one immediately notices that for the $\bm{p}_{hT}$ integration not to affect the FECs, 
it is crucial for the latter to only depend on $z$ and $\qf$, but not explicitly on the transverse momentum $\bm{p}_{hT}$ of $h$.
This is the reason why we have reformulated FECs to be boost-invariant in \sec{ssec:mEEC-strategy}.
Second, upon the integration of $\phi_h$, 
all the modulations in \eq{eq:fsidis-TRc-result} that contain $\phi_h$ naively vanish.
However, this is only true if the $p_{hT}^2$ integral is well behaved, especially as $p_{hT}^2 \to 0$.
As we see from Eqs.~\eqref{eq:fU-coefs-pt}--\eqref{eq:fT-coefs-pt}, 
the hard coefficients $f_{UU, T}$, $f_{LL}$, and $f_{UT}^{(2-)}$ do get singular at $\hat{z} = 1$, or $p_{hT}^2 = 0$.
At this collinear kinematics, $\phi_h$ is undefined, and so the $\phi_h$ integral seems to become ambiguous. 
Addressing this issue is a main focus of this section. 
As it turns out, the singular $p_{hT}^2$ integration will produce a nonzero result to the $\phi_h$ integration, 
which contains collinear singularities that cancel with those from the virtual gluon loop corrections.

\subsection{Factorization with $\vec{p}_{hT}$ integrated}
\label{ssec:fsidis-int-pt-fac}
To properly treat the collinear singularity as $p_{hT} \to 0$ in the $\bm{p}_T$ integral, 
we use dimensional regularization (DR) in $d = 4 - 2\epsilon$ dimensions to regulate both ultraviolet (UV) and IR divergences.
This affects both the loop momenta and final-state phase space.
First, we write the phase space of the observed hadron $h$ as
\beq
\frac{d^{d-1} \bm{p}_h}{(2\pi)^{d-1} 2 |\bm{p}_h|}
= \frac{d z \, d^{d-2} \bm{p}_{hT}}{2 z \, (2\pi)^{d-1}}.
\eeq
This modifies the SIDIS cross section in \eq{eq:sidis-xsec} to
\begin{align}
	\frac{d\sigma}{dx \, dQ^2 \, d\phi_S \, dz}
	= \frac{y^2}{z \, Q^2} \frac{\alpha_e^2 }{4 \, (2\pi)^{4 - 2\epsilon}}
	\int d^{d-2} \bm{p}_{hT} \, \overline{|\M|^2},
	\label{eq:sidis-xsec-d}
\end{align}
where we only change the hadronic subprocess to $d$ dimensions,
while keeping the QED interaction and electron phase space in 4 dimensions;
no ambiguity arises as we only work to the LO in QED.
The $\overline{|\M|^2}$ in \eq{eq:sidis-xsec-d} differs from \eq{eq:amp2-sidis} in two ways:
(1) the phase space $\Pi_X$ of unobserved particles in $X$ and the 4-momentum conservation delta function are in $d$ dimensions, 
and
(2) there is no requirement for $h$ to carry large transverse momentum, 
and thus we allow the sum over $X$ to include states with all particles collinear to the target.

The same modification applies to the energy flow-weighted cross section in \eq{eq:sidis-xsec-e}, which becomes
\begin{align}
	\frac{d\Sigma}{dx \, dQ^2 \, d\phi_S \, dz \, d\eta \, d\phi}
	= \frac{y^2}{z \, Q^2} \frac{\alpha_e^2 }{4 \, (2\pi)^{4 - 2\epsilon}}
	\int d^{d-2} \bm{p}_{hT} \, \overline{|\M_{\E}|^2}.
	\label{eq:sidis-xsec-e-d}
\end{align}
On the right-hand side the $(\eta, \phi)$ of the energy flow are fixed,
which means that the energy flow keeps a fixed configuration 
with respect to the hadron $h$ as its $\bm{p}_{hT}$ is integrated event by event.
We restrict in the $R_C$ region where $\qf$ is small, 
so that the energy flow $\E$ resides in the nonperturbative FEC function 
and does not affect the perturbative calculation of the hard coefficients.
Hence for simplicity, we keep the four-dimensional definition of the energy flow operator and $(\eta, \phi, \qf)$.
The $\overline{|\M_{\E}|^2}$ in \eq{eq:sidis-xsec-e-d} differs from that in \eq{eq:sidis-xsec-e} 
in the same way as noted above.

For the factorization of \eq{eq:sidis-xsec-e-d} in the $R_C$ region, 
the $\int d^{d-2} \bm{p}_{hT} \, \overline{|\M_{\E}|^2}$ contains an additional leading region compared to the $\overline{|\M_{\E}|^2}$ in \eq{eq:sidis-xsec-e},
which has $N = 1$ in \fig{fig:region}(a),
that is, with only two collinear subgraphs associated with the target $p$ and observed hadron $h$, respectively.
The approximations used for collinear and soft lines in \sec{sssec:Rc-factorization}, however, stay unaffected,
and lead to a factorization similar to \eq{eq:Rc-factorize}.
The differences are:
\begin{itemize}
	\item the hard coefficients that describe the partonic scattering in \eq{eq:sidis-hard} now contain $X = {\rm vaccum}$, with $k_X = 0$; and
	\item the integration $\int d^{d-2} \bm{p}_{hT}$ propagates to the partonic level and 
	becomes the integration over $\bm{k}_{2T} = \bm{p}_{hT} / \xi_2$ of the parton $b$ in \eq{eq:sidis-hard} that fragments into $h$,
	\beq[eq:pt-int-to-kt-d]
	\int d^{d-2} \bm{p}_{hT} 
	= \xi_2^{2 - 2\epsilon} \int d^{d-2} \bm{k}_{2T}.
	\eeq
\end{itemize}
The factor $\xi_2^2$ in \eq{eq:pt-int-to-kt-d} cancels the denominator in \eq{eq:Rc-factorize},
and the remaining $\xi_2^{- 2\epsilon}$ is included in the definitions of FECs 
to properly normalize the latter and equip them with probabilistic interpretations in $d$ dimensions~\cite{Collins:2011zzd}.
Also assigning the $(2\pi)^{2\epsilon}$ factor in \eq{eq:sidis-xsec-e-d} to the hard coefficients $C$, $\Delta C$, and $T$, 
we have the factorization formula,
\begin{align}
	&(2\pi)^{2\epsilon} \int d^{d-2} \bm{p}_{hT} \, \overline{|\M_{\E}|^2} \nn\\
	&\hspace{2em} =
	\sum_{a, b} \int \frac{d\xi_1}{\xi_1} \, \int d\xi_2 \, 
	\bigg\{ \unpfec_{1, h/b}(\xi_2, \qf, \mu)
	\bb{ f_{a/p}(\xi_1, \mu) \, \iv{C_{ab} } \pp{ \frac{x}{\xi_1}, \frac{z}{\xi_2}; \frac{Q^2}{\mu^2} }	\right.\nn\\
		&\hspace{19em} \left. {}
		+ P_N \, g_{a/p}(\xi_1, \mu) \, \iv{ \Delta C_{ab} } \pp{ \frac{x}{\xi_1}, \frac{z}{\xi_2}; \frac{Q^2}{\mu^2} }
	}	\nn\\
	&\hspace{6em}  {}
	+ \colfec^{\perp}_{1, h/b}(\xi_2, \qf, \mu) \, h_{a/p}(\xi_1, \mu) \,
	\sum_{i, j = 1}^2 \frac{(\hat{p}_h \times \bm{n}_T)^i}{|\bm{n}_T|} 
	\, \bigiv{ T_{ab}^{ij} }\pp{ \frac{x}{\xi_1}, \frac{z}{\xi_2}; \frac{Q^2}{\mu^2} } s_T^j
	\bigg\} \nn\\
	&\hspace{3em}
	+ \mathcal{O}\bigpp{\qf / Q, \qf / p_{h, {\rm c.m.}}^-}.
	\label{eq:Rc-factorize-d}
\end{align}
The hard coefficients in \eq{eq:Rc-factorize-d} are given by those in \eq{eq:Rc-factorize}
under an extra $\bm{k}_{2T}$ integration,%
\footnote{We label $\bm{k}_{2T}$-integrated quantities $[O]$ by enclosing them by square brackets 
	to distinguish from the unintegrated ones $O$ of the same labels.}
\begin{align}
	&\cc{ \iv{ C_{ab} },  \iv{ \Delta C_{ab} },  \bigiv{ T_{ab}^{ij} } } \pp{ \frac{x}{\xi_1}, \frac{z}{\xi_2}; \frac{Q^2}{\mu^2} }  \nn\\
	&\hspace{4em} 
	= (2\pi)^{2\epsilon} \int d^{d-2} \bm{k}_{2T} 
	\cc{ C_{ab},  \Delta C_{ab},  T_{ab}^{ij} } \pp{ \frac{x}{\xi_1}, \frac{z}{\xi_2}; \frac{\bm{k}_{2T}}{Q}, \frac{Q^2}{\mu^2} }.
\end{align}
Note that although 
the prefactor $\hat{p}_h \times \bm{n}_T$ of the spin transfer coefficient $T$ depends on direction of $\bm{p}_h$, 
we are fixing the orientation of the energy flow {\it relative} to $h$, 
so that $(\hat{p}_h \times \bm{n}_T)^i / |\bm{n}_T| = (-\sin\phi, \cos\phi)^i$ is unaffected by the $\bm{k}_{2T}$ integration in \eq{eq:Rc-factorize-d}.
These hard coefficients can be obtained similarly from 
the $\bm{k}_{2T}$-integrated helicity matrix 
$\iv{ H }_{\lambda_1 \lambda_2, \lambda_1' \lambda_2'}$ as in \eqs{eq:C-delC-def-helicity}{eq:T-def-helicity},
which can be separated
into the same leptonic tensor $L_{\lambda_{\gamma} \lambda_{\gamma}'}$ for the $\gamma^*$ density matrix 
and $\bm{k}_{2T}$-integrated hadronic tensor $\iv{W}_{\lambda_1 \lambda_2, \lambda_1' \lambda_2'}^{\lambda_{\gamma}, \lambda_{\gamma}'}$
for the $\gamma^*$-parton scattering,
yielding the same formulae as \eqs{eq:hard-H-helicity-photon}{eq:hard-H-denmtx-W}.
The integrated $W$ is defined as
\begin{align}
	\iv{W}_{\lambda_1 \lambda_2, \lambda_1' \lambda_2'}^{\lambda_{\gamma}, \lambda_{\gamma}'}\pp{ \frac{x}{\xi_1}, \frac{z}{\xi_2}; \frac{Q^2}{\mu^2} } 
	&= (2\pi)^{2\epsilon} \int d^{d-2} \bm{k}_{2T} 
	\sum\nolimits_X \int d\Pi_X \, (2\pi)^d \delta^{(d)}(q + k_1 - k_2 - k_X) 	\nn\\
	&\hspace{7em} \times
	\Q_{\lambda_1 \lambda_2}^{\lambda_{\gamma}}(k_1, q, k_2; X) \,
	\Q_{\lambda_1' \lambda_2'}^{\lambda_{\gamma}' *}(k_1, q, k_2; X).
	\label{eq:W-def-int}
\end{align}
We will stick to the flavor NS channel, $(a, b) = (q, q)$, and use the shorthand notations
$\iv{W}_{\lambda \lambda'}^{\lambda_{\gamma} \lambda_{\gamma}'}$ and $\Q_{\lambda}^{\lambda_{\gamma}}$
for $\iv{W}_{\lambda \lambda, \lambda' \lambda'}^{\lambda_{\gamma}, \lambda_{\gamma}'}$
and $\Q_{\lambda\lambda}^{\lambda_{\gamma}}$, respectively,
as in \eq{eq:ns-constraint}.
In the following, we distinguish two cases for the calculation of $\iv{W}$,
with $X$ containing nothing, denoted as $X = 0$, or at least one particle, $X > 0$,
and then formulate observables in $\iv{C}$, $\iv{\Delta C}$, and $\iv{T}$.

\subsection{Hard coefficients for $X = 0$}
\label{ssec:fsidis-int-pt-hard1}
When $X$ contains nothing, the partonic scattering is a simple $2 \to 1$ process,
\beq[eq:hard-qq-X0]
q(k_1, \lambda) + \gamma^*(q, \lambda_{\gamma}) \to q(k_2, \lambda),
\eeq
which happens on top of the electron scattering plane, 
with $q(k_1)$ moving along $\hat{z}_B$ and $\gamma^*$ and $q(k_2)$ along $-\hat{z}_B$.

The first question we need to answer is how the hadron coordinate system is defined for this kinematics,
which is apparently a singularity of the definition in \eq{eq:hadron-frame}.
As remarked below \eq{eq:spinor-conversion-T}, 
the $\hat{x}_h$ and $\hat{y}_h$ are determined by the definition of the helicity state $|q(k_2, \lambda)\rangle$,
which is given by a rotation from the state of the same energy and helicity moving along the $\hat{z}_B$ direction~\cite{Weinberg:1995mt}.
However, for $k_2$ moving along the $-\hat{z}_B$ direction, 
there is an ambiguity regarding its azimuthal angle $\phi_0$.
We fix the convention by demanding that the parity relation in \eq{eq:parity-Wbar} directly extrapolate to this case. %
Since a parity transformation relates a state along $\hat{n}(\theta, \phi)$ to one along $\hat{n}(\pi - \theta, \pi + \phi)$, for $\phi \in [0, \pi)$, 
we need $-\hat{z}_B = \hat{n}(\pi, \pi)$ to pair with $\hat{z}_B = \hat{n}(0, 0)$.
Hence, we choose $\phi_0 = \pi$, such that the hadron system is obtained from the Breit system via 
a rotation around $\hat{y}_B$ by $\pi$ followed by a rotation around $\hat{z}_B$ by $\pi$,
which leads to \eq{eq:hadron-system-z}.

Setting $X = 0$ in \eq{eq:W-def-int}, we can use the delta function to integrate over $\bm{k}_{2T}$,
\begin{align}
	\iv{W_{X=0} }_{\lambda \lambda'}^{\lambda_{\gamma} \lambda_{\gamma}'}\pp{ \x, \z; Q^2 / \mu^2 } 
	&= (2\pi)^4 \, \frac{2 \, \delta(1 - \x) \, \delta(1 - \z)}{Q^2} \,
	\Q_{\lambda}^{\lambda_{\gamma}} \,
	\Q_{\lambda'}^{\lambda_{\gamma}' *},
	\label{eq:W-def-int-X0}
\end{align}
which constrains both $\x$ and $\z$ [defined in \eq{eq:partonic-xz}] to be 1.
By angular momentum conservation along the $\hat{z}_B$ axis, 
the amplitude $\Q_{\lambda}^{\lambda_{\gamma}}$ is nonzero 
only when $2\lambda = \lambda_\gamma$,
yielding two nonzero components $\Q_{\pm}^{\pm}$.
Parity further constrains $\Q_{+}^+ = \Q_{-}^-$, 
and thus the four nonzero components of $W$ are all equal and positive,
\begin{align}
	\iv{W_{X=0}}^{++}_{++} &= \iv{W_{X=0}}^{--}_{--} = \iv{W_{X=0}}^{+-}_{+-} = \iv{W_{X=0}}^{-+}_{-+} \nn\\
	& = (2\pi)^4 \, \frac{2 \, \delta(1 - \x) \, \delta(1 - \z)}{Q^2}
	\left| \Q_{+}^{+} \right|^2 
	> 0.
	\label{eq:W-X0-2}
\end{align}
Applied to the integrated version of \eq{eq:hard-H-denmtx-W}, this gives the nonzero components of $\iv{H}$, 
\begin{align}
	\iv{H_{X=0}}_{\pm \pm} 
	&= \frac{2}{Q^2 (1 - \varepsilon)} 
	\pp{ \frac{1 \pm P_e \sqrt{1 - \varepsilon^2}}{2} \iv{W_{X=0}}_{+ +}^{+ +} }, \nn\\
	\iv{H_{X=0}}_{+ -} &= \iv{H_{X=0}}_{- +} %
	= \frac{2}{Q^2 (1 - \varepsilon)} \pp{ -\frac{\varepsilon}{2} \iv{W_{X=0}}_{+ +}^{+ +} },
\end{align}
which then determine the hard coefficients $\iv{C}$, $\iv{\Delta C}$, and $\iv{ T^{ij} }$ by \eqs{eq:C-delC-def-helicity}{eq:T-def-helicity},
\bse[eq:C-del-C-T-X0]\begin{align}
	\iv{ C_{qq, X=0} } &%
	= \frac{2}{Q^2 (1 - \varepsilon)} \pp{ \frac{1}{2} \iv{W_{X=0}}_{++}^{++} }, \\
	\iv{ \Delta C_{qq, X=0} } &%
	= \frac{2}{Q^2 (1 - \varepsilon)} \biggpp{ \frac{P_e \sqrt{1 - \varepsilon^2}}{2} \iv{W_{X=0}}_{++}^{++} }, \\
	\bigiv{ T_{qq, X=0}^{xx} } &= \bigiv{ T_{qq, X=0}^{yy} } %
	= \frac{2}{Q^2 (1 - \varepsilon)} \pp{ -\frac{\varepsilon}{2} \iv{W_{X=0}}_{++}^{++} }, \\
	\bigiv{ T_{qq, X=0}^{xy} } &= \bigiv{ T_{qq, X=0}^{yx} } = 0.
\end{align}\ese
The last two equations of \eq{eq:C-del-C-T-X0} imply that 
the quark transverse spin is transferred just as being ``bent'' on the electron scattering plane, 
\beq[eq:X0-spin-flip]
\bm{s}_{T, q}' \propto -\bm{s}_{T, q}.
\eeq 
Notice that the coordinate systems describing the initial- and final-state quark spins, 
$\bm{s}_{T, q}$ and $\bm{s}_{T, q}'$, respectively,
differ by flipping the $y$ and $z$ axes.
Hence, \eq{eq:X0-spin-flip} means that 
the transverse spin component lying in the electron-scattering plane is flipped, 
while the perpendicular component stays unchanged.

The above discussion applies to all orders of the partonic process in \eq{eq:hard-qq-X0}.
With $X = 0$, the LO is given at tree level, and high-order corrections include only virtual gluon emissions.
In the following, we present these results up to NLO for the $q\to q$ channel.

\subsubsection{LO hard coefficients for $X = 0$}
\label{sssec:fsidis-int-pt-hard1-lo}

\begin{figure}[htbp]
	\centering
	\begin{tabular}{cccc}
		\includegraphics[scale=0.6]{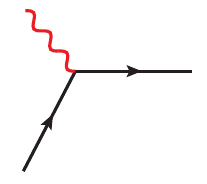} &
		\includegraphics[scale=0.6]{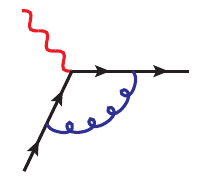} &
		\includegraphics[scale=0.6]{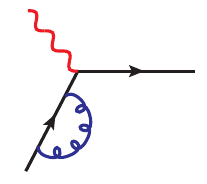} &
		\includegraphics[scale=0.6]{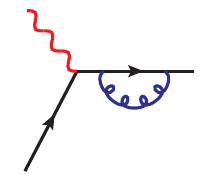} \\
		(a) & (b) & (c) & (d)
	\end{tabular}
	\caption{LO (a) and NLO (b)--(d) diagrams of the $q \to q$ scattering amplitude in \eq{eq:hard-qq-X0}.
		Inclusion of the counterterm diagrams (not shown) for (b)--(d) is understood.}
	\label{fig:q2q-X0}
\end{figure}

The LO amplitude for the scattering \eq{eq:hard-qq-X0} is shown in \fig{fig:q2q-X0}(a).%
\footnote{Note that here ``LO'' refers to the no-radiation case, while ``NLO'' has one virtual or real gluon radiation.
		This differs from the terminology for finite-$p_{hT}$ observables in \sec{sec:fsidis}.}
Using the normalization of \eq{eq:a-q-amplitude},
it can be explicitly computed as
\beq[eq:a-q-amplitude-X0-LO]
\Delta_0 \Q^{+}_{+}
= \Delta_0 \Q^{-}_{-}
= -\sqrt{2} \, i \, e_q \, Q.
\eeq
We use the prefix notation $\Delta_n$ to label the order of $\alpha_s$ in perturbative expansions,
\beq
A = \Delta_0 A + \Delta_1 A + \Delta_2 A + \cdots,
\eeq
with $\Delta_n A \propto \alpha_s^n$.
\eq{eq:a-q-amplitude-X0-LO} gives the $\iv{W}$ matrix in \eq{eq:W-X0-2},%
\footnote{For simplicity, we suppress the subscript $X = 0$ 
	as it is already encoded by the LO prefix $\Delta_0$.}
\beq[eq:W-X0-LO]
\iv{\Delta_0 W}_{++}^{++} = \iv{\Delta_0 W}^{--}_{--} = \iv{\Delta_0 W}^{+-}_{+-} = \iv{\Delta_0 W}^{-+}_{-+}
= 4 (2\pi)^4 \, e_q^2 \, \delta(1 - \x) \, \delta(1 - \z).
\eeq

\subsubsection{NLO hard coefficients for $X = 0$: virtual correction}
\label{sssec:fsidis-int-pt-hard1-nlo}

The NLO corrections are shown in \fig{fig:q2q-X0}(b)--(d).
Due to chiral and parity symmetries, they only contribute an overall factor to \eq{eq:a-q-amplitude-X0-LO}.
In the Feynman gauge, the vertex correction factor is given by the one-loop diagram in \fig{fig:q2q-X0}(b),
\beq
f_v(Q^2) 
= -\frac{\alpha_s C_F}{2\pi} \, \frac{e^{\epsilon( L + \gamma_{\rm E})} }{ \Gamma(1 - \epsilon) }
\bb{ \frac{1}{\epsilon^2}+\frac{3}{2\epsilon}+4+\mathcal{O}(\epsilon)},
\eeq
and its UV counterterm in the $\MS$ scheme,
\beq
\delta f_v = - \frac{\alpha_s C_F}{4\pi}
\bb{ \frac{1}{\epsilon} + \ln(4\pi e^{-\gamma_{\rm E}}) }.
\eeq
Here, we have defined
\beq[eq:def-T]
\bar{\mu}^2 = 4 \pi \mu^2 e^{-\gamma_{\rm E}}, \quad
L = \ln \pp{ \frac{\bar{\mu}^2}{Q^2} }.
\eeq
The external quark leg correction factor shown in \fig{fig:q2q-X0}(c) and (d) 
is given by the residue of the renormalized two-point function,
\beq
f_e = \frac{\alpha_s C_F}{4\pi}
\bb{ \frac{1}{\epsilon} + \ln(4\pi e^{-\gamma_{\rm E}}) }
= -\delta f_v.
\eeq
Combining these together, the vertex counterterm cancels the external quark leg correction, 
leaving just the factor $f_v(Q^2)$ from the bare vertex loop.

Adding the virtual correction to the tree-level amplitude in \eq{eq:a-q-amplitude-X0-LO}, we have
\beq[eq:a-q-amplitude-X0-NLO]
\Delta_1 \Q^{+}_{+}
= \Delta_1 \Q^{-}_{-} 
= -\sqrt{2} \, i \, e_q \, Q f_v(Q^2).
\eeq
This gives the NLO virtual correction to \eq{eq:W-X0-LO},%
\footnote{The prefix notation $\Delta_1^V$ indicating virtual corrections at NLO automatically implies the configuration $X = 0$, 
	so we have suppressed the subscript ``$X = 0$''.}
\begin{align}
	\iv{ \Delta_1^V W }_{+ +}^{++} 
	&= \iv{ \Delta_1^V W }_{--}^{--} 
	= \iv{ \Delta_1^V W }_{+-}^{+-} 
	= \iv{ \Delta_1^V W }_{-+}^{-+} \nn\\
	&= 4 (2\pi)^4 \, e_q^2 \, \delta(1 - \x) \, \delta(1 - \z) \bb{ 2 f_v(Q^2) }.
	\label{eq:W-X0-NLO}
\end{align}

\subsection{Hard coefficients for $X > 0$: real correction}
\label{ssec:fsidis-int-pt-hard2}
The LO contribution to the high-$p_{hT}$ hadron production in \sec{sssec:sidis-lo-qg}
now becomes the real correction to $p_{hT}$-integrated case at NLO.
Most of the calculations are the same as \sec{sssec:sidis-lo-qg}. We concentrate only on their differences.

First, the helicity amplitude $\Q_{\lambda \lambda_g}^{\lambda_{\gamma}}$ of the $2\to 2$ scattering in \eq{eq:lo-q2q} stays the same,
with only the coupling $g$ modified to $g \mu^{\epsilon}$ in DR.
Following \eq{eq:W-def-int}, we supplement \eq{eq:qq-hadronic-tensor-phi} with the factor 
$(4\pi^2 \mu^2)^{\epsilon}$ and the $\bm{k}_{2T}$ integration,
\begin{align}
	\iv{ \Delta_1^R W}_{\lambda\lambda'}^{\lambda_\gamma\lambda_\gamma'}(\x, \z; Q^2 / \mu^2)
	&= (4\pi^2 \mu^2)^{\epsilon} \int d^{d-2} \bm{k}_{2T} 
	\pp{ 8\pi^2 e_q^2 \alpha_s C_F } \z \, \delta\!\pp{ k_{2T}^2 - \z (1 - \z) \s } \nn\\
	&\hspace{10em}
	\times e^{i [(\lambda - \lambda') - (\lambda_\gamma - \lambda_\gamma')] \phi_h}
	\cdot 
	\bar{w}_{\lambda \lambda'}^{\lambda_\gamma \lambda_\gamma'}(\x, \z)	\nn\\
	&\hspace{-9em}= (2\pi)^4 \z \, e_q^2 \, \frac{\alpha_s C_F}{2\pi}
	\frac{ e^{ \epsilon ( L +\gamma_{\rm E} ) } }{ \Gamma(1 - \epsilon) }
	\int_0^{2\pi} \frac{d\phi_h}{2\pi} \biggbb{ 
		e^{i [(\lambda - \lambda') - (\lambda_\gamma - \lambda_\gamma')] \phi_h}
		\frac{\x^{\epsilon} \, \z^{-\epsilon} \, 
			\bar{w}_{\lambda \lambda'}^{\lambda_\gamma \lambda_\gamma'}(\x, \z) 
		}{(1 - \x)^{\epsilon} \, (1 - \z)^{\epsilon} }
	},
	\label{eq:qq-hadronic-tensor-phi-int}
\end{align}
where in the second step we used \eq{eq:kT-to-phi} to separate the $\bm{k}_{2T}$ integration into $k_{2T}^2$ and $\phi_h$,
and used the delta function to integrate over $k_{2T}^2$.
The $\phi_h$-independent factor $\bar{w}$
is the same as that in \eq{eq:qq-hadronic-tensor-phi}
and has been presented in Eqs.~\eqref{eq:fU-coefs-pt}--\eqref{eq:fT-coefs-pt}.

As noted at the beginning of this section, the $\phi_h$ integral is trivial 
if we take the integrations of $k_{2T}^2$ and $\phi_h$ as fully independent,
\beq[eq:naive-phih-integral]
\int_0^{2\pi} \frac{d\phi_h}{2\pi} \, e^{i [(\lambda - \lambda') - (\lambda_\gamma - \lambda_\gamma')] \phi_h}
\overset{?}{=}
\delta_{\lambda - \lambda', \, \lambda_\gamma - \lambda_\gamma'}.
\eeq
If this were indeed the case, the only surviving components in \eq{eq:qq-hadronic-tensor-phi-int} would be
\beq[eq:w-no-phase]
\iv{ \Delta_1^R W}_{++}^{++}, \quad
\iv{ \Delta_1^R W}_{--}^{++}, \quad
\iv{ \Delta_1^R W}_{++}^{0 \, 0}, \quad
\iv{ \Delta_1^R W}_{+-}^{+ \, 0},
\eeq
plus their Hermitian-conjugate and parity counterparts.
The $\iv{W}^{+-}_{+-}$ would not receive any contribution from $X > 0$,
and therefore, the IRC divergences of $\iv{\Delta_1^V W}^{+-}_{+-}$ in \eq{eq:W-X0-NLO} would remain uncanceled.

We rescue this paradox by noticing that the separation of $k_{2T}^2$ and $\phi_h$ is singular as $k_{2T}^2 \to 0$.
Having used the delta function in \eq{eq:qq-hadronic-tensor-phi-int} to set $k_{2T}^2$ to \eq{eq:k2-hT-z},
we converted the singular kinematic point to $\x = 1$ and/or $\z = 1$.
These correspond to three kinematic configurations:%
\footnote{The case with both the final-state quark and gluon being soft does not exist. 
	Otherwise we would have $k_1^+ + q^+ = q^- = 0$, giving $Q^2 = 0$.}
\begin{enumerate}
	\item when $\x = 1$ and $\z \neq 1$, 
	the gluon moves collinearly with the final-state quark in the $-\z_B$ direction;
	\item when $\x \neq 1$ and $\z = 1$, 
	the gluon is collinearly radiated by the initial-state quark in the $\z_B$ direction,
	with the final-state quark moving along $-\z_B$;
	\item when $\x = \z = 1$, the final-state quark moves along the $-\z_B$ direction, with the gluon being soft.
\end{enumerate}
In all these cases, the final-state quark moves in the $-\z_B$ direction, 
so that the kinematic singular point fixes $\phi_h$ to $\pi$,
according to our discussion at the beginning of \sec{ssec:fsidis-int-pt-hard1}. 
This means that if the $\x$-$\z$ integrand in \eq{eq:qq-hadronic-tensor-phi-int} is singular at either $\x = 1$ or $\z = 1$, 
we should set $\phi_h = \pi$ for the singular piece
and thus obtain a non-vanishing $\phi_h$ integral.
This is the case for $\iv{\Delta_1^R W}_{++}^{++}$ and $\iv{\Delta_1^R W}_{+-}^{+-}$, 
exactly corresponding to the helicity structures for $X = 0$ in \eq{eq:W-X0-NLO}.
Their combinations would then cancel the soft divergences.

Below, we first consider the $\phi_h$ integrals for these singular coefficients,
and then for remaining non-singular ones.

\subsubsection{$\phi_h$ integrals for the singular hard coefficients}
\label{sssec:fsidis-int-pt-singular}
For $\iv{\Delta_1^R W}_{++}^{++}$, the phase factor in \eq{eq:qq-hadronic-tensor-phi-int} reduces to unity,
so its $\phi_h$ integral is trivial.
From \eqs{eq:fU-coefs-pt}{eq:fL-coefs-pt}, we have
\begin{align}
	\bar{w}_{++}^{++}(\x, \z) = f_{UU, T} + f_{LL}
	= \frac{4(1 + \x^2 \z^2)}{(1 - \x) (1 - \z)}. %
\end{align}
Plugging this into \eq{eq:qq-hadronic-tensor-phi-int} yields
\begin{align}
	\iv{ \Delta_1^R W}_{++}^{++}
	&= 4 (2\pi)^4 \z \, e_q^2 \, \frac{\alpha_s C_F}{2\pi}
	\frac{ e^{ \epsilon ( L +\gamma_{\rm E} ) } }{ \Gamma(1 - \epsilon) }
	\bb{
		\frac{(1 + \x^2 \z^2) \, \x^{\epsilon} \, \z^{-\epsilon}
		}{(1 - \x)^{1 + \epsilon} \, (1 - \z)^{1 + \epsilon} }
	}.
	\label{eq:qq-W,++,++-phi-int}
\end{align}
The IRC singularities at $\x = 1$ and $\z = 1$ are regulated with $\epsilon < 0$ in DR,
and can be explicitly exposed by using the standard formula,
\begin{align}
	\frac{\x^{\epsilon}}{(1 - \x)^{1 + \epsilon}}
	&= -\frac{\delta(1 - \x)}{\epsilon} + \frac{1}{(1 - \x)_+} 
	+ \epsilon \bb{ \frac{\ln \x}{1 - \x} - \pp{ \frac{\ln (1 - \x)}{1 - \x} }_+ }
	+ \order{\epsilon^2}, \nn\\
	\frac{\z^{-\epsilon}}{(1 - \z)^{1 + \epsilon}}
	&= -\frac{\delta(1 - \z)}{\epsilon} + \frac{1}{(1 - \z)_+} 
	- \epsilon \bb{ \frac{\ln \z}{1 - \z} + \pp{ \frac{\ln (1 - \z)}{1 - \z} }_+ }
	+ \order{\epsilon^2}.
	\label{eq:1/(1-x/z)-expansion}
\end{align}
Then we rewrite \eq{eq:qq-W,++,++-phi-int} as
\begin{align}
	&\iv{ \Delta_1^R W}_{++}^{++}
	= 4 (2\pi)^4 \z \, e_q^2 \, \frac{\alpha_s C_F}{2\pi}
	\frac{ e^{ \epsilon ( L +\gamma_{\rm E} ) } }{ \Gamma(1 - \epsilon) }\,
	\Bigg\{ 
	\frac{ 2 \, \delta(1 - \x) \delta(1 - \z) }{\epsilon^2}
	+ \frac{1 + \x^2 \z^2}{(1 - \x)_+ (1 - \z)_+}
	\nn\\
	&\hspace{2em}
	+ \delta(1 - \z) 
	\bb{ \frac{1}{\epsilon}
		\pp{ 1 + \x - \frac{2}{(1 - \x)_+} }
		- (1 + \x^2) \pp{ \frac{\ln \x}{1 - \x} 
			- \pp{ \frac{\ln (1 - \x)}{1 - \x}}_+
		}
	}
	\nn\\
	&\hspace{2em}
	+ \delta(1 - \x) 
	\bb{ \frac{1}{\epsilon}
		\pp{ 1 + \z - \frac{2}{(1 - \z)_+} }
		+ (1 + \z^2) \pp{ \frac{\ln \z}{1 - \z} 
			+ \pp{ \frac{\ln (1 - \z)}{1 - \z}}_+
		}
	}
	+ \order{\epsilon}
	\Bigg\}.
	\label{eq:qq-W,++,++-phi-int-e}
\end{align}

For $\iv{\Delta_1^R W}_{+-}^{+-}$, we use \eq{eq:fT-coefs-pt} to write its $\phi_h$ integral as
\begin{align}
	\iv{ \Delta_1^R W}_{+-}^{+-}
	&= (2\pi)^4 \z \, e_q^2 \, \frac{\alpha_s C_F}{2\pi}
	\frac{ e^{ \epsilon ( L +\gamma_{\rm E} ) } }{ \Gamma(1 - \epsilon) }
	\int_0^{2\pi} \frac{d\phi_h}{2\pi} \biggbb{ 
		e^{-i \phi_h}
		\frac{ (- 8 \x \z) \, \x^{\epsilon} \, \z^{-\epsilon}
		}{(1 - \x)^{1 + \epsilon} \, (1 - \z)^{1 + \epsilon} }
	},
	\label{eq:qq-W,+-,+--phi-int}
\end{align}
This contains the same IRC singularities and can be expanded in the same way as \eq{eq:qq-W,++,++-phi-int-e}.
The difference, however, lies in the phase $e^{-i \phi_h}$ under the $\phi_h$ integral,
which vanishes unless $\x$ or $\z$ is forced to $1$ by the delta functions in \eq{eq:1/(1-x/z)-expansion}.
Therefore, we retain only terms proportional to $\delta(1 - \x)$ or $\delta(1 - \z)$ and set $\phi_h = \pi$ in the integrand.
This gives
\begin{align}
	&\iv{ \Delta_1^R W}_{+-}^{+-} %
	= 8(2\pi)^4 \z \, e_q^2 \, \frac{\alpha_s C_F}{2\pi}
	\frac{ e^{ \epsilon ( L +\gamma_{\rm E} ) } }{ \Gamma(1 - \epsilon) }
	\Bigg\{ 
	\frac{ \delta(1 - \x) \delta(1 - \z) }{\epsilon^2}
	\nn\\
	&\hspace{4em}
	+ \delta(1 - \z) 
	\bb{ \frac{1}{\epsilon}
		\pp{ 1 - \frac{1}{(1 - \x)_+} }
		- \x \pp{ \frac{\ln \x}{1 - \x} 
			- \pp{ \frac{\ln (1 - \x)}{1 - \x}}_+
		}
	}
	\nn\\
	&\hspace{4em}
	+ \delta(1 - \x) 
	\bb{ \frac{1}{\epsilon}
		\pp{ 1 - \frac{1}{(1 - \z)_+} }
		+ \z \pp{ \frac{\ln \z}{1 - \z} 
			+ \pp{ \frac{\ln (1 - \z)}{1 - \z}}_+
		}
	}
	+ \order{\epsilon}
	\Bigg\}.
	\label{eq:qq-W,+-,+--phi-int-e}
\end{align}

In fact, we can argue to all orders 
that $\iv{ W}_{+-}^{+-}$ is the only nontrivial component 
(together with its conjugate $[W]_{-+}^{-+}$) 
under the $\phi_h$ integral.
The reason is that one needs both a non-unity $\phi_h$ phase 
and a singularity at $\x = 1$ or $\z = 1$ in \eq{eq:qq-hadronic-tensor-phi-int}.
As we noted above, this singularity is of soft or collinear origin 
and happens at the edge of phase space with only one final-state collinear sector associated with the produced quark.
It can therefore be factorized into a PDF for the initial-state quark, an FF for the final-state quark,
and a hard coefficient that is exactly given by \eq{eq:W-def-int-X0} with no other particles in the final state.
Chiral symmetry along with helicity conservation then fix the helicity structures to those in \eq{eq:W-X0-2}, where only $\iv{ W}_{+-}^{+-}$ and $[W]_{-+}^{-+}$ have nontrivial phases.

\subsubsection{$\phi_h$ integrals for non-singular hard coefficients}
\label{sssec:fsidis-int-pt-non-singular}
The other non-vanishing components are given by the last three terms in \eq{eq:w-no-phase}.
They are IRC safe, $\phi_h$ independent, and absent for $X = 0$, 
so we can remove their dimensional regulator and simply integrate over $\phi_h$.
This gives
\begin{align}
	&\cc{ \iv{ \Delta_1 W}_{--}^{++}, \, \iv{ \Delta_1 W}_{++}^{00}, \, \iv{ \Delta_1 W}_{+-}^{+0} }
	= \cc{ \iv{ \Delta_1^R W}_{--}^{++}, \, \iv{ \Delta_1^R W}_{++}^{00}, \, \iv{ \Delta_1^R W}_{+-}^{+0} } \nn\\
	&\hspace{6em}
	= 4 (2\pi)^4 \z \, e_q^2 \, \frac{\alpha_s C_F}{2\pi}
	\cc{ (1 - \x) (1 - \z), \, 2 \x \z, \, \sqrt{\frac{2 \x \z}{(1 - \x) (1 - \z)}} }.
	\label{eq:qq-Wreg-phi-int}
\end{align}

\subsection{Cancellation of IRC singularities}
\label{ssec:fsidis-IRC}
The factorization theorem in \eq{eq:Rc-factorize-d} implies that 
(1) soft divergences cancel between real and virtual gluon emissions,
and
(2) collinear divergences are removed by subtractions associated with the PDFs and FECs.
In this sub-section, we verify these two facts through our NLO calculation of the $q\to q$ channel,
which also provides a consistency check of the perturbative result,
particularly our treatment of 
the singular $\phi_h$-dependent $\bm{k}_{2T}$ integral.

The three components in \eq{eq:qq-Wreg-phi-int} start appearing only at NLO,
so \eq{eq:qq-Wreg-phi-int} is already their final expressions. 
Components related to them by Hermiticity and parity are
\begin{align}
	\iv{ W}_{--}^{++} = \iv{ W }^{--}_{++}, \quad
	\iv{ W }^{00}_{--} = \iv{ W }^{00}_{++}, \quad
	\iv{ W }^{+0}_{+-} = \iv{ W }^{0+}_{-+} = - \iv{ W }^{-0}_{-+} = - \iv{ W }^{0-}_{+-}.
	\label{eq:Ws-int-finite}
\end{align}
The nontrivial components to be dealt with in the following are 
\begin{align}
	\iv{ W}_{++}^{++} = \iv{ W}_{--}^{--}, \quad
	\iv{ W}_{+-}^{+-} = \iv{ W}_{-+}^{-+}.
	\label{eq:Ws-int-div}
\end{align}
All other components of $\iv{W}$ vanish under the $\phi_h$ integration. 

\subsubsection{Combination of real and virtual corrections}
\label{sssec:fsidis-RV-combine}
The $[W]^{++}_{++}$ at NLO is given by the 
virtual contribution in \eq{eq:W-X0-NLO}
and the real contribution in \eq{eq:qq-W,++,++-phi-int-e}.
Combining them together gives
\begin{align}
	[\Delta_1 W]^{++}_{++} 
	&= [\Delta_1^V W]^{++}_{++} + [\Delta_1^R W]^{++}_{++} \nn\\
	&= 4 (2\pi)^4 \z \, e_q^2 \, \frac{\alpha_s C_F}{2\pi}
	\frac{ e^{ \epsilon ( L +\gamma_{\rm E} ) } }{ \Gamma(1 - \epsilon) }\,
	\Bigg\{ \!\!
	-8 \, \delta(1 - \x) \delta(1 - \z)
	+ \frac{1 + \x^2 \z^2}{(1 - \x)_+ (1 - \z)_+}
	\nn\\
	&\hspace{1em}
	- \delta(1 - \z) \bb{ \frac{ P_{qq}(\x) }{\epsilon}
		+ (1 + \x^2) \pp{ \frac{\ln \x}{1 - \x} - \pp{ \frac{\ln (1 - \x)}{1 - \x}}_+ }
	}
	\nn\\
	&\hspace{1em}
	- \delta(1 - \x) \bb{ \frac{K_{qq}(\z) }{\epsilon}
		- (1 + \z^2) \pp{ \frac{\ln \z}{1 - \z} + \pp{ \frac{\ln (1 - \z)}{1 - \z}}_+ }
	}
	+ \order{\epsilon}
	\Bigg\},
	\label{eq:W++,++, NLO}
\end{align}
where the double pole $1/\epsilon^2$ is canceled, signifying the soft gluon cancellation.
The coefficients of the remaining $1/\epsilon$ poles exactly 
organize into the (LO) evolution kernels of PDF and FF,  
$P_{qq}(\x)$ and $K_{qq}(\z)$, respectively~\cite{Altarelli:1977zs},
\beq
P_{qq}(x) = K_{qq}(x) = -1 - x + \frac{2}{(1 - x)_+} + \frac{3}{2} \delta(1-x).
\eeq

Similarly, for $\iv{W}^{+-}_{+-}$, we combine \eqs{eq:W-X0-NLO}{eq:qq-W,+-,+--phi-int-e} and obtain the NLO contribution,
\begin{align}
	[\Delta_1 W]^{+-}_{+-}
	&= [\Delta_1^V W]^{+-}_{+-} + [\Delta_1^R W]^{+-}_{+-} \nn\\
	&= 4(2\pi)^4 \z \, e_q^2 \, \frac{\alpha_s C_F}{2\pi}
	\frac{ e^{ \epsilon ( L +\gamma_{\rm E} ) } }{ \Gamma(1 - \epsilon) }
	\Bigg\{ \!\!
	-8 \, \delta(1 - \x) \delta(1 - \z)
	\nn\\
	&\hspace{2em}
	- \delta(1 - \z) 
	\bb{ \frac{\delta_T P_{qq}(\x)}{\epsilon}
		+ 2 \x \pp{ \frac{\ln \x}{1 - \x} 
			- \pp{ \frac{\ln (1 - \x)}{1 - \x}}_+
		}
	}
	\nn\\
	&\hspace{2em}
	- \delta(1 - \x) 
	\bb{ \frac{\delta_T K_{qq}(\z)}{\epsilon}
		- 2 \z \pp{ \frac{\ln \z}{1 - \z} 
			+ \pp{ \frac{\ln (1 - \z)}{1 - \z}}_+
		}
	}
	+ \order{\epsilon}
	\Bigg\},
	\label{eq:W+-,+-, NLO}
\end{align}
where $\delta_T P_{qq}(\x)$ and $\delta_T K_{qq}(\z)$ are, respectively,
the LO evolution kernels of the transversity PDF and FF~\cite{Artru:1989zv},
\beq
\delta_T P_{qq}(x) = \delta_T K_{qq}(x)
= -2 + \frac{2}{(1 - x)_+} + \frac{3}{2} \delta(1 - x).
\eeq
The cancellation of the $1 / \epsilon^2$ poles in \eq{eq:W+-,+-, NLO} is highly nontrivial:
it shows the consistency of setting $\phi_h$ to $\pi$ in evaluating the singular piece of \eq{eq:qq-W,+-,+--phi-int}.
For the virtual correction $[\Delta_1^V W]^{+-}_{+-}$, 
the framework set up in \eq{eq:W-def-int-X0} relates it to the amplitude of the backward $q\to q$ scattering,
where we have chosen the azimuthal angle of the outgoing quark spinor as $\pi$ 
when explicitly evaluating \eqs{eq:a-q-amplitude-X0-LO}{eq:a-q-amplitude-X0-NLO}.
Had we chosen it as $\phi_0 = \pi + \Delta \phi_0$ with $\Delta \phi_0 \neq 0$, 
the $X = 0$ contribution to $[W]^{+-}_{+-}$ would become, using \eq{eq:W-def-int-X0},
\begin{align}
	[W_{X=0}]^{+-}_{+-}(\phi_0)
	&\propto
	\vv{ q_+(\pi, \phi_0) | S | q_+, \gamma^*_+ }
	\vv{ q_-(\pi, \phi_0) | S | q_-, \gamma^*_- }^*	\nn\\
	&= \vv{ q_+(\pi, \pi) | e^{i J_3 \Delta\phi_0} S | q_+, \gamma^*_+ }
	\vv{ q_-(\pi, \pi) | e^{i J_3 \Delta\phi_0} S | q_-, \gamma^*_- }^* \nn\\
	&= e^{-i \Delta \phi_0}
	\vv{ q_+(\pi, \pi) | S | q_+, \gamma^*_+ }
	\vv{ q_-(\pi, \pi) | S | q_-, \gamma^*_- }^*,
\end{align}
where we used the rotation operator in the second step to relate the quark state $|q_+(\pi, \phi_0)\rangle$ to $|q_+(\pi, \pi)\rangle$,
and in the third step we used $J_3 |q_{\pm}(\pi, \pi)\rangle = \mp (1/2) |q_{\pm}(\pi, \pi)\rangle$ for it moves along the $-\z_B$ direction.
This gives the relation between two different conventions,
\beq[eq:X0-convention]
[W_{X=0}]^{+-}_{+-}(\phi_0)
= e^{-i \Delta\phi_0} [W_{X=0}]^{+-}_{+-}(\pi).
\eeq
Obviously, the helicity diagonal components $[W_{X=0}]^{\pm\pm}_{\pm\pm}$ are unaffected by this convention.
The phase in \eq{eq:X0-convention} directly corresponds to the phase in \eq{eq:qq-W,+-,+--phi-int},
where the backward scattering configuration is reached only as a limit in the integration
and so setting $\phi_h$ to $\pi$ when evaluating \eq{eq:qq-W,+-,+--phi-int} 
was not as obvious as in \eqs{eq:a-q-amplitude-X0-LO}{eq:a-q-amplitude-X0-NLO}.
Now we see that the soft cancellation demands a consistent choice:
if we use $\phi_0 \neq \pi$ in \eqs{eq:a-q-amplitude-X0-LO}{eq:a-q-amplitude-X0-NLO},
we also need to set $\phi_h$ to $\phi_0$ in \eq{eq:qq-W,+-,+--phi-int};
the net effect is an extra overall factor $e^{-i \Delta\phi_0}$
to both $[W_{X=0}]^{+-}_{+-}(\pi)$ and $[W_{X>0}]^{+-}_{+-}(\pi)$.
Of course, our discussion here is only {\it a posteriori}.
Recalling the derivation in Appendix~\ref{sec:hard-coef-helicity} of the hard coefficients in terms of helicity amplitudes, 
the choice of $\phi_0$ is not free at all, but is dictated by the spinor projector 
in the definition of the Collins-type FEC or transversity FF.

\subsubsection{Collinear subtraction}
\label{sssec:fsidis-col-sub}

The $1/\epsilon$ poles in \eqs{eq:W++,++, NLO}{eq:W+-,+-, NLO}
are due to collinear subregions associated with the initial- or final-state quark.
The corresponding subtraction terms can be obtained in a standard way
by factorizing $[W]$ at parton level into the perturbative PDF and FF.
At NLO, for the $q \to q$ NS channel, they are given as
\begin{align}
	\bigiv{\Delta_1^{\rm col} W}^{\lambda_{\gamma} \lambda_{\gamma}'}_{\lambda \lambda'}(\x, \z)
	&= \int\frac{dx'}{x'} \int dz' 
	\bigiv{\Delta_0 W}^{\lambda_{\gamma} \lambda_{\gamma}'}_{\lambda \lambda'}\!\!\pp{ \frac{\x}{x'}, \frac{\z}{z'} } \nn\\
	&\hspace{4em} \times
	\bb{ \Delta_1 f^{\lambda\lambda'}_{q/q}(x') \, \delta(1 - z') + \delta(1 - \x / x') \, \Delta_1 d^{\lambda\lambda'}_{q/q}(z') },
\end{align}
where $\iv{\Delta_0 W}$ is given in \eq{eq:W-X0-LO}, 
and the perturbative quark PDF $\Delta_1 f^{\lambda\lambda'}_{q/q}(x')$ 
and FF $\Delta_1 d^{\lambda\lambda'}_{q/q}(z')$ 
are defined in the helicity basis.
At this order, they are equal to each other, 
given by the square of the $q \to q g$ splitting amplitudes~\cite{Altarelli:1977zs},
\begin{align}
	&\Delta_1 f^{++}_{q/q}(x) = \Delta_1 f^{--}_{q/q}(x)
	= \Delta_1 d^{++}_{q/q}(x) = \Delta_1 d^{--}_{q/q}(x)
	= - \frac{S_{\epsilon}}{\epsilon} \frac{\alpha_s C_F}{2\pi} P_{qq}(x), \nn\\
	&\Delta_1 f^{+-}_{q/q}(x) = \Delta_1 f^{-+}_{q/q}(x)
	= \Delta_1 d^{+-}_{q/q}(x) = \Delta_1 d^{-+}_{q/q}(x)
	= - \frac{S_{\epsilon}}{\epsilon} \frac{\alpha_s C_F}{2\pi} \delta_T P_{qq}(x),
\end{align}
where the factor $S_{\epsilon}$ is inserted to convert them to the $\MS$ scheme~\cite{Collins:2011zzd},
\beq
S_{\epsilon} = \frac{(4\pi)^{\epsilon}}{\Gamma(1 - \epsilon)}
= 1 + \epsilon \ln(4\pi e^{-\gamma_{\rm E} } ) + \order{\epsilon^2}.
\eeq
Using \eq{eq:W-X0-LO}, we get the subtraction terms for $\bigiv{\Delta_1 W}^{++}_{++}$ and $\bigiv{\Delta_1 W}^{+-}_{+-}$,
\begin{align}
	\bigiv{\Delta_1^{\rm col} W}^{++}_{++}
	&= 4 (2\pi)^4 \, e_q^2 \, \z \, \frac{\alpha_s C_F}{2\pi} \bb{ - \frac{S_{\epsilon}}{\epsilon} }
	\bigbb{ P_{qq}(\x) \, \delta(1 - \z) + \delta(1 - \x) \, K_{qq}(\z) }, \nn\\
	\bigiv{\Delta_1^{\rm col} W}^{+-}_{+-}
	&= 4 (2\pi)^4 \, e_q^2 \, \z \, \frac{\alpha_s C_F}{2\pi} \bb{ - \frac{S_{\epsilon}}{\epsilon} }
	\bigbb{ \delta_T P_{qq}(\x) \, \delta(1 - \z) + \delta(1 - \x) \, \delta_T K_{qq}(\z) }.
	\label{eq:nlo-subtraction-term}
\end{align}
Other $[W]$ helicity structures in \eq{eq:Ws-int-finite} do not appear at LO
and have no associated subtraction terms at NLO.

Subtracting \eq{eq:nlo-subtraction-term} from \eqs{eq:W++,++, NLO}{eq:W+-,+-, NLO}
removes the collinear $1 / \epsilon$ poles and converts them to $\ln(\mu^2 / Q^2)$.
After that, we set $\epsilon$ to 0 to restore the finite results in four dimensions,
\bse[eq:Ws.sing.NLO.sub]\begin{align}
	[\Delta_1^{\rm sub} W]^{++}_{++} 
	&= [\Delta_1 W]^{++}_{++} - [\Delta_1^{\rm col} W]^{++}_{++}  \nn\\
	&= 4 (2\pi)^4 \z \, e_q^2 \, \frac{\alpha_s C_F}{2\pi}
	\bigg[\!
	-8 \, \delta(1 - \x) \delta(1 - \z)
	+ \frac{1 + \x^2 \z^2}{(1 - \x)_+ (1 - \z)_+}
	\nn\\
	&\hspace{9em}  
	- A(\x) \, \delta(1 - \z)
	- \delta(1 - \x) \, B(\z)
	\bigg],
	\label{eq:W++,++, NLO, sub} \\
	[\Delta_1^{\rm sub} W]^{+-}_{+-} 
	&= [\Delta_1 W]^{+-}_{+-} - [\Delta_1^{\rm col} W]^{+-}_{+-}  \nn\\
	&= 4(2\pi)^4 \z \, e_q^2 \, \frac{\alpha_s C_F}{2\pi}
	\Big[\!
	-8 \, \delta(1 - \x) \delta(1 - \z)
	\nn\\
	&\hspace{9em}  
	- A'(\x) \, \delta(1 - \z) 
	- \delta(1 - \x) \, B'(\z)
	\Big],
	\label{eq:W+-,+-, NLO, sub}
\end{align}\ese
where we defined
\begin{align}
	A(\x) &= P_{qq}(\x) \ln \frac{\mu^2}{Q^2}
	+ (1 + \x^2) \bb{ \frac{\ln \x}{1 - \x} - \pp{ \frac{\ln (1 - \x)}{1 - \x}}_+ }, \nn\\
	B(\z) &= K_{qq}(\z) \ln \frac{\mu^2}{Q^2}
	- (1 + \z^2) \bb{ \frac{\ln \z}{1 - \z} + \pp{ \frac{\ln (1 - \z)}{1 - \z}}_+ }, \nn\\
	A'(\x) &= \delta_T P_{qq}(\x) \ln \frac{\mu^2}{Q^2}
	+ 2 \x \bb{ \frac{\ln \x}{1 - \x} - \pp{ \frac{\ln (1 - \x)}{1 - \x}}_+ }, \nn\\
	B'(\z) &= \delta_T K_{qq}(\z) \ln \frac{\mu^2}{Q^2}
	- 2 \z \bb{ \frac{\ln \z}{1 - \z} + \pp{ \frac{\ln (1 - \z)}{1 - \z}}_+ }.
\end{align}
We have used the prefix $\Delta_1^{\rm sub}$ in \eq{eq:Ws.sing.NLO.sub} to stand for the collinear-subtracted NLO contribution,
which is to be added to the LO result in \eq{eq:W-X0-LO} to give the final hard coefficients at NLO.

The parity relation in \eq{eq:Ws-int-div} is preserved by the collinear subtraction terms and therefore continues to hold for the subtracted components,
\beq[eq:Ws-int-div-sub]
\bigiv{ \Delta_1^{\rm sub} W }^{++}_{++} = \bigiv{ \Delta_1^{\rm sub} W }^{--}_{--} , \quad
\bigiv{ \Delta_1^{\rm sub} W }^{+-}_{+-} = \bigiv{ \Delta_1^{\rm sub} W }^{-+}_{-+} .
\eeq

\subsection{Observables}
\label{ssec:obs-pt-int}
Up to NLO, all the $\iv{W}$ components have been found to be real.
Combining \eqs{eq:Ws-int-finite}{eq:Ws-int-div} with \eq{eq:hard-H-denmtx-W}, 
we obtain the nonzero $\iv{H}$ components,%
\begin{align}
	\iv{H}_{\pm \pm}
	&= \frac{2}{Q^2 (1 - \varepsilon)} \pp{ \iv{F_{UU, T}} + \varepsilon \iv{F_{UU, L}} \pm P_e \sqrt{1 - \varepsilon^2} \iv{F_{LL}} }, \nn\\
	\iv{H}_{+ -}
	&= \iv{H}_{- +}
	= - \frac{2}{Q^2 (1 - \varepsilon)} \pp{ \varepsilon \bigiv{F_{UT}^{(2-)}} + \sqrt{2\varepsilon (1 + \varepsilon)} \bigiv{F_{UT}^{(1-)}} },
	\label{eq:H-result-int-nlo}
\end{align}
where we used the same $F$ notations as \eqss{eq:F-coef-U}{eq:F-coef-L}{eq:F-coef-T}
to make correspondence of the $\varepsilon$ coefficients,
\begin{align}
	\iv{F_{UU, T}}
	&= \frac{1}{2} \pp{ \iv{W}^{++}_{++} + \iv{W}^{++}_{--} }, 
	\quad
	\iv{F_{UU, L}} = \iv{W}^{00}_{++},
	\nn\\
	\iv{F_{LL}}
	&= \frac{1}{2} \pp{ \iv{W}^{++}_{++} - \iv{W}^{++}_{--} }, 
	\nn\\
	\bigiv{F_{UT}^{(2-)}}
	&= \frac{1}{2} \iv{W}^{+-}_{+-}, 
	\quad
	\bigiv{F_{UT}^{(1-)}}
	= \frac{1}{\sqrt{2}} \iv{W}^{+0}_{+-}.
\end{align}
The bracketed versions here, however, are no longer associated with any $\phi_h$ modulations.
Using \eqs{eq:C-delC-def-helicity}{eq:T-def-helicity}, we obtain the NS hard coefficients at NLO,
\bse[eq:C-del-C-T-NLO]\begin{align}
	\iv{ C_{qq} } 
	&%
	= \frac{2}{Q^2 (1 - \varepsilon)} 
	\Bigpp{ \iv{F_{UU, T}} + \varepsilon \iv{F_{UU, L}} }, \\
	\iv{ \Delta C_{qq} } 
	&%
	= \frac{2}{Q^2 (1 - \varepsilon)} 
	\Bigpp{ P_e \sqrt{1 - \varepsilon^2} \iv{F_{LL}} }, \\
	\bigiv{ T_{qq}^{ij} } 
	&%
	= \frac{2}{Q^2 (1 - \varepsilon)} \, (- \delta^{ij})
	\Bigpp{ \varepsilon \bigiv{F_{UT}^{(2-)}} + \sqrt{2\varepsilon (1 + \varepsilon)} \bigiv{F_{UT}^{(1-)}} }.
\end{align}\ese

Defining the normalization for $[F_X]$,
\beq[eq:F-norm]
\iv{ F_X } = 2 (2\pi)^4 \, e_q^2 \iv{ f_X },
\eeq
we can combine the LO and NLO calculations, in \eqss{eq:W-X0-LO}{eq:qq-Wreg-phi-int}{eq:Ws.sing.NLO.sub}, to get 
\bse\begin{align}
	\bigcc{ \iv{ f_{UU, T} }, \, \iv{ f_{LL} } }
	&= \bb{1 - \frac{4\alpha_s C_F}{\pi} } e_q^2 \, \delta(1 - \x) \, \delta(1 - \z)
	+ e_q^2 \, \frac{\alpha_s C_F}{2 \pi} \z 
	\bigg[ \frac{1 + \x^2 \z^2}{(1 - \x)_+ (1 - \z)_+}\nn\\
	&\hspace{4em} 
	\pm (1 - \x)(1 - \z) - A(\x) \, \delta(1 - \z) - \delta(1 - \x) \, B(\z) 
	\bigg], \\
	\bigiv{ f_{UT}^{(2-)} }
	&= \bb{1 - \frac{4\alpha_s C_F}{\pi} } e_q^2 \, \delta(1 - \x) \, \delta(1 - \z) \nn\\
	&\hspace{4em}
	+ e_q^2 \, \frac{\alpha_s C_F}{2 \pi} \z 
	\bigg[ 
	A'(\x) \, \delta(1 - \z) 
	+ \delta(1 - \x) \, B'(\z)
	\bigg], \\
	\cc{ \iv{ f_{UU, L} }, \, \bigiv{ f_{UT}^{(1-)} } }
	&= e_q^2 \, \frac{\alpha_s C_F}{\pi} \z 
	\cc{ 2 \x \z, \, \sqrt{\frac{\x \z}{(1 - \x) (1 - \z)}} }.
\end{align}\ese

Substituting \eq{eq:C-del-C-T-NLO} into \eq{eq:Rc-factorize-d} and then into \eq{eq:sidis-xsec-e-d}, 
we have the $\bm{p}_{hT}$-integrated differential cross section,
\begin{align}
	\frac{d\Sigma}{dx \, dQ^2 \, d\phi_S \, dz \, d\eta \, d\phi}
	&= \frac{y^2}{z \, Q^4} \frac{\alpha_e^2}{(1 - \varepsilon)}
	\Big\{ \Cint{f}{\unpfec}{ f_{UU, T} } + \varepsilon \, \Cint{f}{\unpfec}{ f_{UU, L} } 	 
	\nn\\
	&\hspace{2em} {}	
	+ s_T \Bigpp{ \varepsilon \, \Cint{h}{\colfec^{\perp}}{f_{UT}^{(2-)} } 
		+ \sqrt{2\varepsilon (1 + \varepsilon)} \, \Cint{h}{\colfec^{\perp}}{f_{UT}^{(1-)} } 
	} \sin(\phi - \phi_S) 
	\nn\\
	&\hspace{2em} {}	
	+ P_N P_e \sqrt{1 - \varepsilon^2} \, \Cint{g}{\unpfec}{f_{LL}} 
	\Big\} \nn\\
	&\hspace{1em}
	+ \mathcal{O}\bigpp{\qf / Q, \qf / p_{h, {\rm c.m.}}^-},
	\label{eq:xsec-int-e-d}
\end{align}
where we defined the notation $\mathcal{C}'$ for the $\bm{p}_{hT}$-integrated convolution,
\begin{align}
	\Cint{f}{\fec}{K} 
	&= \Cint{f}{\fec}{K}(Q^2, x, z; \qf) \nn\\
	&\equiv 
	\int \frac{d\xi_1}{\xi_1} \int d\xi_2 \, f_{q/p}(\xi_1) \, \fec_{1, h/q}(\xi_2, \qf) \, K(x/\xi_1, z/\xi_2; Q^2 / \mu^2).
	\label{eq:conv-fin-pt-int}
\end{align}
Compared with \eq{eq:fsidis-TRc-result} for a finite $p_{hT}$,
we see that \eq{eq:sidis-xsec-e-d} achieves a great simplification.
Unpolarized scattering cross section gives a handle for the unpolarized FEC,
\beq[eq:xsec-unp]
\frac{d\Sigma^{\rm unp.}}{dx \, dQ^2 \, d\phi_S \, dz \, d\eta \, d\phi}
= \frac{y^2}{z \, Q^4} \frac{\alpha_e^2}{(1 - \varepsilon)} \,
\Cint{f}{\unpfec}{ f_{UU, T} + \varepsilon f_{UU, L} },
\eeq
which can also be probed via the double helicity asymmetry,
\beq[eq:DSA-l]
A_{LL}
= \frac{d\Sigma(P_N, P_e) - d\Sigma(P_N, -P_e)}{d\Sigma(P_N, P_e) + d\Sigma(P_N, -P_e)}
= P_N P_e \,  
\frac{ \sqrt{1 - \varepsilon^2} \, \Cint{g}{\unpfec}{f_{LL}} }{ \Cint{f}{\unpfec}{ f_{UU, T} + \varepsilon f_{UU, L} } },
\eeq
along with the polarized PDF $g_{q/p}$.
As for the extraction of the Collins-type FEC, one needs to use
the single transverse spin asymmetry of the target 
along with the $\phi$ modulation,
\beq[eq:SSA-T]
A_{UT}
= \frac{d\Sigma(\phi_S) - d\Sigma(\phi_S + \pi)}{d\Sigma(\phi_S) + d\Sigma( \phi_S + \pi)}
= s_T \sin(\phi - \phi_S) \,  
\frac{ \Cint{h}{\colfec^{\perp}}{\varepsilon f_{UT}^{(2-)} + \sqrt{2\varepsilon (1 + \varepsilon)} f_{UT}^{(1-)} } 
}{ \Cint{f}{\unpfec}{ f_{UU, T} + \varepsilon f_{UU, L} } },
\eeq
where the transverse spin magnitude $s_T$ is held fixed while its direction is flipped in constructing the asymmetry.
In contrast to \eq{eq:fsidis-TRc-result}, there is only one such $\phi$ modulation in this case.
This is because the spin transfer matrix $T$ is an identity, cf.~\eq{eq:C-del-C-T-NLO}.
As noted below \eq{eq:C-del-C-T-X0} for the case of $|X\rangle = |0\rangle$, 
one may heuristically interpret this as that by integrating over $\bm{p}_{hT}$, 
the quark can be thought of as being scattered in the backward direction,
and its transverse spin is bent with respect to the electron scattering plane.%
\footnote{However, one should keep in mind that this picture does not correspond to actual event analysis in experiments,
	where $\phi$ is measured with respect to the hadron production plane which varies from event to event. 
	Only $\phi_S$ is measured with respect to the fixed electron scattering plane in the Breit frame.}

Before closing this section, we would make one final remark.
Unlike \eq{eq:fsidis-TRc-result} , which is valid to all orders, 
the discussion in this section is primarily confined to NLO.
What shall we expect beyond NLO (for the NS channel)? 
First of all, \eqs{eq:Ws-int-finite}{eq:Ws-int-div} contain all the non-vanishing $\iv{W}$ components,
which have all shown up at NLO and entered the cross section in \eq{eq:xsec-int-e-d}.
The only qualitatively new feature that can arise is the imaginary part of $\iv{ W }^{+0}_{+-}$. That is, the last identity in \eq{eq:Ws-int-finite} should be changed to
\begin{align}
	\iv{ W }^{+0}_{+-} = - \iv{ W }^{-0}_{-+} = \bigpp{ \iv{ W }^{0+}_{-+} }^* = - \bigpp{ \iv{ W }^{0-}_{+-} }^*,
	\label{eq:W+0-imag}
\end{align}
while all the other $\iv{W}$ components stay real.
This will induce an imaginary part to $\iv{H}_{\pm\mp}$, which become
\begin{align}
	\iv{H}_{\pm\mp}
	&= - \frac{2}{Q^2 (1 - \varepsilon)} 
	\cc{ \frac{\varepsilon}{2} \iv{ W }^{+-}_{+-} + \sqrt{\varepsilon (1 + \varepsilon)} \Re \iv{W}^{+0}_{+-}
		\pm i \, P_e \sqrt{\varepsilon (1 - \varepsilon)} \Im \iv{W}^{+0}_{+-}
	}	\nn\\
	&= - \frac{2}{Q^2 (1 - \varepsilon)} 
	\cc{ \varepsilon \bigiv{F_{UT}^{(2-)}} + \sqrt{2\varepsilon (1 + \varepsilon)} \bigiv{F_{UT}^{(1-)}}
		\pm i \, P_e \sqrt{2 \varepsilon (1 - \varepsilon)} \bigiv{F_{LT}^{(1-)}}
	},
\end{align}
instead of \eq{eq:H-result-int-nlo}.
The spin transfer matrix $T$ will then generate an off-diagonal element,
\beq
\iv{T}_{qq}
= \begin{pmatrix}
	\Re\iv{H}_{+-} & \Im\iv{H}_{+-}\\
	-\Im\iv{H}_{+-} & \Re\iv{H}_{+-}
\end{pmatrix}
\eeq
which will contribute an additional modulation $\cos(\phi - \phi_S)$ to \eq{eq:xsec-int-e-d}.
It requires a longitudinal beam polarization and consequently leads to a double spin asymmetry $A_{LT}$, following \eq{eq:SSA-T}.
This imaginary part, $\Im \iv{W}^{+0}_{+-}$, can arise at NNLO from a threshold loop,
so is expected to be suppressed by $\order{\alpha_s^2}$.
There are no other observables expected to arise beyond NNLO;
higher orders only induce quantitative corrections.

\section{Summary}
\label{sec:summary}

Parton fragmentation dynamics in the transverse direction 
encodes a large amount of the nonperturbative hadronization information,
and has been extensively studied in the literature with various types of observables.
In this paper, we give a detailed study of the FEC
in the SIDIS.
As a hybrid of nonperturbative parton FF
and IRC-safe energy flow operator, 
the FEC provides a powerful observable to probe the transverse fragmentation dynamics.

We have introduced a modification of the original definition in \refcite{Liu:2024kqt}
to equip the FEC with a boost-invariant property,
which makes the global extraction of FECs from different experiments more consistent 
and allows us to study two types of SIDIS observables in this paper: 
	one with a large transverse momentum of the observed hadron
	and the other with the transverse momentum inclusively integrated over.
After arguing for an all-order factorizations of the FEC in both cases, 
we performed a leading- or next-to-leading-order calculation and confirmed the consistency.
Especially, we have given a detailed discussion of an azimuthal singularity in the perturbative calculation 
in the second type of observables,
which is characteristic of the transverse spin transfer phenomenon in SIDIS and is presented here for the first time.

A great advantage of FEC lies in its collinear factorization
which shares a parallel structure as the normal SIDIS.
Theoretically, this allows one to directly copy the hard coefficient calculations from SIDIS.
Experimentally, this provides an observable that can be directly obtained 
via a straightforward analysis or re-analysis of the SIDIS experiments that have been or are being analyzed.
One only needs to record the energy flow next to the measured hadron, without any extra measurements.
The same SIDIS kinematics applies, whether one integrates over the hadron transverse momentum or not.
The extraction of FEC can also be done in a parallel way to the FF.

Without measuring the spin of the final state hadron, 
the azimuthal asymmetry of the energy flow around the hadron provides an extra handle for the transverse spin of the fragmenting quark.
This asymmetry, encoded in the Collins-type FEC, is a particular focus of this paper.
It is an analogy of the transverse-momentum-dependent Collins function, 
both being sensitive to the chiral symmetry breaking effect in QCD,
but the FEC is more easily measured and reveals complementary dynamic information.
We have given a thorough discussion of the azimuthal modulations 
that are associated with the measurement of such FEC. 
Our formalism also applies to lepton-lepton and hadron-hadron collisions in a similar way.
The universality of FECs allows them to be extracted in a global analysis.
We will present our phenomenology study in a separate paper.

\acknowledgments
We thank Xiaohui Liu, Ian Moult, Aditya Pathak, and Feng Yuan for helpful discussions.
This work is supported in part by the National Natural Science Foundation of China under contract No. 12425505 and 12235001, and The Fundamental Research Funds for the Central Universities, Peking University.
Zhite Yu is supported in part by the U.S. Department of Energy (DOE) Contract No.~DE-AC05-06OR23177, 
under which Jefferson Science Associates, LLC operates Jefferson Lab,
and C.-P.~Yuan is supported by the U.S. National Science Foundation under Grant No.~PHY-2310291.

\appendix
\input{appendix}

\newpage

\bibliographystyle{JHEP}
\bibliography{biblio.bib}

\end{document}

%% file: appendix.tex
\section{Hard coefficients in terms of helicity amplitudes}
\label{sec:hard-coef-helicity}

 Here, we present the definitions of the hard coefficients 
$C_{qq}$, $\Delta C_{qq}$, and $T_{qq}^{ij}$ in \eq{eq:Rc-factorize} 
for the quark-to-quark channel, $(a, b) = (q, q)$.

\begin{figure}[htbp]
	\centering
	\includegraphics[scale=0.6]{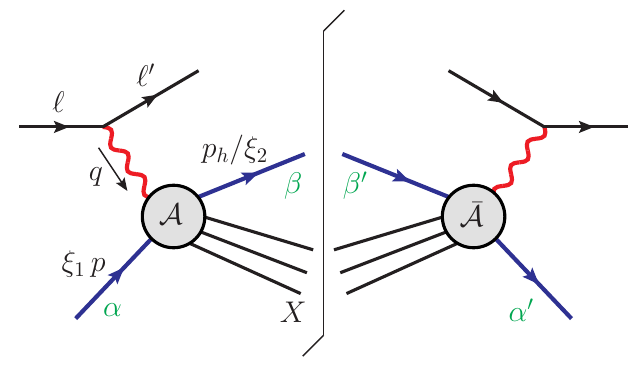}
	\caption{Hard scattering of the factorization formulae in \eqs{eq:sidis-factorization}{eq:Rc-factorize}.}
\label{fig:sidis-hard}
\end{figure}

Denote $\A_{\beta\alpha}(\l, k_1, \l, k_2; k_X)$ as the amplitude of the partonic scattering,
\beq[eq:partonic-scattering-qq]
	e(\ell) + q(k_1) \to e(\ell') + q(k_2) + X(k_X), 
\eeq
amputated on the external quark legs, 
whose momenta $k_1$ and $k_2$ have been projected on shell as in \eq{eq:partonic-momenta}.
(Recall that we use massless hadron approximation.)
Then the hard factor resulted from the factorization of region $R_C$ in \fig{fig:region}(a) is given by
\beq
	H_{\beta\alpha, \alpha'\beta'}
	= H_{\beta\alpha, \alpha'\beta'}(\ell, k_1, \ell', k_2)
	= \int d\Pi_{X} (2\pi)^4 \delta^{(4)}(q + k_1 - k_2 - k_X)
		\A_{\beta\alpha}
		\bar{\A}_{\alpha'\beta'},
\eeq
which inclusively sums over $X$ and its phase space $\Pi_X$, and where $\bar{\A} = \gamma^0 \A^{\dag} \gamma^0$.
This is shown in \fig{fig:sidis-hard}.
$H_{\beta\alpha, \alpha'\beta'}$ also  includes the subtraction of collinear singularities associated with the initial- and final-state quarks;
however, this will not affect  the discussion that follows.

The hard coefficients $C_{qq}$, $\Delta C_{qq}$, and $T_{qq}^{ij}$ 
are obtained from the projectors in \eq{eq:spinor-projector},
given by $H_{\beta\alpha, \alpha'\beta'}$ traced with certain spinor matrices,
\bse[eq:sidis-hard-coef-def]\begin{align}
	C_{qq}(x / \xi_1, z / \xi_2, Q^2, \bm{p}_{hT}) &\equiv 
		\bigpp{ \gamma \cdot k_2 }_{\beta'\beta} \, 
		H_{\beta\alpha, \alpha'\beta'} 
		\biggpp{ \frac{\gamma\cdot k_1}{2} }_{\alpha\alpha'},	\\
	\Delta C_{qq}(x / \xi_1, z / \xi_2, Q^2, \bm{p}_{hT}) &\equiv 
		\bigpp{ \gamma \cdot k_2 }_{\beta'\beta} \,
		H_{\beta\alpha, \alpha'\beta'} 
		\pp{ \frac{\gamma_5 \gamma\cdot k_1}{2} }_{\alpha\alpha'}, \\
	T_{qq}^{ij}(x / \xi_1, z / \xi_2, Q^2, \bm{p}_{hT}) &\equiv 
		\bigpp{ \gamma\cdot k_2 \gamma^i \gamma_5 }_{\beta'\beta} \,
		H_{\beta\alpha, \alpha'\beta'} 
		\pp{ \frac{\gamma\cdot k_1 \gamma^j \gamma_5}{2} }_{\alpha\alpha'},
	\label{eq:sidis-hard-coef-def-T}
\end{align}\ese
where repeated spinor indices are summed over.
They all take real values by Hermiticity.
As remarked below \eq{eq:transversity}, the transverse indices $i$ and $j$ are defined in different coordinate systems,
which complicates actual calculations.
This can be avoided by expressing the hard coefficients in terms of helicity amplitudes,
and then $C_{qq}$, $\Delta C_{qq}$, and $T^{ij}$ will be manifested as different helicity structures.

We first note the following identities for any on-shell spinors 
$u(k, \lambda)$ and $\bar{u}(k, \lambda)$ of a lightlike momentum $k$ and helicity $\lambda$, 
\begin{align}
	\gamma \cdot k 
		&= u(k, +) \, \bar{u}(k, +) + u(k, -) \, \bar{u}(k, -),	\nn\\
	\gamma_5 \gamma \cdot k 
		&= u(k, +) \, \bar{u}(k, +) - u(k, -) \, \bar{u}(k, -).
\label{eq:spinor-conversion}
\end{align}
This motivates us to define the helicity amplitude for \eq{eq:partonic-scattering-qq},
\beq[eq:partonic-scattering-amp-qq]
	\A_{\lambda_1 \lambda_2}(\l, k_1, \l, k_2; k_X)
	\equiv 
	\bar{u}_{\beta} (k_2, \lambda_2) \, \A_{\beta\alpha}(\l, k_1, \l, k_2; k_X) \, u_{\alpha}(k_1, \lambda_1)
\eeq
and the corresponding hard factor in the helicity space,
\begin{align}
	H_{\lambda_1 \lambda_2, \lambda_1'\lambda_2'} 
	& \equiv 
	\int d\Pi_{X} (2\pi)^4 \delta^{(4)}(q + k_1 - k_2 - k_X)
		\A_{\lambda_1 \lambda_2}
		\A^*_{\lambda_1' \lambda_2'} \nn\\
	&= 
	u_{\beta'} (k_2, \lambda_2') \bar{u}_{\beta} (k_2, \lambda_2) 
		\, H_{\beta\alpha, \alpha'\beta'} \,
		u_{\alpha}(k_1, \lambda_1) \bar{u}_{\alpha'}(k_1, \lambda_1').
\label{eq:hard-H-helicity-space}
\end{align}
By \eq{eq:spinor-conversion}, we can then rewrite $C_{qq}$ and $\Delta C_{qq}$ as
\bse[eq:C-delC-def-helicity]\begin{align}
	C_{qq} &= \frac{1}{2} \pp{ H_{++, ++} + H_{--, --} + H_{+-, +-} + H_{-+, -+} }, \\
	\Delta C_{qq} &= \frac{1}{2} \pp{ H_{++, ++} - H_{--, --} + H_{+-, +-} - H_{-+, -+} }.
\end{align}\ese
Clearly, the $C_{qq}$ sums over both initial- and final-state spins,
while $\Delta C_{qq}$ only sums over the final-state spin 
and incorporates the helicity difference of the initial-state quark.

The Dirac projectors in $T_{qq}$ are chirally odd, and therefore mix spinors of different helicities.
Their exact expressions, like \eq{eq:spinor-conversion}, depend on the relative phase between $u(k, +)$ and $\bar{u}(k, -)$.
We adopt the convention of the $\gamma$ matrices as given in \refcite{Peskin:1995ev} 
and take as part of the definition for $u(k, +)$ and $\bar{u}(k, -)$ the identity,
\begin{align}
	u(k, +) \, \bar{u}(k, -) = \frac{1}{2}\gamma \cdot k \pp{ \gamma^x + i \gamma^y } \gamma_5,
	\quad
	u(k, -) \, \bar{u}(k, +) = \frac{1}{2}\gamma \cdot k \pp{ \gamma^x - i \gamma^y } \gamma_5,
\label{eq:spinor-conversion-T}
\end{align}
the two of which are related by Hermitian conjugation.
\eq{eq:spinor-conversion-T} can be readily checked if $k$ moves along the $z$ direction.
For $k$ moving along an arbitrary direction described by $(\theta, \phi)$, 
its spinors $u(k, \pm)$ are defined by rotating both sides of \eq{eq:spinor-conversion-T}
first by $\theta$ around the $y$ axis and then by $\phi$ around the $z$ axis,
in exactly the same way as for single-particle states~\cite{Weinberg:1995mt}.
This also shows how the $x$ and $y$ indices on the right-hand sides of \eq{eq:spinor-conversion-T}
are defined,
\beq
	\hat{y}_k = \frac{\hat{z} \times \bm{k}}{|\hat{z} \times \bm{k}|}, \quad
	\hat{x}_k = \hat{y}_k \times \frac{\bm{k}}{|\bm{k}|},
\eeq
where $\hat{z}$ is the unit vector along the lab-frame $z$ axis, 
and $\hat{x}_k$ and $\hat{y}_k$ are the unit vectors defining the $x$ and $y$ indices associated with the momentum $k$.
See also \refcite{Yu:2023shd} for a detailed review of this.
Using \eq{eq:spinor-conversion-T}, we can then rewrite the four components of \eq{eq:sidis-hard-coef-def-T} as
\bse[eq:T-def-helicity]\begin{align}
	T_{qq}^{xx}
		&= \frac{1}{2} \pp{ H_{++, --} + H_{--, ++} + H_{+-, -+} + H_{-+, +-} }, \\
	T_{qq}^{yy}
		&= \frac{1}{2} \pp{ H_{++, --} + H_{--, ++} - H_{+-, -+} - H_{-+, +-} }, \\
	T_{qq}^{xy}
		&= \frac{-i}{2} \pp{ H_{++, --} - H_{--, ++} + H_{+-, -+} - H_{-+, +-} }, \\
	T_{qq}^{yx}
		&= \frac{i}{2} \pp{ H_{++, --} - H_{--, ++} - H_{+-, -+} + H_{-+, +-} }.
\end{align}\ese

By the simple Hermiticity property encoded in \eq{eq:hard-H-helicity-space},
\beq
	H_{\lambda_1 \lambda_2, \lambda_1'\lambda_2'} 
	= H^*_{\lambda_1'\lambda_2', \lambda_1 \lambda_2},
\eeq
we easily see that the expressions in \eqs{eq:C-delC-def-helicity}{eq:T-def-helicity} are all real functions,
in agreement with the remark below \eq{eq:sidis-hard-coef-def}.

\section{Angular integral in $d$ dimensions}
\label{sec:angular-d-dim}
The discussion here follows Appendix B.3.2 of Schwartz~\cite{Schwartz_2013} and is an analytic continuation from integer-dimensional space.

A vector $\bm{r}_d$ in $d$ dimensions can be written as
\beq
	\bm{r}_d = r \, \bm{n}_d 
		= r \, (n_d^1, n_d^2, \cdots, n_d^{d}),
\eeq
where $r = \sqrt{\bm{r}_d \cdot \bm{r}_d} \geq 0$ and $\bm{n}_d$ is a unit vector in $d$ dimensions.
Since $|\bm{n}_d|^2 = 1$, it can be parametrized by $(d-1)$ angular variables $(\phi_1, \cdots, \phi_{d-1})$ that span the hypersphere $S_{d-1}$.
The first few nontrivial examples are
\begin{align}
	\bm{n}_2 &= (\cos\phi_1, \, \sin\phi_1), \nn\\
	\bm{n}_3 &= (\sin\phi_2 \cos\phi_1, \, \sin\phi_2 \sin\phi_1, \, \cos\phi_2), \nn\\
	\bm{n}_4 &= (\sin\phi_3 \sin\phi_2 \cos\phi_1, \, \sin\phi_3 \sin\phi_2 \sin\phi_1, \, \sin\phi_3 \cos\phi_2, \, \cos\phi_3 ), 
	\quad \ldots
\end{align}
from which one can easily deduce the iterative formula,
\beq[eq:nd-iterate]
	\bm{n}_d = \pp{ \sin\phi_{d-1} (\bm{n}_{d-1}), \, \cos\phi_{d-1} }, \quad
	\mbox{with} \quad
	d \geq 2 
	\mbox{ and }
	\bm{n}_2 = (\cos\phi_1, \, \sin\phi_1).
\eeq
For $d = 2$, the angle $\phi_1$ should go from $0$ to $2\pi$ to cover the whole circle, 
such that the two components $n_2^1$ and $n_2^2$ can have all possible signs: $(n_2^1, n_2^2) = (+, +), \, (-, -), \, (+, -)$, and $(-, +)$.
As one builds $\bm{n}_d$ upon the iteration in \eq{eq:nd-iterate}, 
since the internal $\bm{n}_{d-1}$ has covered the whole hypersphere $S_{d-2}$, 
one only needs to require $\cos\phi_{d-1}$ to cover $[-1, 1]$ while keeping $\sin\phi_{d-1} \geq 0$,
i.e., $\phi_{d-1} \in [0, \pi]$ (otherwise having $\phi_{d-1} \in [0, 2\pi)$ gives double-counted directions).
That is, we have
\beq
	\phi_1 \in [0, \, 2\pi), \quad
	\phi_i \in [0, \, \pi], \quad
	\mbox{ for }
	i = 2, \cdots, d-1.
\eeq
From \eq{eq:nd-iterate}, one can get a general explicit expression for $\bm{n}_d = (n_d^1, n_d^2, \cdots, n_d^{d})$,
\beq[eq:nd-explicit]
	n_d^1 = \cos\phi_1 \prod_{k = 2}^{d-1} \sin\phi_k, \quad
	n_d^2 = \sin\phi_1 \prod_{k = 2}^{d-1} \sin\phi_k, \quad
	n_d^j = \cos\phi_{j-1} \prod_{k = j}^{d-1} \sin\phi_k, \quad
	j = 3, \cdots, d,
\eeq
with the convention that $\prod_{k = a}^b c_k = 1$ for any quantity $c$ if $a > b$.%
\footnote{Note that the convention for $\phi_1$ is different from all the other $\phi_j$. 
If we defined $\bm{n}_2 = (\sin\phi_1, \, \cos\phi_1)$, then \eq{eq:nd-explicit} would be simpler,
\begin{equation*}
	n_d^j = \cos\phi_{j-1} \prod_{k = j}^{d-1} \sin\phi_k, \quad
	j = 1, \cdots, d.
\end{equation*}
with the understanding that $\phi_0 = 0$.
}

The integration over $\bm{r}_d$ can be written as an integration over $r$ and $(d-1)$-dimensional integration over the angles,%
\footnote{Note that the notation for $\Omega_d$ differs from Schwartz's~\cite{Schwartz_2013} in that here $\Omega_d$ refers to the solid angle in $S_d$
whereas in~\cite{Schwartz_2013} the $\Omega_d$ refers to the solid angle in $S_{d-1}$. 
}
\begin{align}
	\int d^d \bm{r}_d 
		&= \int dr \, r^{d-1}  \int d\Omega_{d-1} 	\nn\\
		&= \int dr \, r^{d-1}  \int_0^{2\pi} d\phi_1 \int_0^{\pi} d\phi_2 \cdots \int_0^{\pi} d\phi_{d-1} \, J_{d-1}(\phi_1, \cdots, \phi_{d-1}).
\end{align}
Here $J_{d-1}$ is the Jacobian associated with the $(d-1)$-dimensional angular integral.
It can be obtained by an explicit calculation,
\begin{align}
	 J_{d-1}
		& = \frac{1}{r^{d-1}}{\rm abs} 
			\begin{vmatrix}
				\partial \bm{r}_d / \partial r \\
				\partial \bm{r}_d / \partial \phi_{d-1} \\
				\partial \bm{r}_d / \partial \phi_{d-2} \\
				\cdots \\
				\partial \bm{r}_d / \partial \phi_1
			\end{vmatrix}
		= {\rm abs} 
			\begin{vmatrix}
				\sin\phi_{d-1} (\bm{n}_{d-1}) 	& \cos\phi_{d-1} \\
				\cos\phi_{d-1} (\bm{n}_{d-1})	& -\sin\phi_{d-1} \\
				\sin\phi_{d-1} \partial \bm{n}_{d-1} / \partial \phi_{d-2} & 0 \\
				\cdots & \cdots \\
				\sin\phi_{d-1} \partial \bm{n}_{d-1} / \partial \phi_{1} & 0 \\
			\end{vmatrix}	,
\end{align}
where we have singled out the last column in the determinant.
Factorizing all the coefficients of $\bm{n}_{d-1}$ in the other columns gives
\begin{align}
	 J_{d-1}
		& = \abs{ \sin^{d-1}\!\phi_{d-1} \cos\phi_{d-1}}  \, {\rm abs} 
			\begin{vmatrix}
				\bm{n}_{d-1} 	& \cot\phi_{d-1} \\
				\bm{n}_{d-1}	& -\tan\phi_{d-1} \\
				\partial \bm{n}_{d-1} / \partial \phi_{d-2} & 0 \\
				\cdots & \cdots \\
				\partial \bm{n}_{d-1} / \partial \phi_{1} & 0 \\
			\end{vmatrix}	,
\end{align}
which allows us to subtract the second row from the first to obtain a recursion relation,
\begin{align}
	 J_{d-1}
		= \abs{ \frac{\sin^{d-1}\!\phi_{d-1} \cos\phi_{d-1}}{ \sin\phi_{d-1} \cos\phi_{d-1}} }  \, \, {\rm abs} 
			\begin{vmatrix}
				\bm{n}_{d-1} \\
				\partial \bm{n}_{d-1} / \partial \phi_{d-2} \\
				\cdots \\
				\partial \bm{n}_{d-1} / \partial \phi_{1} \\
			\end{vmatrix}		
		= \sin^{d-2}\!\phi_{d-1} J_{d-2},
\end{align}
where 
\beq
	J_{d-1} = J_{d-1}(\phi_1, \cdots, \phi_{d-1}), \quad
	J_{d-2} = J_{d-2}(\phi_1, \cdots, \phi_{d-2}).
\eeq
The initial value of this recursion can be easily calculated to be $J_1 = 1$.
Then, we obtain the Jacobian
\beq
	J_{d-1}(\phi_1, \cdots, \phi_{d-1})
		= \sin^{d-2}\!\phi_{d-1} \sin^{d-3}\!\phi_{d-2} \cdots \sin\phi_{2},
\eeq
and hence the angular integration in $S_{d-1}$,
\beq[eq:Omega-d-1]
	\int d\Omega_{d-1} 
		= \int_0^{2\pi} d\phi_1 \int_0^{\pi} d\phi_2 \cdots \int_0^{\pi} d\phi_{d-1} \, 
			\sin^{d-2}\!\phi_{d-1} \sin^{d-3}\!\phi_{d-2} \cdots \sin\phi_{2}.
\eeq

For the purpose of studying azimuthal modulation in dimensional regularization,
we keep $\phi_1 \equiv \phi$ and integrate over all other angles.
This gives
\beq[eq:Omega-d-to-1]
	\int d\Omega_{d-1} = \frac{ \pi^{\frac{d-2}{2}} }{\Gamma\pp{ \frac{d}{2} }} \int_{0}^{2\pi} d\phi.
\eeq
Applying this to the transverse momentum $\bm{p}_T$ integration,
we note that $\bm{p}_T$ has $(d-2) = (2-2\epsilon)$ dimensions in dimensional regularization, 
and contains $(d-3) = (1-2\epsilon)$ angular variables.
Therefore, we have
\begin{align}
	\int d^{d-2} \bm{p}_T 
		&= \int_0^{\infty} d p_T \, p_T^{d-3} \int d\Omega_{d-3} 
		= \int_0^{\infty} d p_T \, p_T^{d-3} \bb{ \frac{ \pi^{\frac{d-4}{2}} }{\Gamma\!\pp{ \frac{d-2}{2} }} \int_{0}^{2\pi} d\phi }	\nn\\
		&= \frac{\pi^{-\epsilon}}{2 \Gamma(1 - \epsilon)}
			\int_0^{\infty} d p_T^2 \, (p_T^2)^{-\epsilon} \int_{0}^{2\pi} d\phi.
\label{eq:kT-to-phi}
\end{align}
This yields the formula used in \eq{eq:qq-hadronic-tensor-phi-int}.